%% file: 3year_paper_v2.6.tex
\documentclass[aps,prd,reprint,superscriptaddress,nofootinbib]{revtex4-1}

\usepackage{graphicx}   %% this helps recognize png's
\usepackage{epstopdf}   %% this helps recognize eps's
\usepackage[T1]{fontenc}
\usepackage{color}

\usepackage{sidecap}
\sidecaptionvpos{figure}{c} 

\usepackage{subfigure}   

\usepackage{multirow}
\usepackage{booktabs}

\usepackage{mdwlist}

\usepackage{amsmath,amssymb} 
\usepackage{placeins}
\usepackage{rotating}
\usepackage{mathtools}

\usepackage{wrapfig}

\usepackage{setspace}

\usepackage{comment}
\usepackage{hyperref}

\usepackage{array}

\input{define}

\begin{document}

\title{Cosmic Ray Spectrum and Composition from PeV to EeV 
Using 3 Years of Data From IceTop and IceCube}

\date{\today:~~VERSION 2.65} % delete this line to display the current date

\begin{abstract}

We report on measurements of the all-particle cosmic ray energy spectrum and composition in the PeV to EeV energy range using three years of data from the IceCube Neutrino Observatory.
The IceTop detector measures cosmic ray induced air showers on the surface of the ice, from which the energy spectrum of cosmic rays is determined by making additional assumptions about the mass composition.  
A separate measurement is performed when IceTop data are analyzed in coincidence with the high-energy muon energy loss information from the deep in-ice IceCube detector. 
In this measurement, both the spectrum and the mass composition of the primary cosmic rays are simultaneously reconstructed using a neural network trained on observables from both detectors.
The performance and relative advantages of these two distinct analyses are discussed, including the systematic uncertainties and the dependence on the hadronic interaction models, and both all-particle spectra as well as individual spectra for elemental groups are presented.

\end{abstract}


\author{}

%\thanks{}
%\affiliation{}
%\collaboration{IceCube Collaboration}
\noaffiliation

%\keywords{}


\affiliation{III. Physikalisches Institut, RWTH Aachen University, D-52056 Aachen, Germany}
\affiliation{Department of Physics, University of Adelaide, Adelaide, 5005, Australia}
\affiliation{Dept. of Physics and Astronomy, University of Alaska Anchorage, 3211 Providence Dr., Anchorage, AK 99508, USA}
\affiliation{Dept. of Physics, University of Texas at Arlington, 502 Yates St., Science Hall Rm 108, Box 19059, Arlington, TX 76019, USA}
\affiliation{CTSPS, Clark-Atlanta University, Atlanta, GA 30314, USA}
\affiliation{School of Physics and Center for Relativistic Astrophysics, Georgia Institute of Technology, Atlanta, GA 30332, USA}
\affiliation{Dept. of Physics, Southern University, Baton Rouge, LA 70813, USA}
\affiliation{Dept. of Physics, University of California, Berkeley, CA 94720, USA}
\affiliation{Lawrence Berkeley National Laboratory, Berkeley, CA 94720, USA}
\affiliation{Institut f{\"u}r Physik, Humboldt-Universit{\"a}t zu Berlin, D-12489 Berlin, Germany}
\affiliation{Fakult{\"a}t f{\"u}r Physik {\&} Astronomie, Ruhr-Universit{\"a}t Bochum, D-44780 Bochum, Germany}
\affiliation{Universit{\'e} Libre de Bruxelles, Science Faculty CP230, B-1050 Brussels, Belgium}
\affiliation{Vrije Universiteit Brussel (VUB), Dienst ELEM, B-1050 Brussels, Belgium}
\affiliation{Dept. of Physics, Massachusetts Institute of Technology, Cambridge, MA 02139, USA}
\affiliation{Dept. of Physics and Institute for Global Prominent Research, Chiba University, Chiba 263-8522, Japan}
\affiliation{Dept. of Physics and Astronomy, University of Canterbury, Private Bag 4800, Christchurch, New Zealand}
\affiliation{Dept. of Physics, University of Maryland, College Park, MD 20742, USA}
\affiliation{Dept. of Astronomy, Ohio State University, Columbus, OH 43210, USA}
\affiliation{Dept. of Physics and Center for Cosmology and Astro-Particle Physics, Ohio State University, Columbus, OH 43210, USA}
\affiliation{Niels Bohr Institute, University of Copenhagen, DK-2100 Copenhagen, Denmark}
\affiliation{Dept. of Physics, TU Dortmund University, D-44221 Dortmund, Germany}
\affiliation{Dept. of Physics and Astronomy, Michigan State University, East Lansing, MI 48824, USA}
\affiliation{Dept. of Physics, University of Alberta, Edmonton, Alberta, Canada T6G 2E1}
\affiliation{Erlangen Centre for Astroparticle Physics, Friedrich-Alexander-Universit{\"a}t Erlangen-N{\"u}rnberg, D-91058 Erlangen, Germany}
\affiliation{Physik-department, Technische Universit{\"a}t M{\"u}nchen, D-85748 Garching, Germany}
\affiliation{D{\'e}partement de physique nucl{\'e}aire et corpusculaire, Universit{\'e} de Gen{\`e}ve, CH-1211 Gen{\`e}ve, Switzerland}
\affiliation{Dept. of Physics and Astronomy, University of Gent, B-9000 Gent, Belgium}
\affiliation{Dept. of Physics and Astronomy, University of California, Irvine, CA 92697, USA}
\affiliation{Dept. of Physics and Astronomy, University of Kansas, Lawrence, KS 66045, USA}
\affiliation{SNOLAB, 1039 Regional Road 24, Creighton Mine 9, Lively, ON, Canada P3Y 1N2}
\affiliation{Department of Physics and Astronomy, UCLA, Los Angeles, CA 90095, USA}
\affiliation{Dept. of Astronomy, University of Wisconsin, Madison, WI 53706, USA}
\affiliation{Dept. of Physics and Wisconsin IceCube Particle Astrophysics Center, University of Wisconsin, Madison, WI 53706, USA}
\affiliation{Institute of Physics, University of Mainz, Staudinger Weg 7, D-55099 Mainz, Germany}
\affiliation{Department of Physics, Marquette University, Milwaukee, WI, 53201, USA}
\affiliation{Institut f{\"u}r Kernphysik, Westf{\"a}lische Wilhelms-Universit{\"a}t M{\"u}nster, D-48149 M{\"u}nster, Germany}
\affiliation{Bartol Research Institute and Dept. of Physics and Astronomy, University of Delaware, Newark, DE 19716, USA}
\affiliation{Dept. of Physics, Yale University, New Haven, CT 06520, USA}
\affiliation{Dept. of Physics, University of Oxford, Parks Road, Oxford OX1 3PQ, UK}
\affiliation{Dept. of Physics, Drexel University, 3141 Chestnut Street, Philadelphia, PA 19104, USA}
\affiliation{Physics Department, South Dakota School of Mines and Technology, Rapid City, SD 57701, USA}
\affiliation{Dept. of Physics, University of Wisconsin, River Falls, WI 54022, USA}
\affiliation{Dept. of Physics and Astronomy, University of Rochester, Rochester, NY 14627, USA}
\affiliation{Oskar Klein Centre and Dept. of Physics, Stockholm University, SE-10691 Stockholm, Sweden}
\affiliation{Dept. of Physics and Astronomy, Stony Brook University, Stony Brook, NY 11794-3800, USA}
\affiliation{Dept. of Physics, Sungkyunkwan University, Suwon 16419, Korea}
\affiliation{Dept. of Physics and Astronomy, University of Alabama, Tuscaloosa, AL 35487, USA}
\affiliation{Dept. of Astronomy and Astrophysics, Pennsylvania State University, University Park, PA 16802, USA}
\affiliation{Dept. of Physics, Pennsylvania State University, University Park, PA 16802, USA}
\affiliation{Dept. of Physics and Astronomy, Uppsala University, Box 516, S-75120 Uppsala, Sweden}
\affiliation{Dept. of Physics, University of Wuppertal, D-42119 Wuppertal, Germany}
\affiliation{DESY, D-15738 Zeuthen, Germany}

\author{M. G. Aartsen}
\affiliation{Dept. of Physics and Astronomy, University of Canterbury, Private Bag 4800, Christchurch, New Zealand}
\author{M. Ackermann}
\affiliation{DESY, D-15738 Zeuthen, Germany}
\author{J. Adams}
\affiliation{Dept. of Physics and Astronomy, University of Canterbury, Private Bag 4800, Christchurch, New Zealand}
\author{J. A. Aguilar}
\affiliation{Universit{\'e} Libre de Bruxelles, Science Faculty CP230, B-1050 Brussels, Belgium}
\author{M. Ahlers}
\affiliation{Niels Bohr Institute, University of Copenhagen, DK-2100 Copenhagen, Denmark}
\author{M. Ahrens}
\affiliation{Oskar Klein Centre and Dept. of Physics, Stockholm University, SE-10691 Stockholm, Sweden}
\author{C. Alispach}
\affiliation{D{\'e}partement de physique nucl{\'e}aire et corpusculaire, Universit{\'e} de Gen{\`e}ve, CH-1211 Gen{\`e}ve, Switzerland}
\author{K. Andeen}
\affiliation{Department of Physics, Marquette University, Milwaukee, WI, 53201, USA}
\author{T. Anderson}
\affiliation{Dept. of Physics, Pennsylvania State University, University Park, PA 16802, USA}
\author{I. Ansseau}
\affiliation{Universit{\'e} Libre de Bruxelles, Science Faculty CP230, B-1050 Brussels, Belgium}
% ...

%\date{\today}

\collaboration{IceCube Collaboration}
\email{analysis@icecube.wisc.edu}
\noaffiliation

\maketitle

Measurements of the cosmic ray energy spectrum and mass composition in the PeV to EeV energy range provide key information about the origin and propagation of cosmic rays in what is commonly considered to be a transition region from galactic to extragalactic cosmic rays.  In the present paper we report on composition and energy spectrum measurements using three years of data from the IceCube Neutrino Observatory.

The IceCube Neutrino Observatory is a versatile particle detector located at the geographic South Pole, with both surface and deeply-buried components.
The latter, or InIce, component (described in detail in \cite{IceCube_detector}) consists of 5160 Digital Optical Modules (DOMs) \cite{Abbasi:2009} deployed with 17~m spacing on 86 strings in a 125~m triangular grid formation, at depths from 1450~m to 2450~m below the surface in transparent ice.  Each DOM contains a 10~inch Hamamatsu photomultiplier tube (PMT) and electronics for signal processing and readout \cite{Abbasi:2010}. 

The surface component, IceTop (described in detail in \cite{icetop_technical}), is an array of pairs of tanks filled with water that has frozen and containing two DOMs each, operating at different PMT gains for increased dynamic range; the pairs of tanks are called stations, and each station is located above a string of the InIce detector (see Figure~\ref{f:detector_schematic}).

As Cherenkov detectors, both components are sensitive to the charged particles in the extensive air showers (EAS) produced by cosmic rays from the Southern Hemisphere sky, and both are used to extract the cosmic ray energy spectrum in different ways.
In this paper, we will discuss two methods for extracting this information: the first uses IceTop only and will be referred to as \emph{the IceTop-alone analysis}, while the other uses both the IceTop and InIce detectors in tandem, and will be referred to as \emph{the coincident analysis}.  The coincident analysis also provides a new measurement of the composition of cosmic rays.

The IceTop detector can be used alone to measure the core position, direction, and  size of the air showers at the surface.  These observables are utilized to extract a cosmic ray energy spectrum: in particular the shower size, discussed in Section~\ref{sec:IT_reco}, is strongly correlated with the energy of the incident cosmic ray primary.  
It is important to note that as the shower size depends slightly on the mass of the cosmic ray primary, the composition must be assumed in order to extract the energy spectrum when using IceTop-alone. 
Since IceTop collects a large number of showers, rare high-energy events are collected in sufficient numbers to extend the analysis into the EeV range.
This analysis was performed most recently in \cite{it73_icetopalone_spectrum}, in which one year of 73-station data was used to measure a spectrum from a few PeV to EeV.

The InIce detector measures the energy loss of the high-energy muons in the deep ice, which is strongly dependent upon the mass of the incident cosmic ray primary, as discussed in Section~\ref{sec:InIce_reco}.  
Thus, with the IceTop and InIce detectors working in tandem, the high-energy muon component of the air showers are measured in coincidence with the electromagnetic %and low-energy muon components;
component; thus both the energy spectrum \emph{and} mass composition are measured without making assumptions about one to determine the other.
However, this method requires \emph{coincident} events between the two components of IceCube: due to the long lever arm between the two arrays, only zenith angles of approximately 0-$30^\circ$ on the sky pass the coincident selection criteria, which then yield 
fewer events from the same data sample as the IceTop-alone analysis.
Thus, the coincident analysis cannot reach as high an energy as IceTop-alone.
A coincident analysis like this was performed in \cite{ic40_coincidence}, in which one month of data from the half-completed 40-station, 40-string detector was used to measure a spectrum and average logarithmic mass (\lna) from 1 to 30~PeV;
that analysis was extended with improved reconstruction techniques in \cite{it73_coincidence_icrc2013, Feusels:thesis} using one year of data from the nearly complete 73-station, 79-string array, achieving better resolution and reaching to 1~EeV.

In this paper, we will present for the first time the coincident analysis in detail.  In addition, the one-year analyses of IceTop-alone \cite{it73_icetopalone_spectrum} and coincident \cite{it73_coincidence_icrc2013, Feusels:thesis} data are extended to three years, and we report these improved and updated results.

\section{Data, Simulation, and Reconstruction}

\subsection{The 3-year Data Set} 

\begin{figure}[h]\begin{center}
\includegraphics[width=0.42\textwidth]{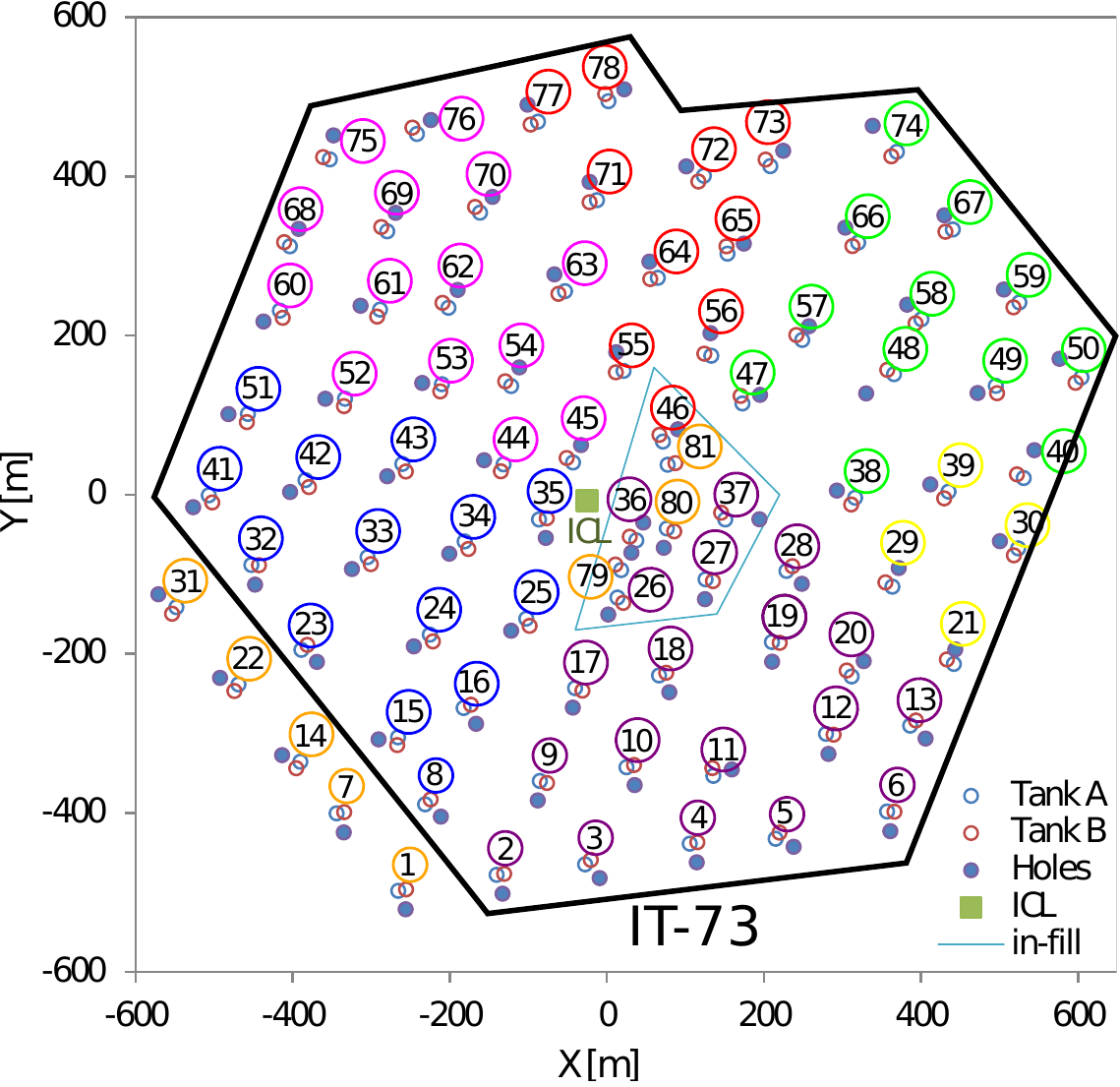}
\caption{A top view of the IceTop surface array. 
Colors indicate the construction periods for the strings and tanks. 
This work will focus on IceTop-73 (IT-73) and IceCube-79 (IC-79), which includes all detectors {\it except} those from the last construction year colored orange.
Strings 79 and 80 and stations 79, 80, and 81 are excluded from this work, even though they lie inside the border of IT-73 shown in black. \cite{Feusels:thesis}
} %based on a fig from \cite{IceTop-NIM}
\label{f:detector_schematic}
\end{center}\end{figure}

The analyses described here use three years of data, from June 1, 2010, through May 2, 2013. 
The IceTop-alone and the coincident analyses use the same dataset, and thus have both a total livetime of 977.6 days (with a negligible uncertainty of less than half a percent).
The IceCube Neutrino Observatory was not running in its complete configuration (81 stations, 86 strings) until May 14, 2011.
Therefore, the first year of data included here was taken in the incomplete 73-station, 79-string configuration (IT-73/IC-79).  The simulation used in these analyses was also produced using the IT-73/IC-79 detector configuration.
Thus, in order to handle the two following years of data in the same analysis, all information from the final 7 deployed strings and 8 stations has been removed from data processing.  The configuration used is shown in Figure~\ref{f:detector_schematic}.

\subsection{Simulation}
\label{s:simulation}

%%----- CORSIKA --------
Both of the analyses presented here use the same set of Monte Carlo simulations of cosmic ray events to establish relationships between detector observables and cosmic ray energy and mass.  (This simulated dataset was also used in \cite{it73_icetopalone_spectrum}.)
30000 air showers of four primary types (protons, helium, oxygen, and iron) were simulated using the \textsc{CORSIKA} air shower generator \cite{CORSIKA_Heck} with an $E^{-1}$ spectrum between \lte\,=\,5.0 and 8.0.
Additionally, thinned \textsc{CORSIKA} showers \cite{CORSIKA_manual, Hillas:thinning} were generated at higher energies: 
12000 showers of each type, again with an $E^{-1}$ spectrum between \lte\,=\,7.0 and 9.5. 
High-energy muons are not thinned in the algorithm.  
The range of energy overlap between \lte\,=\,7.0 and 8.0 allows for verification of the un-thinning algorithm \cite{icetop_technical}.

Two hadronic interaction models were used: FLUKA \cite{FLUKA:2006} below 80~GeV, and Sibyll 2.1 \cite{Ahn:2009} above 80~GeV. 
EGS4 \cite{EGS4:1985} was used to model the electromagnetic interactions. 
Other high energy hadronic interaction models are used for systematic studies, as discussed in Section~\ref{sec:syst_hadintmodel}. 
The zenith angle from the primary particles is sampled from a $\cos(\theta)\sin(\theta)$ distribution between 0$^\circ$ and 40$^\circ$, while the azimuth angle is drawn from a uniform distribution over the whole 2$\pi$ azimuth range. 
To be able to study atmospheric changes and apply corrections where needed, a reference atmosphere was chosen based on the MSIS-90-E parametrization \cite{MSIS:1991} of the South Pole Atmosphere on July 1, 1997, which has a ground pressure of 692.2 g/cm$^2$ at the South Pole altitude (CORSIKA atmosphere 12, \cite{CORSIKA_manual}).
To make more efficient use of the CORSIKA showers available, each shower is copied, or \emph{resampled}, 100 times, and thrown at random locations within a circle of radius $R$ centered on the center of the IceTop array.  The resampling radius is the largest possible for the shower to trigger the array \cite{icetop_technical}.

%% ------- Detector simulation
Next, the particles generated by CORSIKA are propagated into the detectors.
The individual responses of the IceTop tanks are simulated using a detailed Geant4 \cite{GEANT4:2003, GEANT4:2006} model which takes into account the individual IceTop tank properties, including snow and air above the tanks, and the detector electronics.  The resulting signal is converted into units of photoelectrons using constants unique to each tank \cite{icetop_technical}.  In this way the same calibration procedure (described in Section~\ref{sec:Calib}) 
can be performed on both simulated events and experimental data.

For the coincident analysis, the high-energy muons in the CORSIKA air showers must also be propagated through the Antarctic ice and through the deep InIce detector.
Muons with energy above 273~GeV\footnote{273~GeV is the energy at which 0.1\% of the muons are expected to reach the top of the InIce detector and could create a detectable amount of Cherenkov light within the array \cite{Feusels:thesis}} are propagated \cite{Chirkin:2004} through the Antarctic ice to the bottom of the InIce array.  Propagating the Cherenkov photons from the muons through the South Pole ice to the DOMs by directly tracking each one is computationally prohibitive.  Therefore, light profiles for~GeV emitters (including both pure Cherenkov emission and more diffuse emission from cascade light sources) are tabulated in a software package \cite{Lundberg:2007}, which includes a model of the full structure of the ice properties \cite{SPICE:2013}. The expected number of photoelectrons and their arrival times at each DOM are retrieved from the tables for each muon and for electromagnetic and hadronic cascades.  Simulated noise hits are then added.
Finally, the simulated photoelectrons are fed into a simulation of the InIce readout electronics and the detector trigger.
The simulated DOM signals then follow the same processing chain as the experimentally measured DOM signals.

\subsection{Pulse cleaning and Calibration}
\label{sec:Calib}
% Latest try
Only events which pass the cosmic ray filters~\cite{icetop_technical} and which contain at least 6 hard local coincidence (HLC\footnote{HLC hits occur when both tanks in one station are hit within 1~$\mu$s}) DOMs within 6~$\mu$s are processed further.   
In a first signal cleaning of selected events, signals from all DOMs which were determined to be unreliable at the time of data taking are removed.
The remaining signals %waveforms
are calibrated using the %DOMCal 
procedure described in~\cite{IceCube_detector}, which returns the number of photoelectrons.
At this point the IceTop and InIce data are split.  

The IceTop data is cleaned and calibrated as described in detail in~\cite{icetop_technical}. 
At this stage, only HLC hits are preserved. 
All signals within one large trigger window are then split into clusters of hits that are likely related to one air shower. 
Next, an additional noise cleaning procedure is applied to the calibrated HLC signals.  This procedure begins with a cluster of 3 hit stations and adds more tanks to the event using a simple distance-per-time requirement \cite{Feusels:thesis}. This extra cleaning procedure improves the removal of noise hits, of tanks with time-fluctuations, and of muon-like signals at large lateral distances from the shower core.  These, in turn, improve the IceTop reconstruction procedures described in Section~\ref{sec:IT_reco}.  
The specific tank response is then taken into account using a calibration procedure which transforms the signal into that expected from vertical (equivalent) muons (VEMs).

For the coincident analysis, the InIce hits causally connected to the event seen in IceTop are selected based on an allowed time difference window between IceTop and InIce hits. 
Therefore, before the IceTop and InIce events can be connected, InIce noise hits that affect the trigger time must first be removed in a procedure based on distance-per-time requirements (similar to that in IceTop).  Then, InIce triggers matching the IceTop triggers and pulses are selected, and the pulses not connected to the selected trigger are removed.  
A final noise removal procedure in the InIce detector uses the reconstruction of the track by IceTop whereby only hits connected in time to the track, and in a cylinder around it, are preserved. These steps effectively remove random coincidences between the two detectors, and events that are very close in time. 

Finally, after the above event cleaning and calibration procedure, events are required to have hits in at least 5 IceTop (IT-73) stations and, in the coincident analysis, hits in at least 8 InIce (IC-79) DOMs.

\subsection{IceTop Reconstruction}
\label{sec:IT_reco}
Cleaned data from IceTop tanks are processed by a reconstruction software package called \emph{Laputop}, which has been described in detail in \cite{icetop_technical}. 
For each event, Laputop finds the best-fit shower core position ($x_c$, $y_c$, $z_c$) and direction ($\theta$, $\phi$), as well as
two parameters describing the shape of the lateral distribution function (LDF) of deposited charge (\s, $\beta$).  
The functional form of the charge LDF is a double logarithmic parabola:

\begin{equation*}
S = S_{\mathrm{125}}\cdot\left(\frac{r}{125\,m}\right)^{-\beta-\kappa \log_{10}\left(\frac{r}{125\,m}\right)},
\end{equation*}
where $\beta$ is a measure of the steepness of the LDF and \s~is the signal expectation at a perpendicular reference distance of 125 meters from the shower axis \cite{Klepser:thesis}.  \s~will be referred to throughout this paper as the \emph{shower size}.  
$\kappa$ scales with the curvature of the parabola which is approximately constant for all hadronic showers; thus $\kappa$ is presently set as a default to 0.303, while the other two parameters are allowed to vary event by event.
The best-fit parameters are found using a 3-step maximum-likelihood technique, which compares the timing and charge of the hits to both the expected charge LDF and an expected timing LDF. %(``curvature'').  
Both saturation of the tanks, as well as stations which are \emph{not} hit, are taken into account in the likelihood.

\begin{figure*}[htb] \begin{center}
\includegraphics[width=1.0\textwidth]{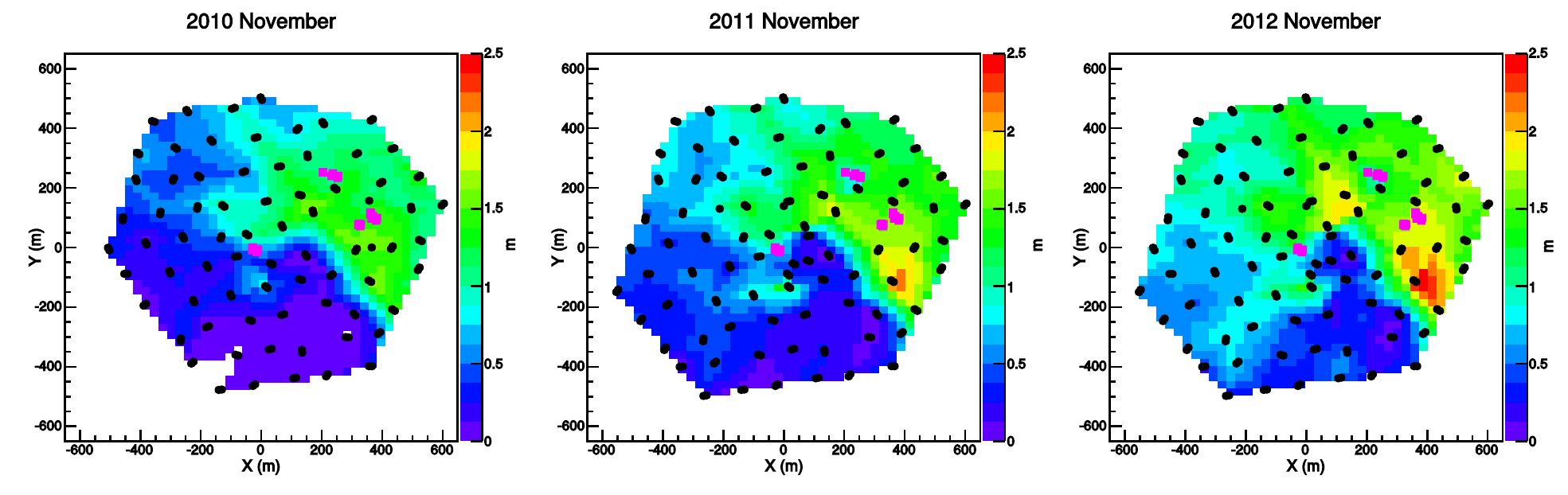}
\caption{Depths of snow covering the IceTop tanks, measured during each of the three years studied in this work.
The pink dots indicate the positions of buildings at the site. 
The z-axis of these plots represents the snow depth, measured in meters.\fixka{These plots need a z axis label!!}}
\label{f:snowmaps}
\end{center} 
\end{figure*}

%\subsubsection{Snow in IceTop}
%\label{sec:snow}
The Laputop reconstruction also takes the actual snow depths on top of the tanks into account. 
The frozen water tanks of IceTop were deployed flush with the surface of the snow at the site.
However, wind-blown snow continuously drifts over the array and covers the tanks with an overburden that increases over time and varies from tank to tank.
Figure~\ref{f:snowmaps} shows the depths of snow covering the array's tanks in each of the three years analyzed in this work: 2010, 2011 and 2012.  The site accumulates about 20~cm of snow per year on average.

The total signal $S$ observed by the two DOMs in the IceTop tank consists of snow attenuated electromagnetic ($e^\pm/\gamma$) particles and unattenuated muons.  
If the attenuation of the snow is not taken into account, the reduction in total signal amplitude will make an event look less energetic, or ``smaller'', than it really is.
To take snow attenuation into account, a simple exponential reduction S$_{red}$ is applied to the expected $S$:
\begin{equation}
S_{red} = S \cdot e^{\frac{-d}{\lambda \cos \theta}},
\label{eqn:sred}
\end{equation}
which only depends on the slant depth of snow overburden for the tank ($d/\cos \theta$), and an effective attenuation length $\lambda$, which takes into account absorption/generation in the shower and particle-type specific absorption behavior.
The same attenuation length $\lambda$ is applied throughout the array, in the same way for all showers of any size.  
However, because of the increasing snow load from year-to-year, the optimal $\lambda$ also changes from year-to-year.  
Thus each of the three years of data was optimized separately to find the $\lambda$ which best creates agreement in the \s~spectrum across different regions in the array (deeply-buried, and sparsely-buried).
These best-fit values of $\lambda$ are: 2.1 meters for 2010/11, 2.25 meters for 2011/12, and 2.25 meters for 2012/13.  
Furthermore, the snow depth, $d$, has two sources of uncertainty \cite{DeRidder:thesis}.  First, sastrugi\footnote{Sastrugi are irregular waves formed on the surface of the snow by wind erosion.} cause variations in the snow depth across a single tank.  Sastrugi heights measured in-situ were observed to follow a Gaussian distribution with a standard deviation of 4~cm.  
Second, the depth at each IceTop tank is measured twice per year, in February and in November, with an occasional third measurement in January; for all other times, a linear interpolation between these measurements is used to estimate the daily snow depth.  
Finer measurements of accumulation at the South Pole (made monthly by the Antarctic Meteorological Research Center at a site near IceTop) 
exhibit variations around a smooth interpolation with a $\sigma$ of 27\% of the difference between the November and February measurements ($\Delta H_{snow}$, in centimeters).  
The two sources of uncertainty are combined and the snow depth $d$ in Eqn.~\ref{eqn:sred} (the signal reduction due to the snow) is smeared by a Gaussian with a sigma of $\sqrt{(4\,\text{cm})^{2}+(0.27\cdot\Delta H_{snow})^{2} }$ \cite{DeRidder:thesis}.

\subsection{InIce Reconstruction}
\label{sec:InIce_reco}
\internalcite{Tom's thesis, mostly}
The IceCube detector observes the Cherenkov light pattern from the high-energy muon bundles that 
propagate through the Antarctic ice as well as their accompanying energy losses (which create Cherenkov-light-emitting cascades).  

These measured observables provide a handle on the primary composition: iron-induced showers are more muon-rich than proton-induced showers of the same energy, due to the superposition model of nucleons (detailed in e.g. \cite{KAMPERT2012660}). Therefore, iron-induced showers are more muon-rich than proton-induced showers of the same energy, therefore iron-induced showers will result in a greater \emph{overall} deposit of Cherenkov light in IceCube than proton-induced showers of the same primary energy.
Furthermore, 
proton-induced showers are more likely than iron-induced showers to have extremely high-energy muons in the bundle, which can create larger \emph{local} energy depositions from Bremsstrahlung.  Consequently, proton-induced showers are expected to create fewer but higher-energy stochastic losses in the detector than iron-induced showers (of the same primary energy).  On the other hand, since iron-induced showers have more total muons than those induced by protons (of the same primary energy), an iron-induced shower is likely to undergo more lower-energy stochastic losses in the detector than a proton-induced shower (of the same primary energy).  Therefore, counting these stochastic fluctuations (henceforth ``stochastics'') gives us additional composition-sensitive information.

In order to measure these composition-sensitive parameters for coincident events, the track position and direction as reconstructed by IceTop (with Laputop) is used, and an energy loss reconstruction algorithm (\textsc{Millipede}) uses the timing and charge information of the observed Cherenkov light in the ice to create 
a profile of energy losses along the track as a function of slant depth, with 20-meter segmentation (as discussed in detail in \cite{energy_reco_paper}).  

\begin{figure}[htb]\begin{center}

\includegraphics[width=0.48\textwidth]{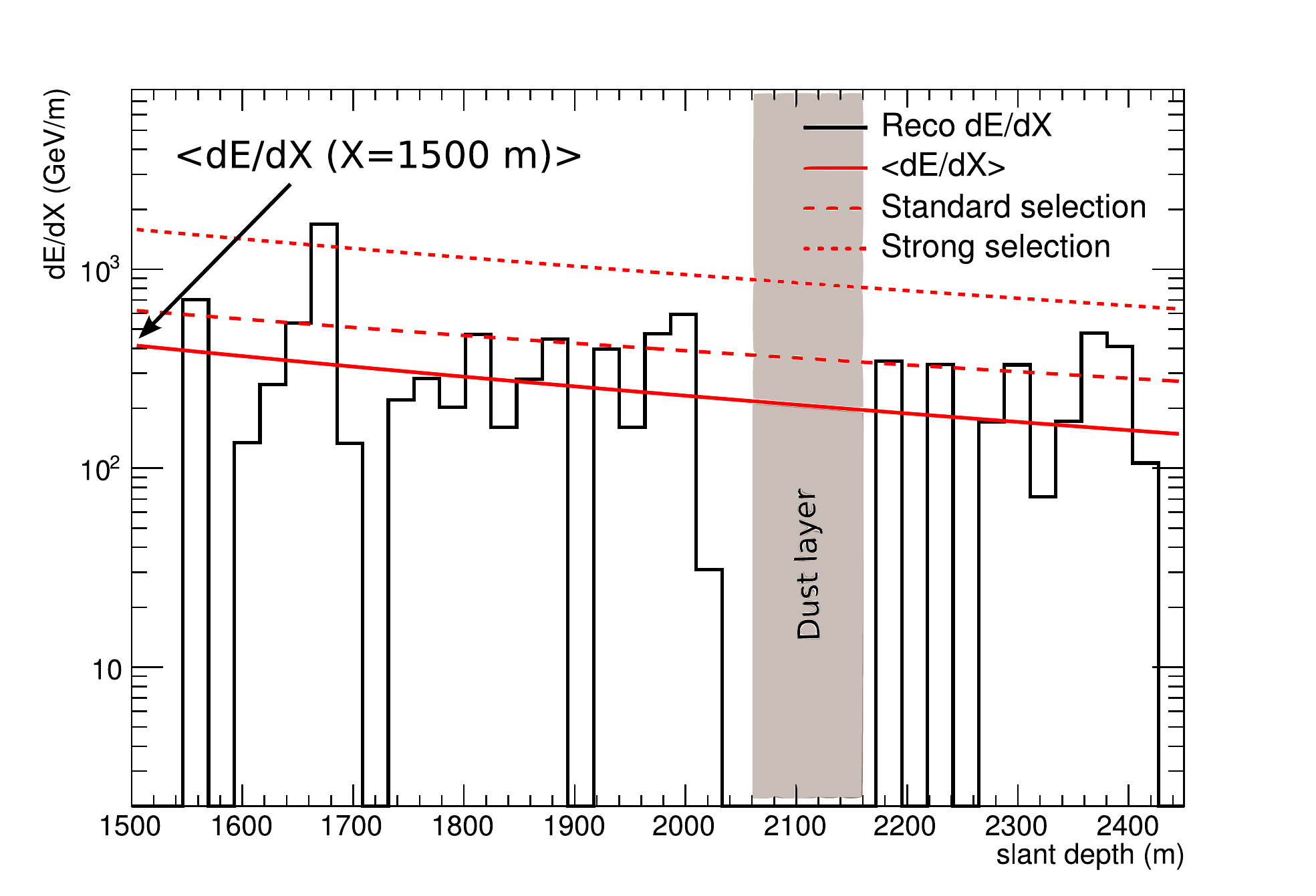}%tomsthesis/Big
\caption{
Example of the energy loss reconstruction of a large event, where the solid red line demonstrates the average energy loss fit, the dashed red line  represents the standard stochastics selection, and the dotted red line indicates the strong stochastics selection, as noted in the legend.  Note that when successive bins exceed the selection criteria they are counted individually.  (For example, around slant depth of 2350~m two bins exceed the standard selection.  These are therefore counted as two high energy stochastics.)  The gray band is the approximate location of the dust layer for the slant depth of this particular event.
\cite{Feusels:thesis, DeRidder:thesis} %event is inclined to 15 degrees, per Sam's email April 9, 2019
} 
\label{f:energy_loss_profile}
\end{center}\end{figure}

Figure~\ref{f:energy_loss_profile} shows an example energy loss profile from a cosmic ray event.
Note that sections of the energy loss profile corresponding to IceCube's dust layer\footnote{The ``dust layer'' is a distinctive thick layer of dust deposited several millennia ago and currently located $\sim$1950-2050~m beneath the surface of the ice \cite{SPICE:2013}.} (where there are consequently few photons) and near the boundaries of IceCube's volume (where reported energy losses can be irregular) are removed (as shown in gray in the figure).
The gaps seen in the energy loss profile are not segments on the muon bundle track where no energy was lost, but rather are pieces of the track that are not well-sampled by the detector.

%% ------DESCRIBE HOW WE'RE GOING TO USE INICE FOR COSMIC RAY ANALYSIS
The energy loss profile is then fit to extract two composition-sensitive parameters: a) the average energy loss behavior, which is indicated as the red line in Figure~\ref{f:energy_loss_profile}, and b) the size and quantity of deviations from that average behavior (the stochastics).  
The energy loss observable ($dE_\mu/dX$) is defined as the value of the fit to the energy loss profile at a fixed slant depth of X=1500~m, which corresponds roughly to the top of the IceCube detector (marked on the left side of Figure~\ref{f:energy_loss_profile}).

Two methods of selecting a number of high-energy stochastics from a energy loss profile are used in this work: a {\it standard selection} (marked as the red dashed line in Figure~\ref{f:energy_loss_profile} and a {\it strong selection} requiring higher stochastic energy loss (marked as the red dotted line in Figure~\ref{f:energy_loss_profile} \cite{Feusels:thesis, DeRidder:thesis}.  
The selection criterion for the stochastics is given by:

\begin{equation*}
\frac{\mathrm{d}E_\mu}{\mathrm{d}X}(X) > a \cdot  \bigg(\frac{\mathrm{d}E_\mu}{\mathrm{d}X}(X)\bigg)_{\mathrm{reco}}^{b},
\end{equation*}

\noindent where $a = 5$ and $b = 0.8$ for the standard selection, and $a = 7$ and $b = 0.9$ for the strong selection (with appropriate dimensions for $a$ and $b$).

\begin{figure*}[htp]\begin{center}
\includegraphics[width=1.0\textwidth]{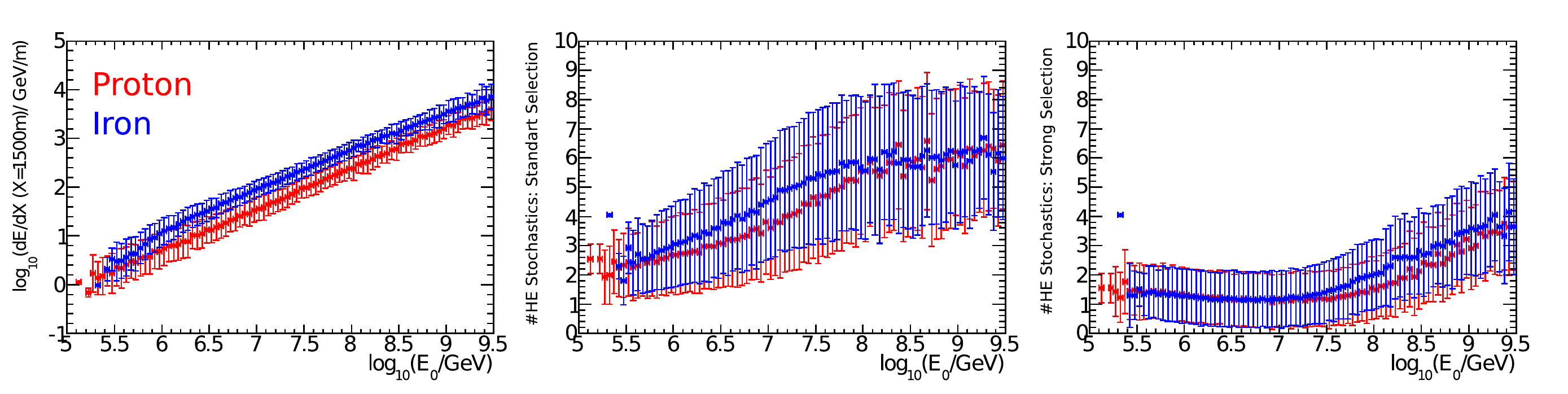}
\caption{Composition sensitivity in MC simulations (Sibyll2.1) of three InIce variables:
energy loss $\mathrm{d}E/\mathrm{d}X$ (left), the number of high energy stochastics standard selection (middle), and the number of high energy stochastics strong selection (right) as discussed in Section~\ref{sec:InIce_reco}. 
Error bars represent RMS spread of the distribution.
As the standard stochastics count begins to lose sensitivity at 100 PeV, the strong stochastics count begins to be sensitive \cite{Feusels:thesis}.  It is clear that the energy loss is the primary composition-sensitive parameter.}
%\internalcite{Tom's thesis, figures 4.26 and 4.27}}
\label{f:millipede_sensitivity}
\end{center}\end{figure*}

Figure~\ref{f:millipede_sensitivity} demonstrates the composition-sensitivity of the InIce observables reconstructed by \textsc{Millipede}.
The energy loss parameter is directly comparable to the number of high-energy muons in the air shower and is therefore the primary composition-sensitive observable, as shown in Figure~\ref{f:millipede_sensitivity}, left.
The stochastics provide additional composition-sensitivity, as shown in Figure~\ref{f:millipede_sensitivity}, center and right: 
Iron bundles have a larger number of stochastic losses since they have more muons, but the energy losses from proton bundles can be more extreme since the same total energy is transferred to fewer muons.  
The number of stochastics with the standard selection can separate masses best at low muon multiplicities, below 100 PeV.
From about 30 PeV and above, for bundles containing at least 100 muons, the stronger selection performs better.
In both cases there are more high energy cascades selected for iron bundles than for proton.

% The seasonal variation stuff.
The muon multiplicity in air showers detected by IceCube shows a seasonal variation, which is due to the semi-annual alternation between the polar day and night, and the accompanying temperature changes in the atmosphere.
The measured variation of $\log_{10}(\mathrm{d}E_\mu/\mathrm{d}X)$ is found to be 10-15\% of the proton-iron difference.
Since simulations are only performed with one atmosphere, the July,1997 atmosphere, all other months of data need to be corrected with respect to July.
This correction correlates the changing temperature profile of the entire atmosphere, weighted with the muon production depth profile, with the measured variation of $\log_{10}(\mathrm{d}E_\mu/\mathrm{d}X)$.
This energy dependent correlation factor is used as a correction.
A small but symmetric variation of $\pm3\%$ in the proton iron-space remains, which smears the data slightly.
For more details, see \cite{SeasMuon}.

\subsection{Quality Cuts}

Quality cuts are used to ensure a sample of events which will be well reconstructed.
The cuts for the IceTop-alone analysis are described in \cite{it73_icetopalone_spectrum} while the cuts for the coincident analysis are detailed here.  

%third version: Karen wrote this
\emph{IceTop Selections:}
Many of the IceTop selections revolve around the success of the Laputop reconstruction algorithm, which does an excellent job of reconstructing \emph{contained} events (those with a shower core inside the area of the array) but its performance suffers for uncontained events. 
\begin{itemize}
\setlength\itemsep{0 pt}
\item{The number of stations after cleaning is required to be $\ge 5$.}  
\item{The largest snow-corrected charge measured in any tank is required to be at least 6~VEM.}
\item{The station with the highest deposited charge is not allowed to be at the edge of the detector.%loudest station
}
\item{The neighboring tank in the same station as the tank with the largest signal must have at least 4~VEM.}
\item{The fraction of hit stations within a circle centered on the center of gravity of the shower with outer radius at the furthest hit station must be greater than 0.2.}
\item{The reconstruction algorithm Laputop is required to converge.}
\item{The LDF slope parameter $\beta$ is required to be between 1.4 and 9.5.}
\item{The core location of the air shower must be reconstructed within a scaled factor of 0.96 of the area of the array.}
\end{itemize}

\emph{InIce Selections:}
The InIce reconstruction begins with a fixed track position and direction from Laputop.  Therefore the InIce quality selections focus on ensuring an accurate reconstruction of the energy loss from \textsc{Millipede}.
\begin{itemize}
\setlength\itemsep{0 pt}
\item{The track position and direction calculated by Laputop is required to pass within the In-Ice instrumented volume.}
\item{A minimum of 8 InIce DOMs are required to be hit.} 
\item{The \textsc{Millipede} energy loss reconstruction must succeed with log$_{10}(rlogl) < 2.0$ 
and the total charge predicted (QTOT) must be at least 90\% of that measured ( log$_{10}(\frac{QTOT_{predicted}}{QTOT_{measured}}) > -0.03$ ) .}
\item{At least 3 reconstructed cascades remain after all previous selections and after removal of cascades in the dust layer and at the edge of the detector (as discussed above).}
\end{itemize}

\emph{A note about the zenith angle:}
As in \cite{it73_icetopalone_spectrum}, the IceTop-alone analysis presented here is divided into four bins of zenith angle, the steepest of which is limited to $cos(\theta) \ge 0.80$. 
No explicit cut in zenith angle is applied for the coincident analysis, since the solid angle acceptance of the two detectors together limits the zenith angle range to $cos(\theta) \sim 0.85$.

\subsection{Performance}
\label{sec_perf}
\begin{figure}
\bc
\includegraphics[width=0.98\linewidth]{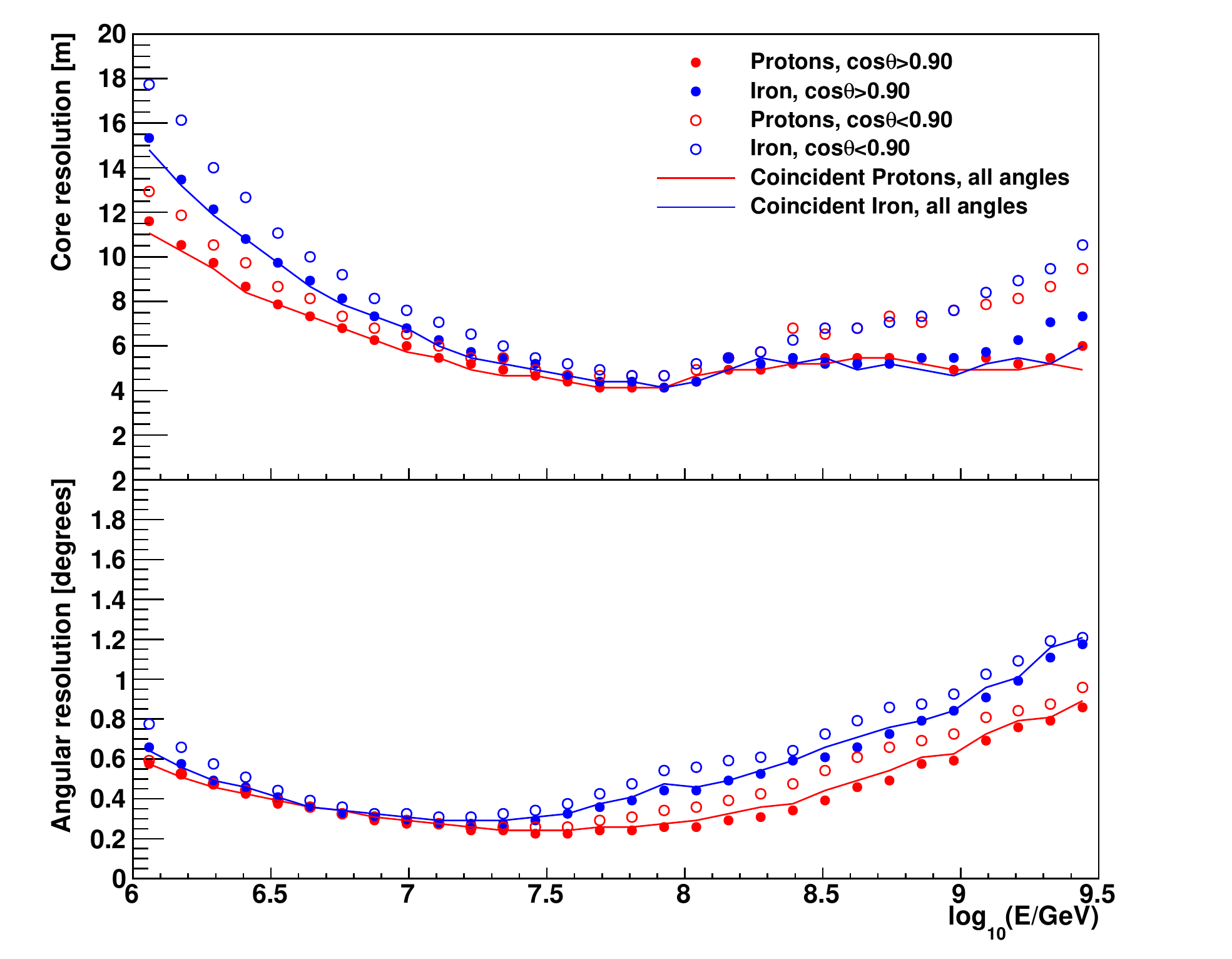}
\caption{Core position resolution (upper) and angular resolution (lower) of the reconstructed air shower (both
defined as containing 68\% of the events) as a function of the primary energy, after quality cuts.
\internalcite{Kath, see tex for details}}
\label{f:core_ang_resolution}
\includegraphics[width=0.98\linewidth]{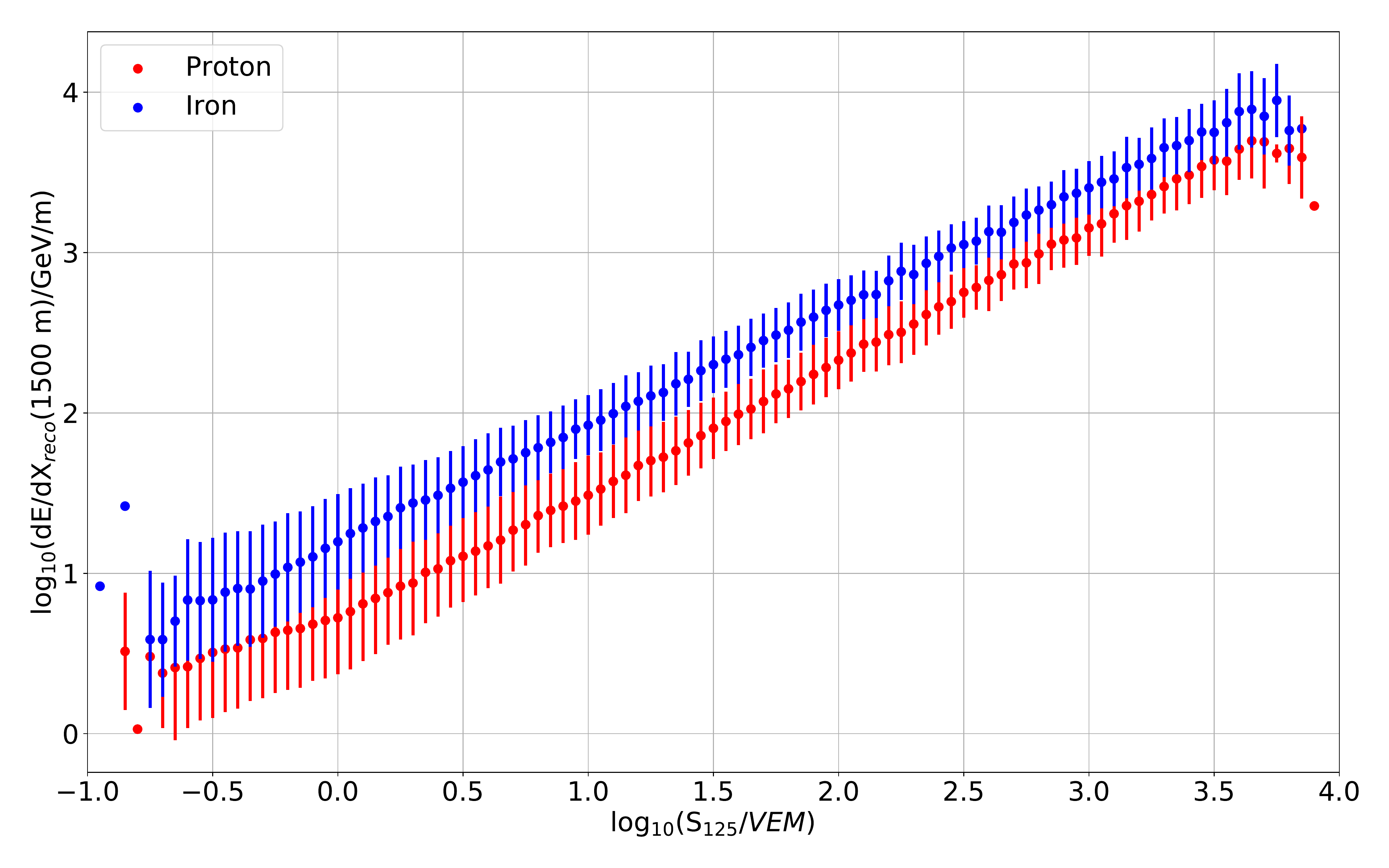}
\caption{
Reconstructed energy loss as a function of \s~for proton (red) and iron (blue) 
simulations, with the standard deviation indicated as error bars.}
\label{millipede_performance}
\ec
\end{figure}

The reconstruction procedure and quality cuts described above yield a set of events with a core position resolution of 6-20 meters, and a track direction resolution of 0.3-1.0 degrees.  
These values depend on energy: Figure~\ref{f:core_ang_resolution} shows the position and angular resolutions as a function of energy, for the coincident sample and the IceTop-alone sample (divided into high-zenith and low-zenith angles).  Note that the detector reaches 100\% efficiency for all particle types at $\sim$3~PeV (\cite{Feusels:thesis}). 
As the size of the air showers approach and exceed the size of the IceTop array and the tanks become saturated, the reconstruction becomes less precise.  

Figure~\ref{millipede_performance} shows the reconstructed energy loss from IceCube compared to the reconstructed \s~ parameter from IceTop:
in this parameter space there is clearly a strong separation between proton and iron primaries.

\section{The IceTop-alone analysis}
\label{sec:IT_alone_ana}
\begin{figure}[h]
\begin{centering}
\includegraphics[width=.48\textwidth]{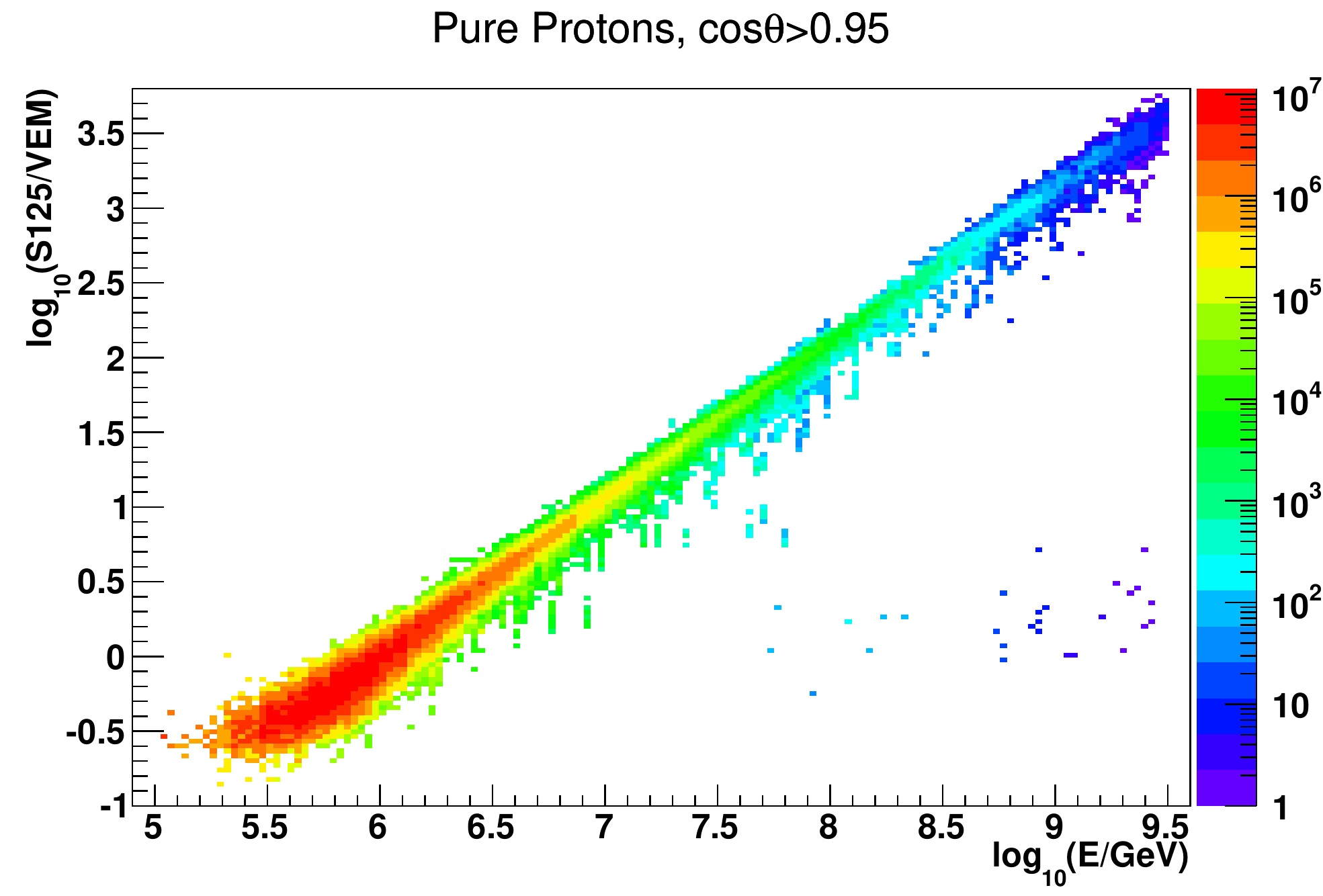}
\caption{Relationship between \s~and primary energy: 
updated version of Figure~4 from \cite{it73_icetopalone_spectrum}, for primary protons at high ($\cos(\theta) \ge 0.95$) zenith angles. 
\fix{Do we want to show the H4a-mixed version instead?}
\fix{Should label z axis.}   
\internalcite{Kath, see tex for details}}
\label{f:s_to_e_conversion}
\end{centering}
\end{figure}

The IceTop-alone analysis is sensitive to the energy spectrum of cosmic rays up to the EeV energy range.  The reconstructed shower size parameter (\s) in particular is highly correlated to the  energy of the primary air shower.  Figure~\ref{f:s_to_e_conversion} shows the relationship between \logs~to \lte.  In \cite{it73_icetopalone_spectrum}, a function relating \logs~to \lte~was derived using Monte Carlo simulations divided into four ranges of zenith angles.  For a given zenith range, the distributions of true energy for each slice of 0.05 in \logs~were weighted using the H4a model from \cite{Gaisser_H4a} as a composition assumption, and fitted with a Gaussian.  (Since neither silicon nor magnesium were simulated, but both are included in the H4a model, simulated oxygen was weighted by the sum of CNO and MgSi model components.)  The fitted mean of the Gaussian was then used as the energy estimate for that slice in \logs.  The functional form of the conversion is:
\begin{equation}
    \mathrm{log}_{10}(E/\mathrm{GeV}) =  p_0 + p_1 \mathrm{log}_{10}(S_{125}/\mathrm{VEM}) .
    \label{eqn:unfolding}
\end{equation}

Using updated simulation and reconstruction algorithms, the mapping of \logs~to \lte~from \cite{it73_icetopalone_spectrum} has been re-optimized and applied to the 3-year data set.  The updated fit parameters for Eqn.~\ref{eqn:unfolding} are in the following table:

\begin{table}[h]
\begin{tabular}{l|l|l}
\hline \hline
~Zenith range			& ~$p_0$	    & ~$p_1$	\\ \hline
~$0.95 < \cos(\theta) \le 1.0$~	& ~6.011~   & ~0.933~ \\ \hline
~$0.90 < \cos(\theta) \le 0.95$~	& ~6.055~   & ~0.924~ \\ \hline
~$0.85 < \cos(\theta) \le 0.90$~	& ~6.110~   & ~0.915~ \\ \hline
~$0.80 < \cos(\theta) \le 0.85$~	& ~6.177~   & ~0.907~ \\ \hline \hline
\end{tabular}
\caption{Fit parameters for converting IceTop shower size \s to energy in Eqn.~\ref{eqn:unfolding}, using the ``H4a'' composition assumption.  
Errors are on the order of 0.006 for $p_0$, and 0.0035 for $p_1$. }
\label{t:s_e_conversion}
\end{table}
% KR: here are all the errors, in case anyone is curious: 
% 95: 0.0060 and 0.0035
% 90: 0.0060 and 0.0035
% 85: 0.0059 and 0.0034
% 80: 0.0059 and 0.0034

%% ABOUT THE COLORFUL S-VS-E PLOT:
%Files used:
%IceTop-alone: /data/user/tfeusels/L3_L4/Level3b_ITalone/7***
%              copied to AK: /wasabi/sim_it73/Level3b_ITalone
%Cuts applied:
%IceTop-alone: graphical containment, loudest-station-not-on-edge, Q1>6, beta, cos(theta)>=0.95
%              (note: logS125>0.0 cut was NOT done, since you can clearly see them on the plot)
%See script in: sandbox/kath/composition_3year/for_3year_paper/colorful_S125_vs_E.C and colorful_S125_vs_E_finish.C
% (There was something wrong with ``SetLogz()'' on my desktop, so I had to finish the plot on a different machine.)

\begin{figure}[h]
\begin{centering}
\includegraphics[width=.48\textwidth]{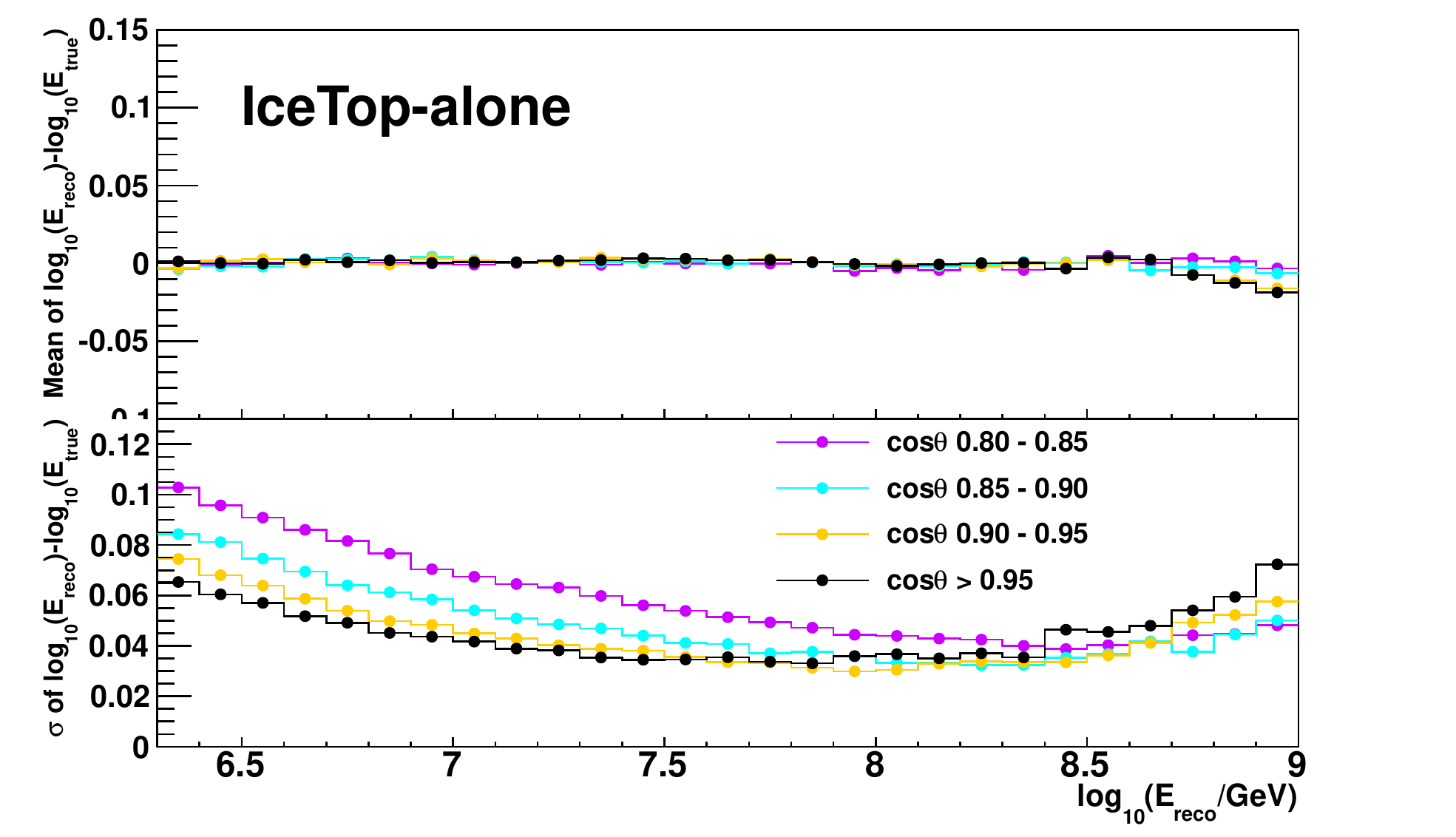}
\caption{Energy bias (upper) and resolution (lower) of the conversion functions in Table~\ref{t:s_e_conversion}, for simulated events mixed according to the H4a model in the four zenith bins.
(These are updated versions of Figures 7 and 8 from \cite{it73_icetopalone_spectrum}.)
\internalcite{Kath}}
\label{f:e_resolution_bias}
\end{centering}
\end{figure}

The energy bias and resolution of this technique  
are shown in Figure~\ref{f:e_resolution_bias}.  The reduced precision beyond 8.0 (100~PeV) is related to the reduced angular and position resolution shown in Figure~\ref{f:core_ang_resolution}, which creates an extra smearing effect in \s. 

\begin{figure}[htbp]
\begin{centering}
\includegraphics[width=.5\textwidth]{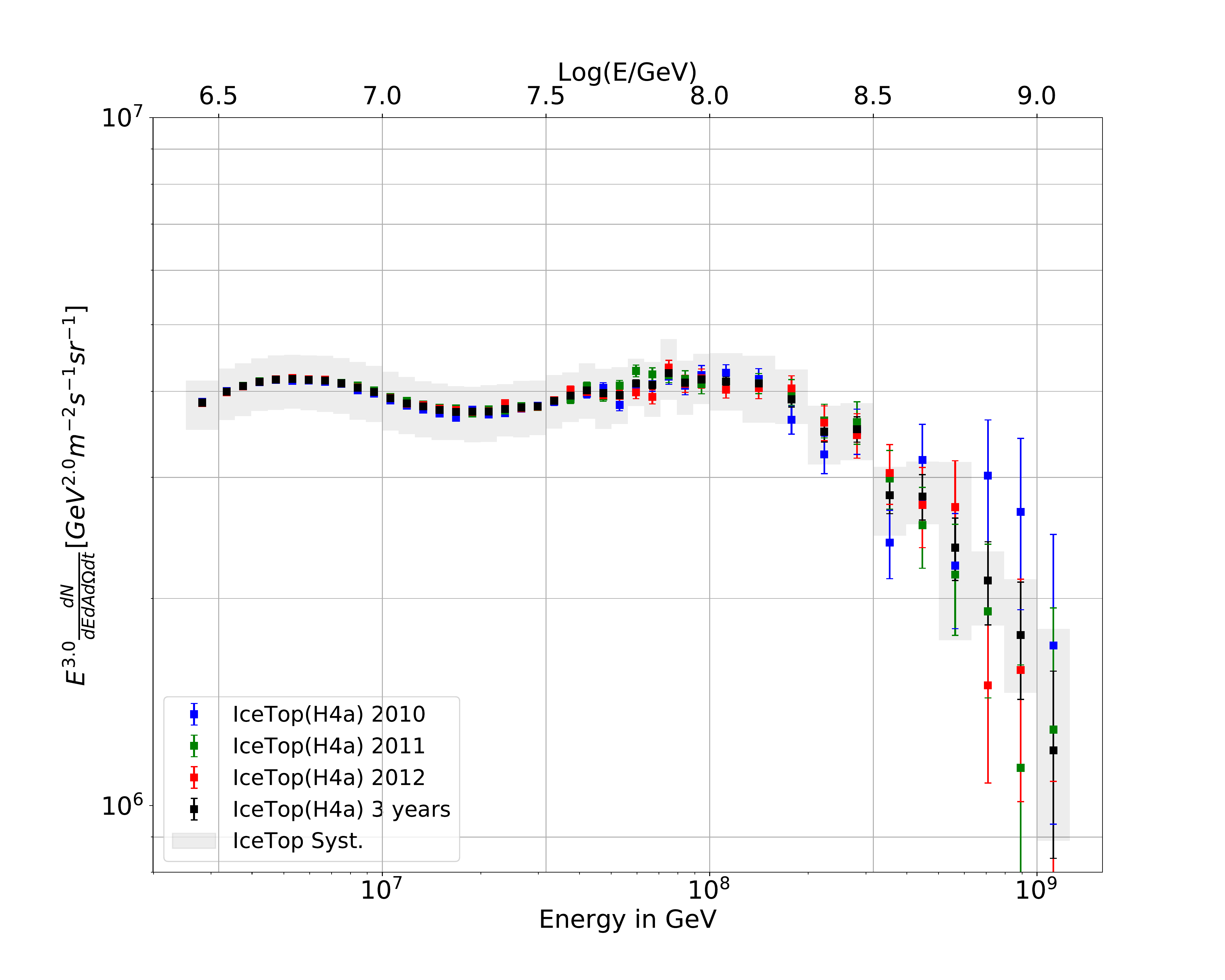} 
\caption{All-particle energy spectrum from the IceTop-alone analysis from each of the three years, and the three years together. \internalcite{Kath, January 2018}
}
\label{f:espec_italone}
\end{centering}
\end{figure}

Figure~\ref{f:espec_italone} shows the result of the IceTop-alone 3-year analysis for each year individually and combined, multiplied by a factor of $E^{3}$ to highlight the details.  The gray band represents the total systematic uncertaintity of the IceTop detector, as described in \cite{it73_icetopalone_spectrum}.  These results are consistent between the three years and are also consistent with those previously published from one year of data \cite{it73_icetopalone_spectrum}.

\section{The Coincident analysis}
\label{coinc_anal}
When the surface observables from IceTop are combined with the additional observables from the InIce detector, 
the high-energy muon component of the shower is measured in coincidence with the electromagnetic component of the EAS.  Using this coincident configuration, a mass-independent primary energy spectrum and individual elemental spectra are measured.  This technique was developed for the measurement of the cosmic ray composition of one month of IT-40/IC-40 data in the energy range between 1~PeV to 30~PeV~\cite{Andeen:thesis,ic40_coincidence} using two input variables.
Building on this experience, the technique was extended to five input variables over a wider primary energy range, optimized over a larger scan of different network types, and trained on more Monte Carlo simulated events.  This updated technique was applied to a single year of data from the nearly complete IT-73/IC-79 detector in \cite{it73_coincidence_icrc2013,Feusels:thesis}.  Here, the one year analysis is improved and further expanded to include three years of data.

\subsection{Neural Network Mapping Technique}
\label{sec:nn}

This analysis includes five variables which depend on primary energy and primary mass in a non-linear fashion:
the shower size in IceTop (\s), the zenith angle ($\cos(\theta)$), the muon energy loss in the ice ($\mathrm{d}E/\mathrm{d}X$), and the number of high-energy stochastics under two selections (standard and strong).  
There is no theoretical analytical expression that relates our input variables to primary mass and primary energy; thus, an artificial neural network (NN)\footnote{In particular, a feed-forward multilayer-perceptron\,(MLP) neural network is used from the \textsc{TMVA}~\cite{TMVA} machine learning package.} is trained on simulation to determine the relationships between the five inputs and the two outputs.  The network is strongly dependent on the two primary parameters, \s~and $\mathrm{d}E/\mathrm{d}X$, but the three other parameters do contribute to the energy and mass reconstruction.

The final high-quality sample of simulated Monte Carlo data is split into three parts. 
Half of the sample is used to generate the neural network (the network sample). 
The other half (the verification sample) is used for comparisons of data and simulation in the final analysis steps.  
The network sample is again split in two: 74357 events are used to train the network (the training sample), the remaining 67399 events (the test sample) serve to test the network and to select the network architecture and optimal activation function based on the network performance. 
Networks were trained on unweighted events; however, every Monte Carlo sample mentioned above is chosen in such a way that it contains an equal mixture of each of the four primary types (p, He, O and Fe) and covers the full energy range.  

During the first 5000 of 10000 minimizer iterations (also called cycles or epochs), only a random selection of 60\% of the training data is utilized. 
After the training converged on this random selection, 
the training continues on the full training set.

\subsection{Optimizing the Neural Network}
Many different neural network architectures were evaluated for performance before analyzing any data, as discussed in \cite{Feusels:thesis}.  In addition to networks with 5 inputs as described above, alternative networks with the 2 primary inputs (\logs\,and $\log_{10}(\mathrm{d}E/\mathrm{d}X)$\,only), 3 inputs (adding $\cos(\theta)$) and 4 inputs (adding the standard selection of high energy stochastics only) were tested.
Three groups of network structures were explored: with one, two, and three hidden layers, and the number of neurons was varied within the hidden layers.
Two activation functions (a sigmoid, and a tanh) were explored.
In total, 207 networks for each of the two activation functions and for each number of inputs (1656 networks in total) were trained on the simulations.

\begin{figure}[htbp]
\begin{center}
\includegraphics[width=0.48\textwidth]{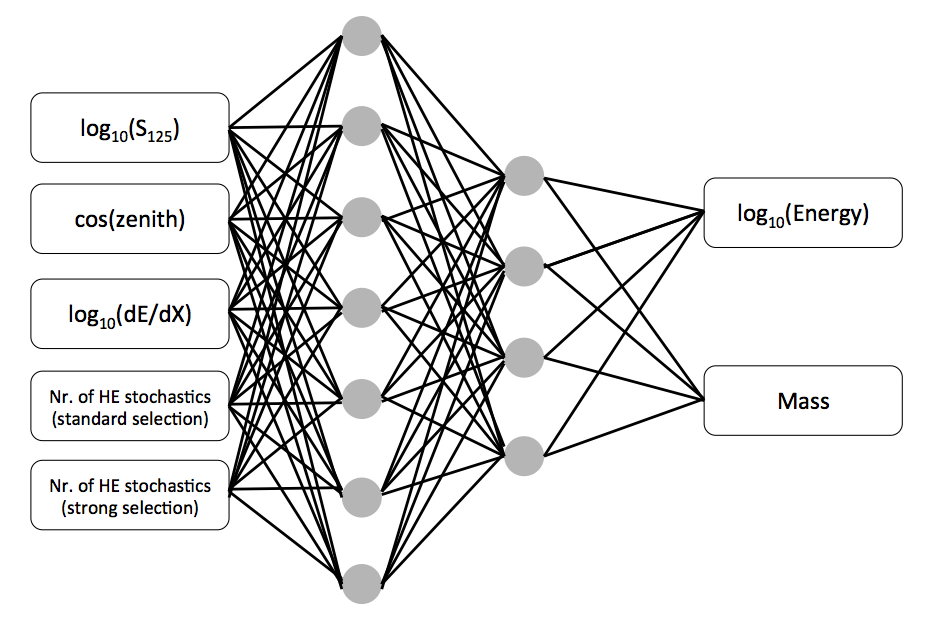}
\caption{The neural network architecture of the best performing neural network.
This network maps five input variables onto two output variables using two hidden layers with respectively seven and four neurons using a tanh activation function. 
It is therefore called a 5-7-4-2 network.}
\label{f:nn-arch}
\end{center}
\end{figure}

The performance of each network was assessed according to how well it reconstructed primary energy and primary mass.  The assessment process was optimized to find the network with the smallest and most consistent RMS spread and bias over all energies, and which had mass groups that were best-separated and most distinctive (i.e. ``peaky'').  
The final optimized network has 5 inputs, 7 neurons in a first hidden layer, 4 neurons in a second hidden layer, and 2 outputs, with a tanh activation function connecting the neurons and a linear mapping from the last layer to the output neurons.  
A schematic of this network is shown in Figure~\ref{f:nn-arch}.

It is important to note that this neural network has two target outputs which are very different in nature: the first output is a continuous energy distribution, the second target output is instead is composed of four discrete numbers corresponding to four elemental masses simulated.  Therefore, the \emph{neural network energy output} (E$_{0,reco}$) is also a continuous distribution which is expected to reproduce the true primary energy (within some bias and resolution) for each event, as discussed below in Sec. \ref{nn_ereco}.  On the other hand, the \emph{neural network mass output} results in smeared distributions around the four discrete mass numbers, which require further analysis in order to decompose the primary mass.  The mass is therefore not reconstructed on an event-by-event basis but is determined statistically for the entire data set, as discussed below in Sec. \ref{kde_explan}.
 
\subsection{Neural Network Primary Energy Reconstruction}
\label{nn_ereco}

\begin{figure}[t]\begin{center}
\includegraphics[width=0.48\textwidth]{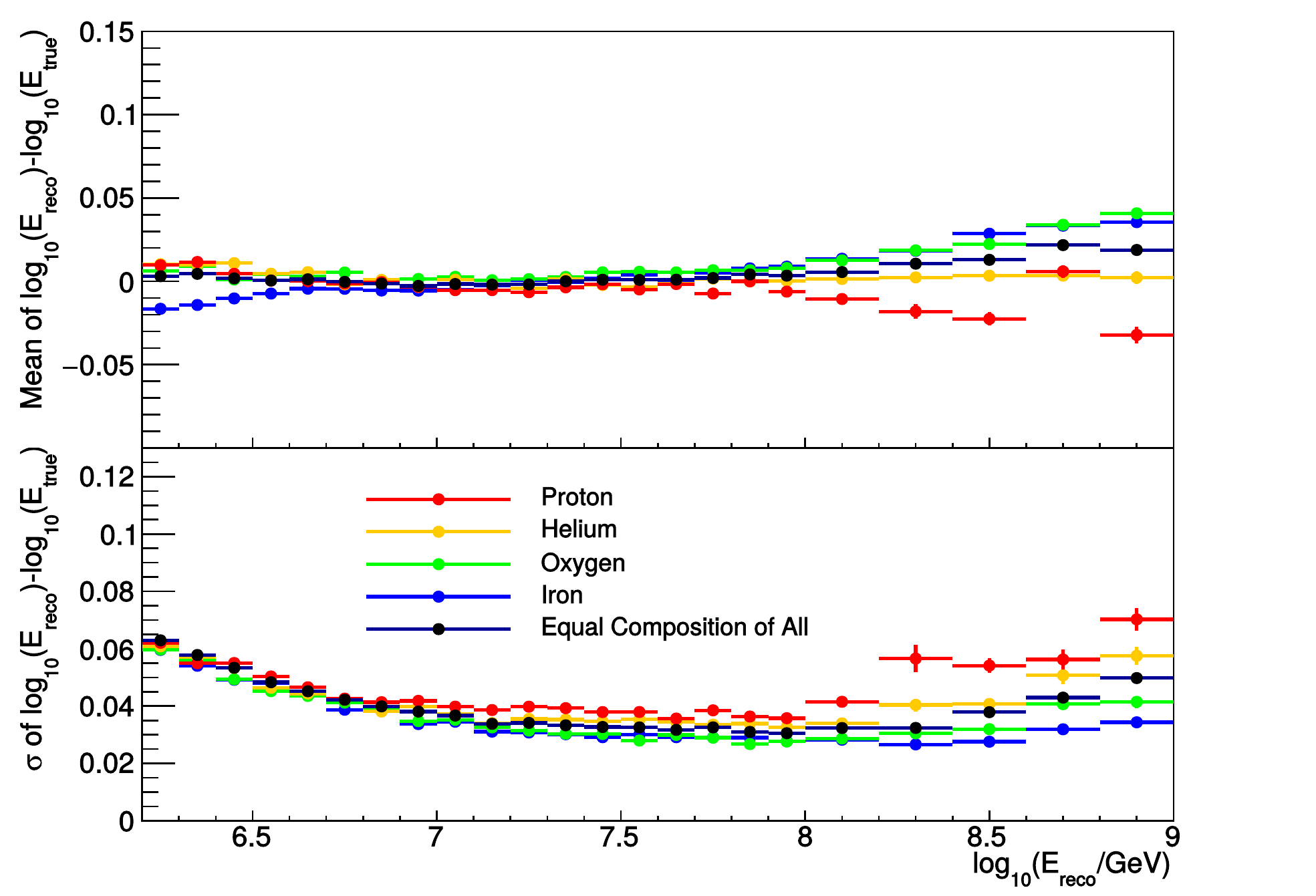}
\caption{Energy reconstruction bias (upper) and resolution (lower) as a function of the reconstructed energy for the different primary types and for an equal mixture of each type.}
\label{f:nn_bias_resolution}
\end{center}\end{figure}

The energy dependence of the primary energy bias and resolution as reconstructed by the NN are shown in Figure~\ref{f:nn_bias_resolution}. 
The energy resolution (Figure~\ref{f:nn_bias_resolution}, lower) ranges from 9\% (for iron showers at around 30 PeV) and 18\% with the worst resolutions below the energy threshold of $\sim$3~PeV and at the highest energies due to the worsening core position and angular resolution (as discussed in Section~\ref{sec_perf}).
Heavier primaries can be reconstructed more precisely %have a better energy resolution 
because of their lower intrinsic shower fluctuations. 
As mentioned in Section~\ref{sec:IT_alone_ana}, the overall decrease in precision  beyond $\sim$100~PeV is partially caused by the decrease in precision in angular and position resolution shown in Figure~\ref{f:core_ang_resolution}, which creates an extra smearing effect in \s. 

In this analysis, events are divided into energy bins of width 0.1 in \lte\,, which is larger than both the energy bias and the energy resolution as shown in Figure~\ref{f:nn_bias_resolution}.
However, due to the decrease in accuracy, precision, and available statistics at high energies (\lte\,>\,8.0), bins of width 0.2 are used in this region.
Above 1~EeV the energy bias dependence on the primary type becomes too large and limits the energy range over which this analysis is optimal.

\begin{figure}[t]
\begin{centering}
\includegraphics[width=.5\textwidth]{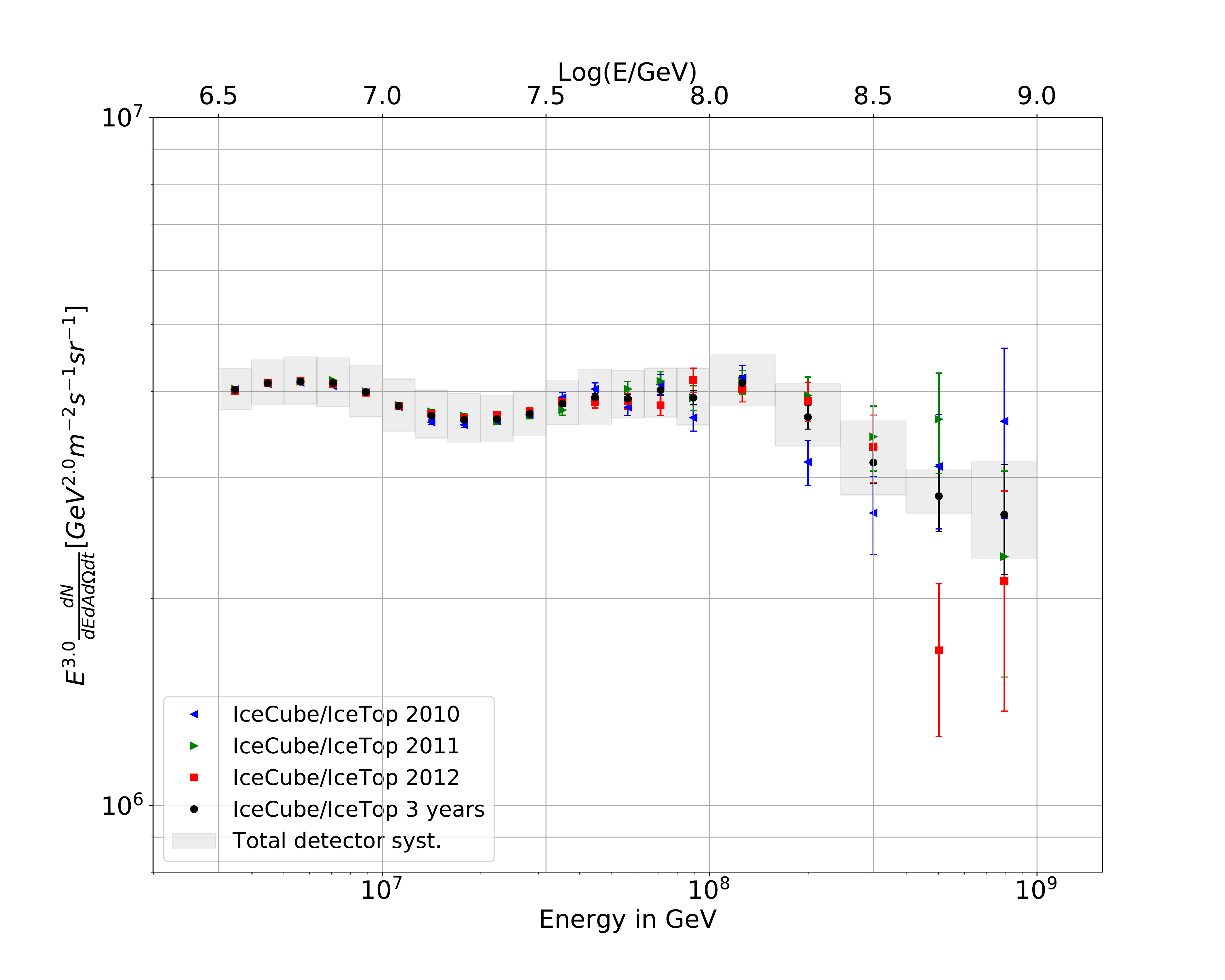}
\caption{All-particle energy spectrum from the coincident analysis from each of the three years analyzed individually compared to the combined result. The gray band represents the total detector uncertainty from both the IceTop and InIce arrays, as discussed in Section~\ref{sec:totaldetuncertainty}. \internalcite{IceTop\_3year\_update}}
\label{f:espec_coinc}
\end{centering}
\end{figure}

Figure~\ref{f:espec_coinc} shows the all-particle energy spectrum results for the coincident analysis for the three years individually and combined, multiplied by a factor of $E^{3}$ to highlight the details: the results are consistent between the years. The gray band represents the combined systematic uncertainties of the IceTop and InIce detectors for the coincident analysis, as discussed in Section~\ref{sec:totaldetuncertainty}.  These results are included in Table~\ref{table:coinc_all_particle} in Appendix~\ref{appendix:tables}.

\subsection{Composition Reconstruction using kernel density estimation to fit neural network templates}
\label{kde_explan}

\begin{figure}[t]\begin{center}
\includegraphics[width=0.48\textwidth]{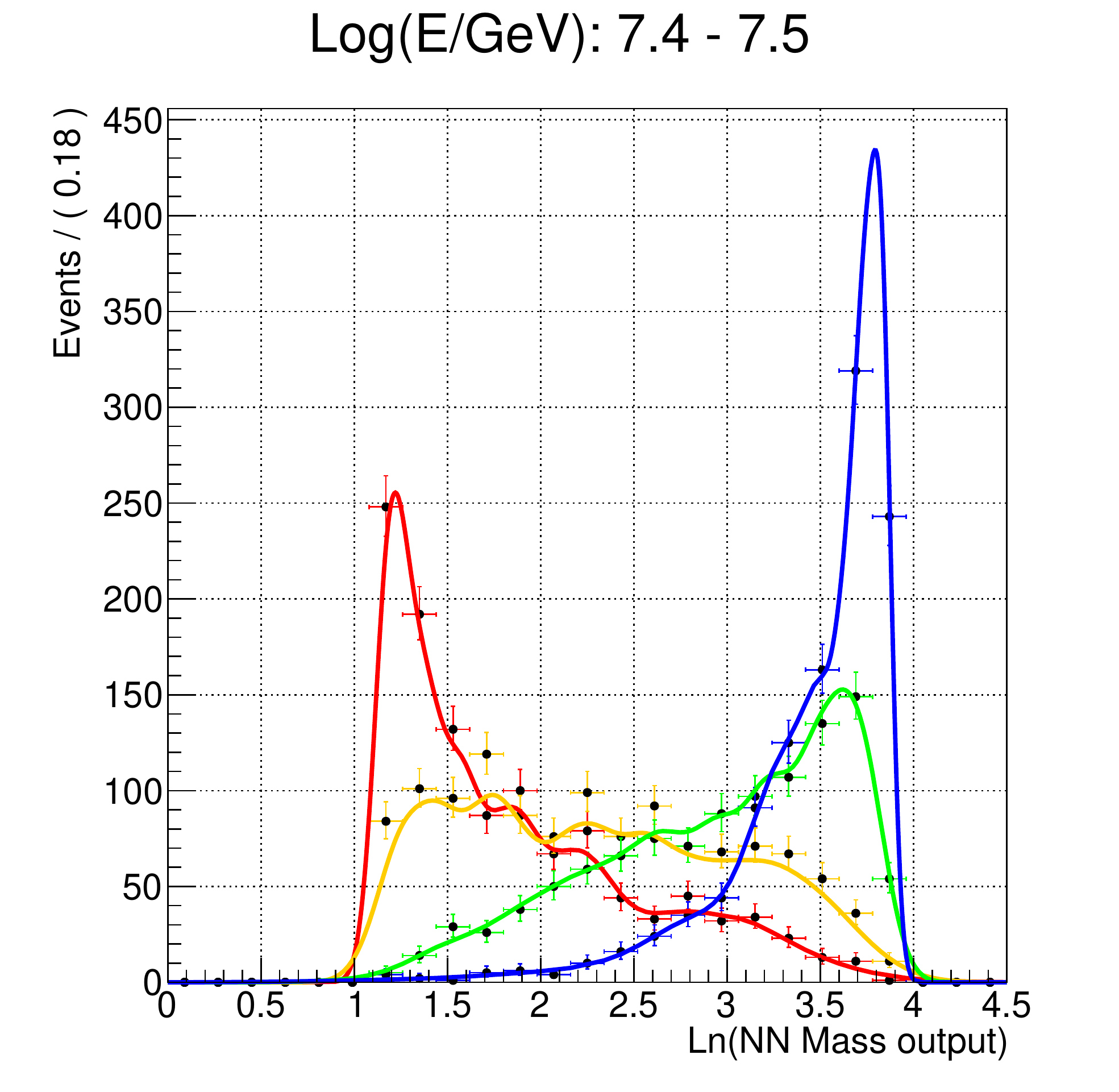}
\caption{Example distribution of the natural logarithm of the neural network mass output in the energy range between 7.4 and 7.5 in log$_{10}$(E$_{0,reco}$/GeV).
The y-axis represents the number of simulated events for proton (red), helium (orange), oxygen (green) and iron (blue). 
The solid lines represent the probability density function (p.d.f.) found by the adaptive KDE method.}
\label{f:nn_template_example}
\end{center}\end{figure}

%Karen try:
Figure~\ref{f:nn_template_example} shows histograms for each simulated element (proton, helium, oxygen and iron) in the natural logarithm of the neural network mass output for one slice in reconstructed energy.  (The four simulated types (proton, helium, oxygen and iron) are equidistant in \lna, but not in $A$. Thus, the histograms are expected to be more distinct in logarithmic space.)  In every slice in energy, the histogram for each primary element is converted into a template probability density function (p.d.f.) using an adaptive kernel density estimation (KDE) method \cite{KDE_Cramer}.  The template p.d.f.'s are shown as the solid lines in Figure~\ref{f:nn_template_example}.  
The template p.d.f.'s for all energy slices used in this analysis are given in Appendix~\ref{appendix:templates} in Figure~\ref{f:nn_templatehisto}.  
The four primary types exhibit four very distinctive shapes in each slice in energy over the whole energy range.  At \lte\,=\,8.0, the template p.d.f.'s  begin to exhibit greater overlap due to the limited statistics in the Monte Carlo sets, reducing the composition sensitivity of the analysis.  Beyond \lte\,=\,9.0 (1 EeV), the analysis becomes unreliable due to the overlap and reduction in data statistics.

\begin{figure}[t]\begin{center}
\begin{minipage}[c]{0.48\textwidth}
\includegraphics[width=1\textwidth]{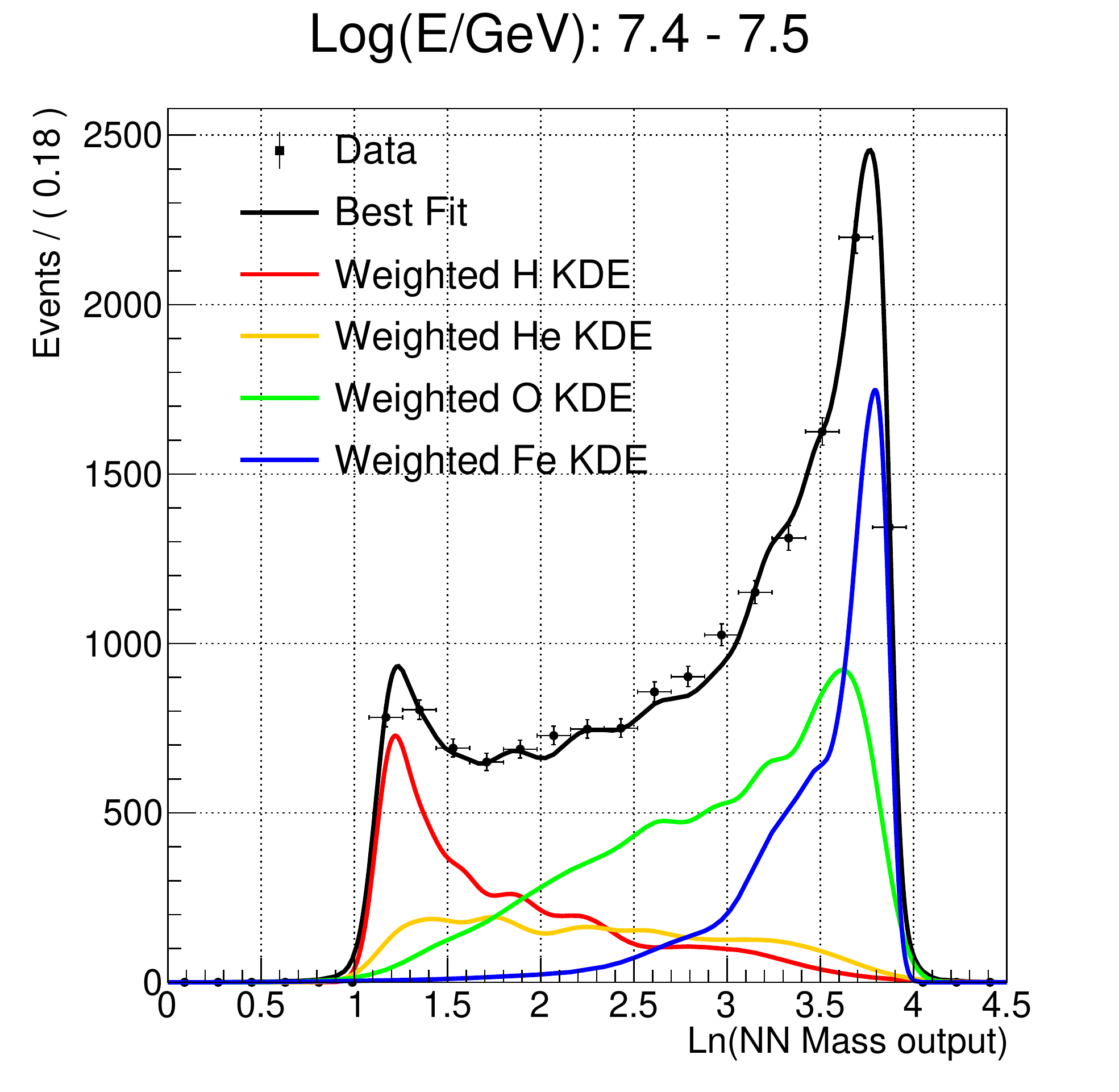}
\end{minipage}
\begin{minipage}[c]{0.48\textwidth}
\includegraphics[width=1\textwidth]{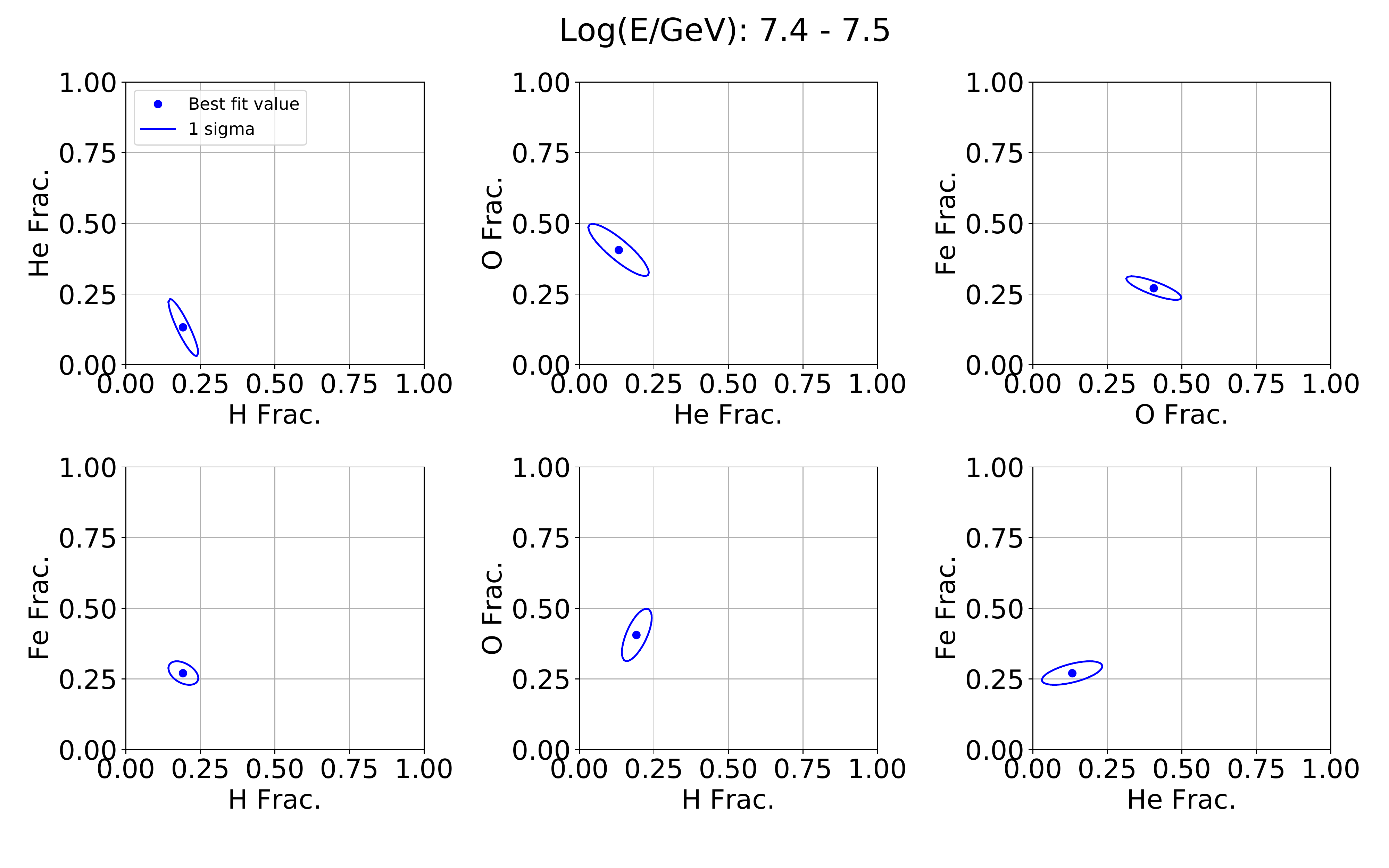}
\end{minipage}
\caption{\textbf{upper}: Example data distribution fit of the natural logarithm of the neural network mass output in the energy range between 7.4 and 7.5 in log$_{10}$(E$_{0,reco}$/GeV) and \textbf{lower}: corresponding contour plot with the best fit values.  Data distributions, contour plots, and complete correlation coefficient matrices for all energy bins are included in Appendix \ref{appendix:templates}.  
}
\label{f:fit_example}
\end{center}
\end{figure}

Using the Roofit package \cite{roofit_2003}, the set of four template p.d.f.'s were then weighted to find the \emph{fractions} (which were constrained to add to unity) which best fit the NN mass output for the experimental data in the same slice in reconstructed energy\footnote{This serves a similar purpose to the chi-squared minimization approach described in \cite{Andeen:thesis, ic40_coincidence}; however, the new unbinned extended likelihood technique improves on the previous method by correctly taking into account the Poisson fluctuations of the bin contents in both the data and the templates, which is particularly relevant in bins with few events.}.
The result of this method applied to the experimental data for the same energy bin as in Figure~\ref{f:nn_template_example} is shown in Figure~\ref{f:fit_example}~(upper). Additionally, the correlation between the fitted weights is shown in the form of uncertainty elipses in Figure~\ref{f:fit_example}~(lower). (The fit plots for all energy bins are given in Figure \ref{f:nn_fithisto} in Appendix~\ref{appendix:templates} and the corresponding correlation coefficients are shown in Table~\ref{table:appendix_corelation_coef} in Appendix~\ref{appendix:templates}.) The resulting fractions of neighboring elements (i.e. protons and helium) are anti-correlated, while those from distant elements (i.e. protons and iron) are virtually uncorrelated: this means a proton primary is more likely to be confused for a helium primary than for an iron primary, which is expected.

\begin{figure}[t]
\begin{centering}
\includegraphics[width=1.\linewidth]{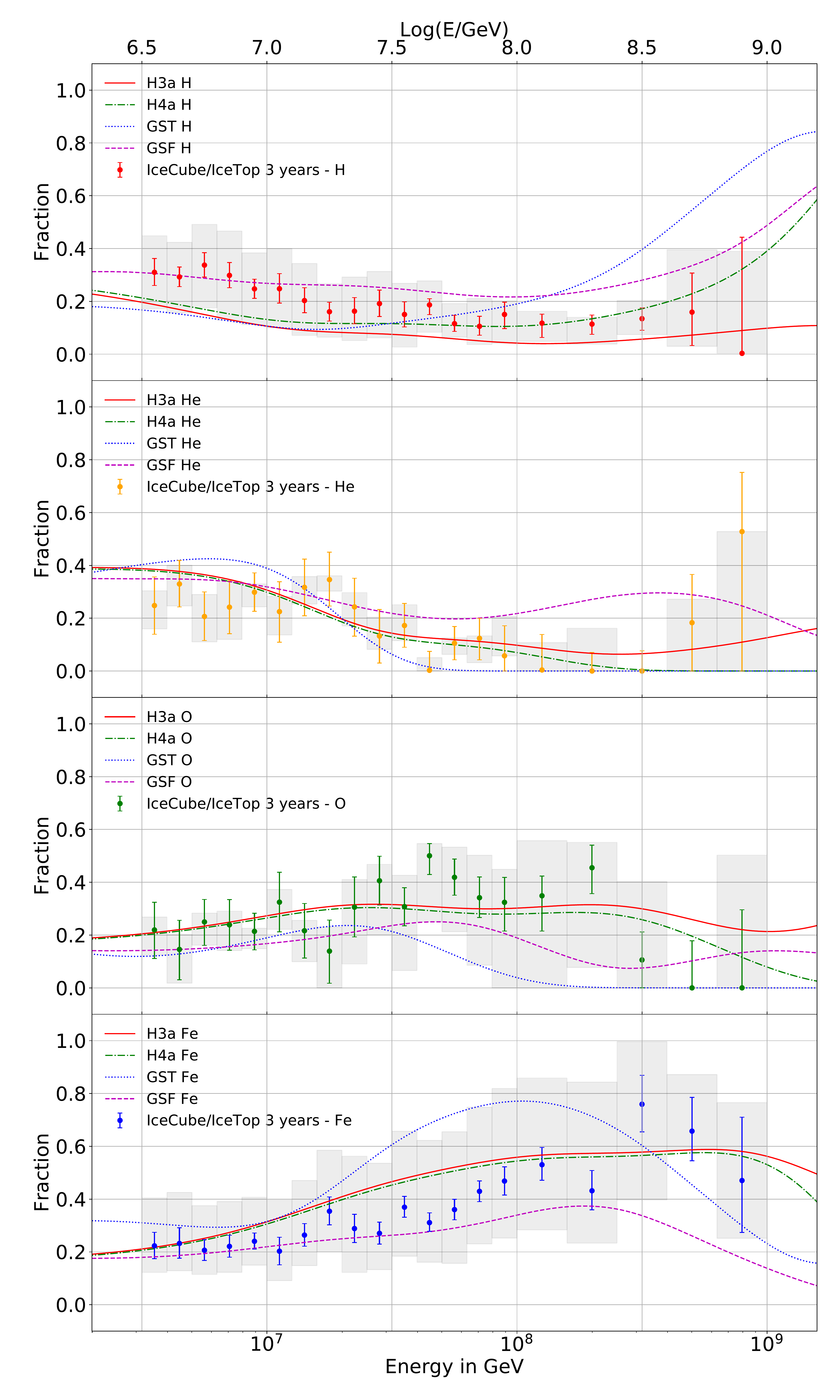}
\caption{Fractions for the four mass groups (protons in red, helium in yellow, oxygen in green, and iron in blue) including the total detector systematic compared with various cosmic ray models (H3a and H4a~\cite{Gaisser_H4a}) and phenomenological experimental fits (GST~\cite{GST_2013} and GSF~\cite{Dembinski:2017zsh}).  Sibyll~2.1 was used for the hadronic interaction model in the simulated dataset.
\internalcite{IceTop\_3year\_update}}
\label{f:individual_fractions}
\end{centering}
\end{figure}

\begin{figure}[t]
\begin{centering}
\includegraphics[width=1.\linewidth]{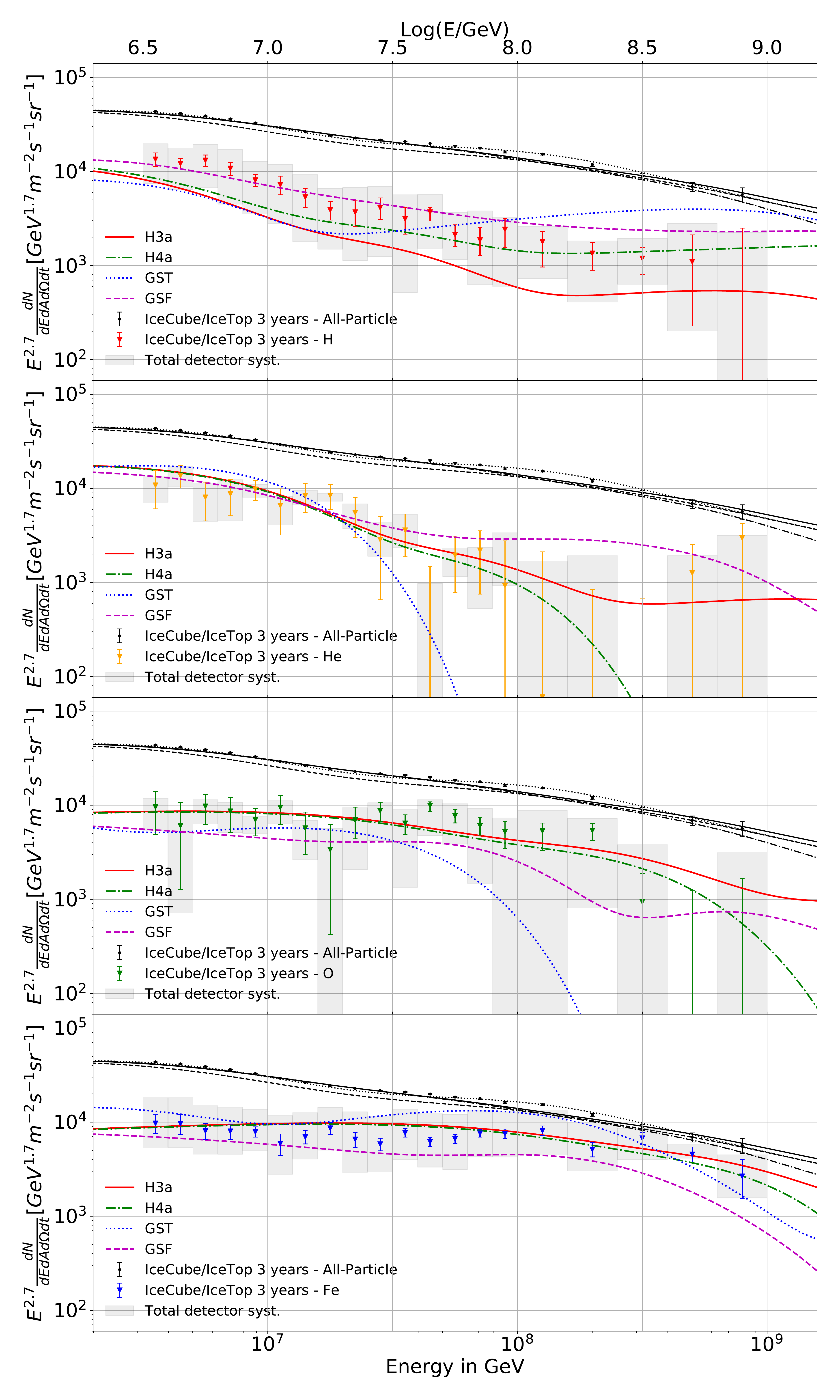}
\caption{Individual spectra for the four mass groups (protons in red, helium in yellow, oxygen in green, and iron in blue) including total detector systematic compared with various cosmic ray models (H3a and H4a~\cite{Gaisser_H4a}) and phenomenological experimental fits (GST~\cite{GST_2013} and GSF~\cite{Dembinski:2017zsh}).  Sibyll~2.1 was used for the hadronic interaction model in the simulated dataset.
\internalcite{IceTop\_3year\_update}}
\label{f:individual_spectra}
\end{centering}
\end{figure}

The KDE template-fitting procedure %discussed in Section~\ref{kde_explan} above 
yields a measurement of the fractions of each of the four nuclear mass groups (represented by H, He, O, and Fe), for each bin in energy.
The fractions are shown in Figure~\ref{f:individual_fractions}.  Each of these four individual fractions is then translated into an individual spectrum for the corresponding elemental group, as shown in Figure~\ref{f:individual_spectra} (colors) compared to the all-particle spectrum (black).  Recent model predictions are also included in Figures~\ref{f:individual_fractions} and \ref{f:individual_spectra}, which will be discussed in Section~\ref{sec:results}.  In both figures, the gray band represents the total coincident detector uncertainty from both the IceTop and InIce arrays, which will be discussed in Section~\ref{sec:totaldetuncertainty}.  These results are included in Tables~\ref{table:proton_helium}-\ref{table:oxygen_iron} in Appendix~\ref{appendix:tables}.

Intermediate elements, not part of the four groups listed above, are expected to produce neural network outputs in between the adjacent groups, so will partially contribute to the flux of the groups that bracket it.   In order to test this, a small sample of silicon was passed through the NN~+~KDE chain and treated as ``data''.  The natural log of the mass of silicon is approximately midway between that of oxygen and that of iron; therefore, as expected (due to the regression-style neural network mass output), the silicon is reconstructed as a nearly 50/50 mixture of oxygen and iron across all energies.

\begin{figure}[t]
\begin{centering}
\includegraphics[width=.48\textwidth]{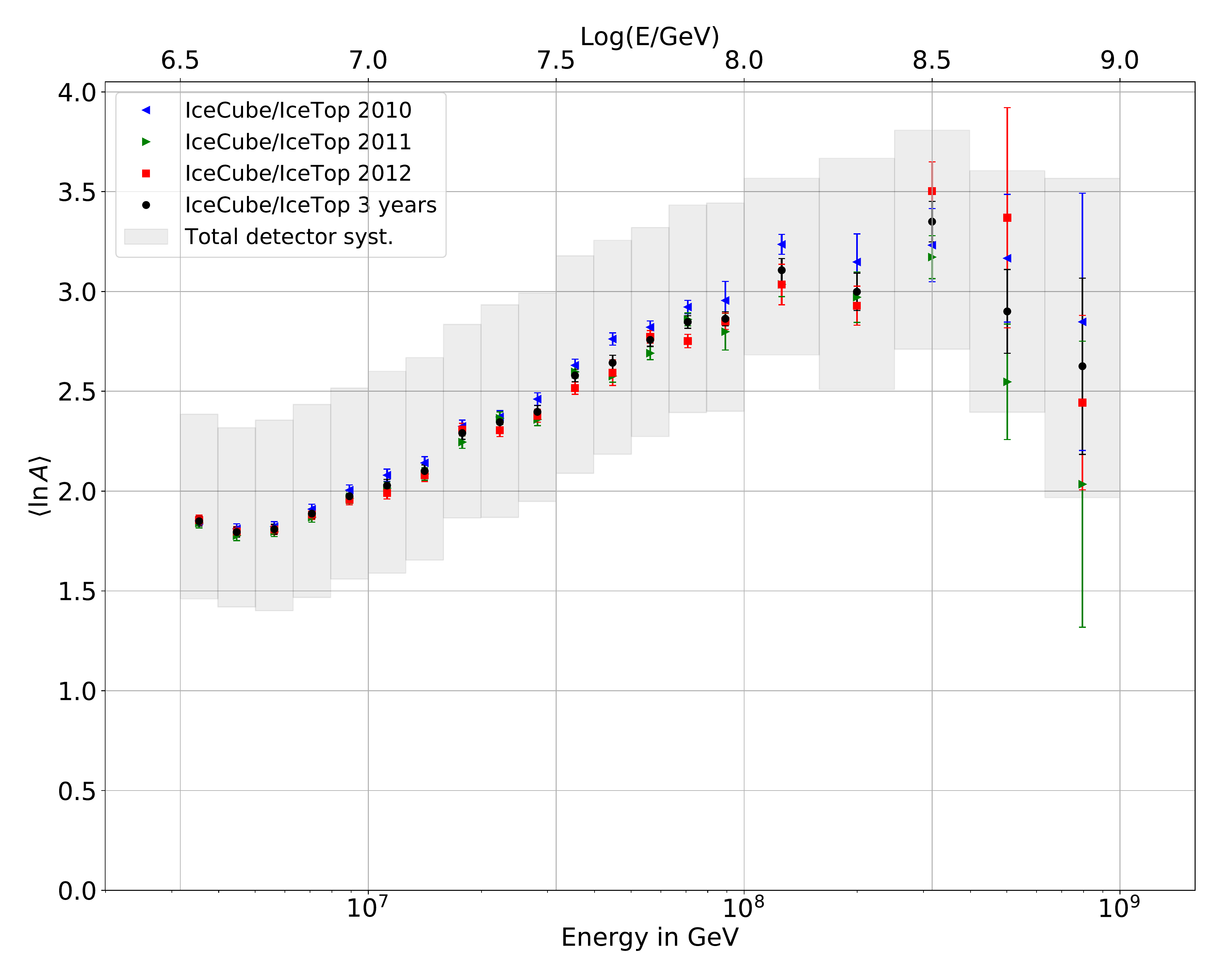}
\caption{Mean log mass (\lna) for the three individual years, and the three years combined.  The gray band represents the combined systematic uncertainties of the IceTop and InIce detectors for the coincident analysis, as discussed in Section~\ref{sec:totaldetuncertainty}.  Sibyll~2.1 was used for the hadronic interaction model in the simulated dataset.
\internalcite{IceTop\_3year\_update}}
\label{f:lnA}
\end{centering}
\end{figure}

Figure~\ref{f:lnA} shows the mean log mass, which is derived from the individual fractions shown in Figure~\ref{f:individual_fractions}.  Again, the gray band represents the total coincident detector uncertainty from both the IceTop and InIce arrays, which will be discussed in Section~\ref{sec:totaldetuncertainty}.   Each of the three years of data are again shown both separately and combined and agree very well within the statistical and systematic uncertainties.

\section{Systematic Uncertainties}
The uncertainties in the coincident analysis reported here can be grouped into three categories: analysis method, detector effects, and the hadronic interaction model. 

\subsection{Analysis Method}
As described above, the shape of the Monte Carlo templates is derived from an adaptive KDE method, which determines the \textit{optimal} width of the Gaussian kernel function.  
To check the robustness of the composition fitting results, the optimal kernel width is artificially modified by a factor of $\pm90\%$, resulting in either very jagged or very smooth templates. 
These artificially modified templates are then used in place of the optimal templates for the remainder of the analysis in order to measure the effect of the vastly different templates on the results.   
The variation in the final results due to the modified template shapes is so small that it is not visible in a figure; however, the values for uncertainty are included in the tables in Appendix~\ref{appendix:templates}.

\subsection{Detector Uncertainties}
\label{sec:totaldetuncertainty}
Three main detector effects contribute to the uncertainty in the composition results: the snow correction, the absolute energy scale of IceTop, and the light yield in the ice.

\begin{figure}[h]\begin{center}
\includegraphics[width=0.48\textwidth]{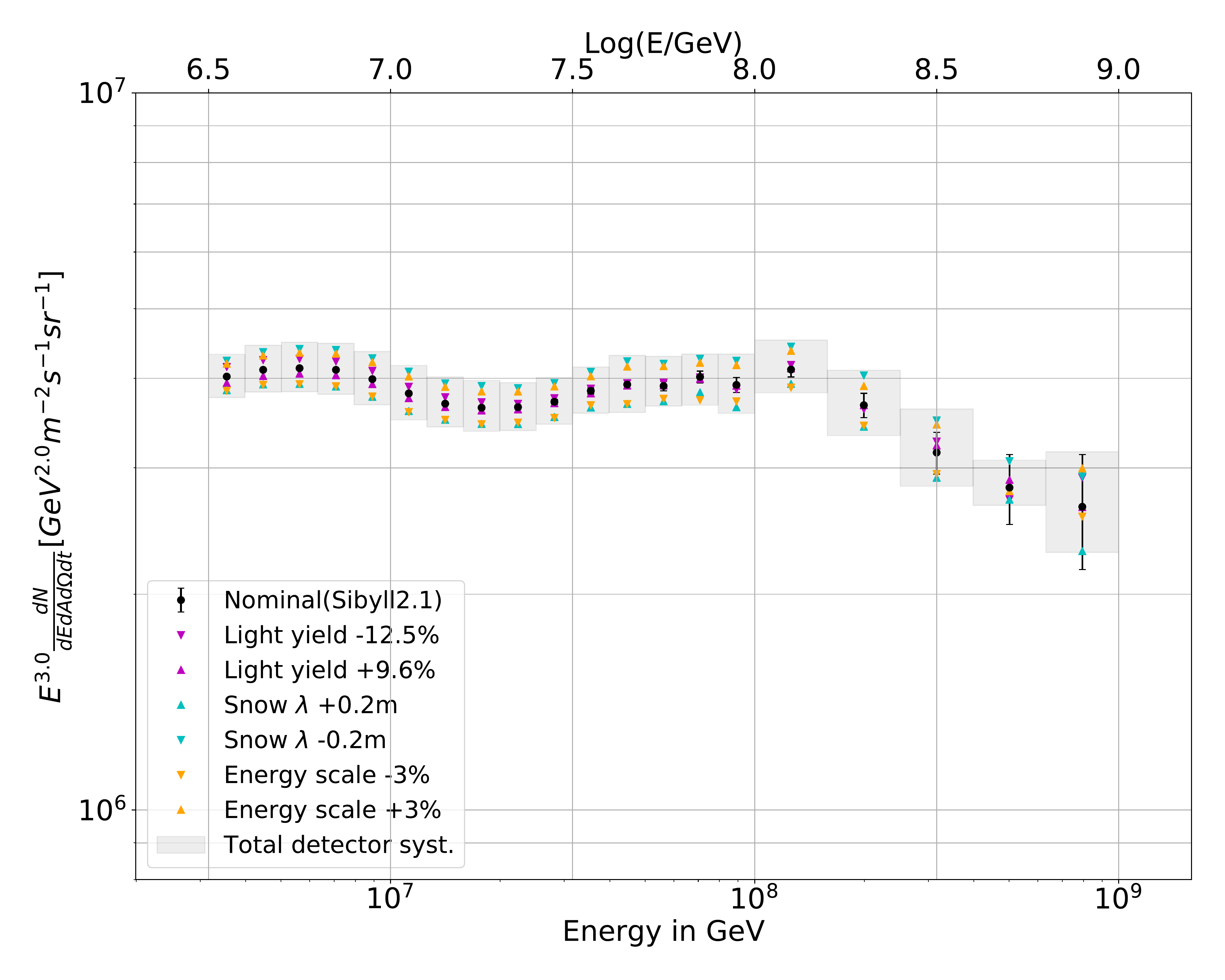}
\caption{Total detector uncertainty (gray) on all-particle energy spectrum from the combination of light yield (magenta), snow correction (cyan) and energy scale uncertainty (orange).}
\label{f:sys_det_single_spec}
\end{center}\end{figure}

\begin{figure}[h]\begin{center}
\includegraphics[width=0.48\textwidth]{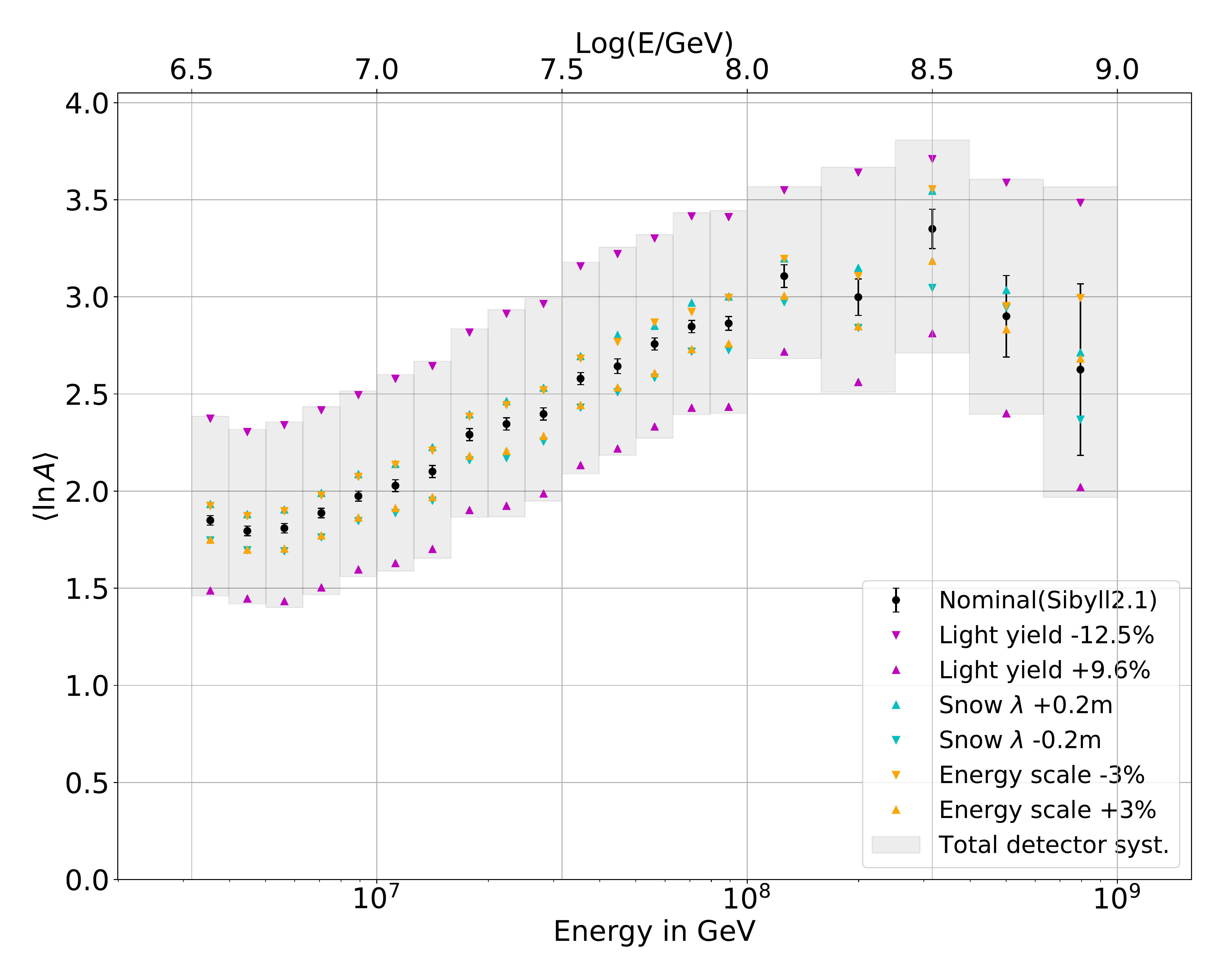}
\caption{Total detector uncertainty (gray) on \lna\,from the combination of light yield (magenta), snow correction (cyan) and energy scale uncertainty (orange). }
\label{f:sys_det_lnA}
\end{center}\end{figure}

\begin{figure}[h]\begin{center}
\includegraphics[width=1.0\linewidth]{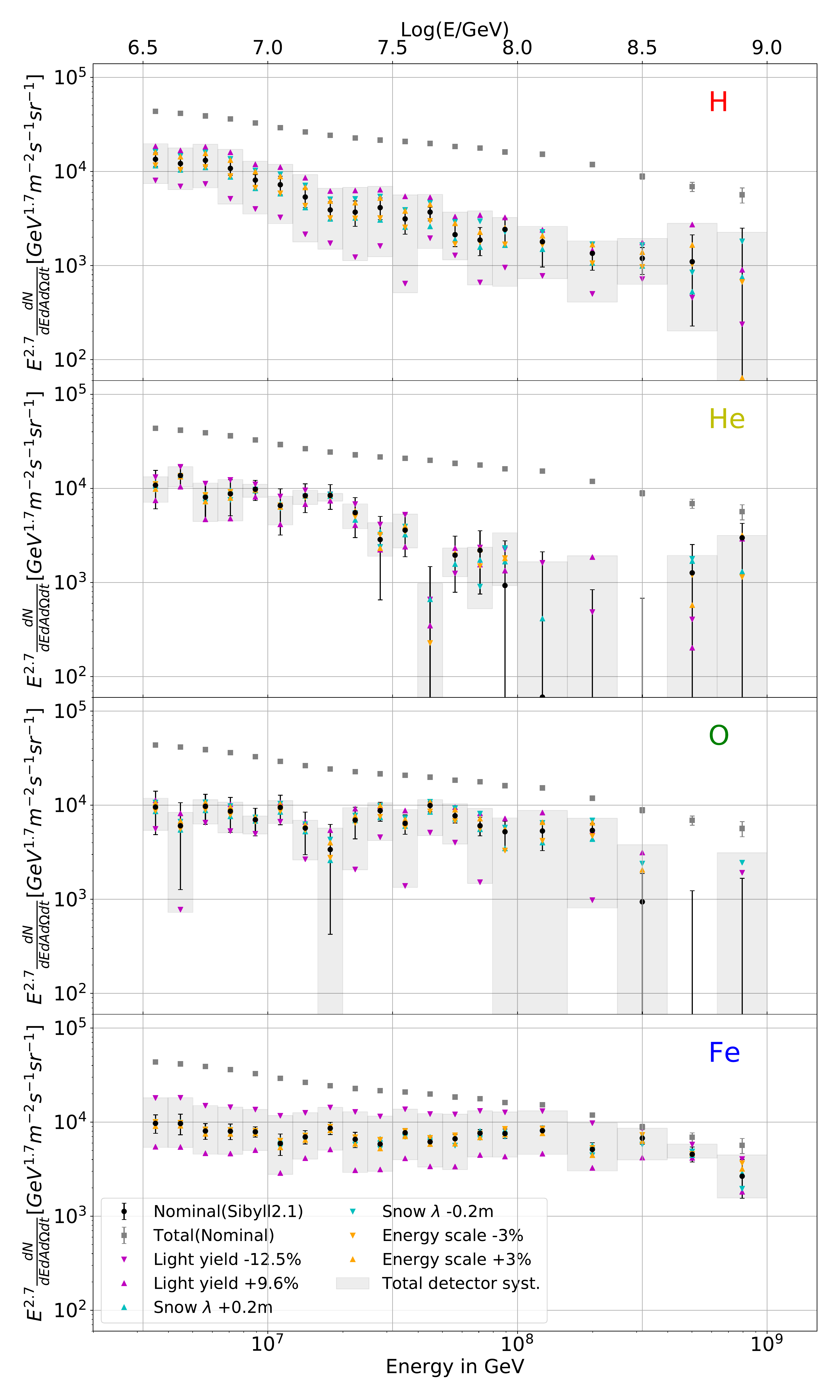}
\caption{Total detector uncertainty (gray) on elementary energy spectra from the combination of light yield (magenta), snow correction (cyan) and energy scale uncertainty (orange).}
\label{f:sys_det_elem_spec}
\end{center}\end{figure}

\subsubsection{Snow correction} 
Although the method for snow correction described in Section~\ref{sec:IT_reco} works well \emph{on average} over the entire energy and zenith angle region, it is not perfect. 
This is predominantly due to its inability to distinguish between the electromagnetic and muonic component of the air shower. 
A systematic uncertainty of $\pm$0.2~m was assigned to $\lambda$, which covers the variations due to the unknown composition, energy and zenith angle dependence, as discussed in the Appendix of \cite{it73_icetopalone_spectrum}.

\subsubsection{Absolute energy scale}

To obtain a reliable energy scale, the simulated response of each IceTop tank is calibrated using the signal from atmospheric muons~\cite{icetop_technical}.
This procedure has proven to be very reliable, but due to the unknown composition, atmospheric conditions, etc, a final 3\% uncertainty on the charge calibration, and thus on the absolute energy scale, needs to be taken into account\cite{VEM_Arne}. This translates directly to a 3\% shift in S$_{125}$.

\subsubsection{Light Yield}
The uncertainty on the photon detection efficiency by the DOM (referred to as the \emph{DOM efficiency}) is $\pm$3\%, which implies a $\pm$3\% possible variation of the total light yield observed. 
The uncertainty caused by the photon propagation in the ice includes scattering and absorption coefficients in the \emph{bulk ice} and an effective scattering length of the \emph{hole ice}, the ice in the drill hole around the DOM \cite{HoleIce}. 
For the bulk ice scattering and absorption coefficient, 3 points are taken on the 1 $\sigma$ error ellipse around the nominal values, shown in \cite{SPICE}: a +10\% scattering coefficient, a +10\% absorption coefficient and a -7.1\% scattering and absorption coefficient. 
Alternate effective scattering lengths of 30~cm and 100~cm were used for the hole ice model. 
For all these individual systematic uncertainties, simulations were produced and their effect was studied. 
It was found that the main combined effect is to influence the light yield in the DOM, and that this effect is rather independent of the initial light yield and zenith angle of the muon bundle.\fixka{Sam had ... here--more things?}
This means that all those systematic uncertainties can be combined and modeled as a shift on the observed light yield.
Furthermore, the systematic uncertainties are (nearly) uncorrelated; thus the various errors and shift on light yield are added in quadrature, which gives a total light yield uncertainty of +9.6\% and -12.5\%. 
The individual contribution to the observed light yield shifts, as well as the total light yield uncertainty are given in Table~\ref{table:lightyield_shift}.
\begin{table}[h]
\caption{Sytematic light yield shift}
\begin{center}
\begin{tabular}{c|c}
\hline
Effect	&	Light yield shift \\ 
\hline \hline
+10\% scattering & +3.6\%\\
+10\% absorption & $-$11.8\%\\
$-$7.1\% scattering and absorption & +7\%\\
30~cm hole ice scattering & +4.5\%\\
100~cm hole ice scattering & $-$2.9\% \\
DOM efficiency & $\pm3\%$\\
\hline
Total Light Yield Effect & +9.6\%,$-$12.5\%\\
\hline
\end{tabular}
\label{table:lightyield_shift}
\end{center}
\end{table}

\subsubsection{Total Detector Uncertainties}
The three detector uncertainties discussed above (snow correction, absolute energy scale, and light yield) are added together in quadrature into a total detector uncertainty.  Figures~\ref{f:sys_det_single_spec},~\ref{f:sys_det_lnA}, and~\ref{f:sys_det_elem_spec} show individual and combined contributions to the uncertainty in the all-particle energy spectra, mean log mass, and the individual elemental spectra.  In all figures, the gray bands are the total combined detector uncertainty which match the gray bands shown in Figures \ref{f:espec_coinc}, \ref{f:lnA}, and \ref{f:individual_spectra}, respectively, and are included in the tables in Appendix~\ref{appendix:tables}.

\subsection{Hadronic Interaction Model}
\label{sec:syst_hadintmodel}

The influence of the hadronic interaction model on the measurement of the cosmic ray composition arises mainly from variations between the models in the predicted number of high-energy muons. 
Sibyll~2.1 was used as the hadronic interaction model for the baseline simulation datasets, while the three post-LHC models--QGSJET~II-04\cite{QGSJETII}, Sibyll~2.3\cite{sibyll2.3} and EPOS-LHC\cite{EposLHC}--are used as alternate models. 
The variation between the models in the high-energy ($>300$~GeV) muon number is smaller than for the surface (GeV) muons, but is still of order 15\% at maximum (between QGSJet~II-04 and EPOS-LHC).   \footnote{Here it is important to emphasize that the muon energy loss, $dE_\mu/dX$, as measured in IceCube is a proxy for the number of high energy muons in the muon bundle.  These are muons in excess of 300 GeV which are created near the first interaction of the cosmic ray primary with the atmosphere.  Recently, a number of studies (summarized here: \cite{Dembinski:UHECR2018}) show a discrepancy between the experimentally measured number of muons (from many experiments), and the simulated number of muons from latest post-LHC hadronic interaction models.  This discrepancy is in the number of lower energy muons, which are produced late in shower development, after many interactions.  This discrepancy presently seems to be due to the accumulation of small discrepancies at each interaction; thus, the impact is large for the muons produced after many interactions, and is small/non-existent for those muons produced near the first interaction (\cite{Riehn:ICRC2019}).  Furthermore, this discrepancy has only been measured at cosmic ray primary energies above 1 EeV and it seems to increase with primary cosmic ray energy.  ALICE (the only other experiment which has measured the high energy muons in the range between PeV and EeV) reports relatively good agreement between the number of muons in the bundles produced by PeV to EeV cosmic rays and QGSJET~II-04 (\cite{ALICE:2016}), although the number of events is small.  Thus, we presently expect little impact on the results here.  However, this ``muon puzzle'', is still under intense investigation by the community.}

It is important to note that the main Sibyll~2.1 dataset included full samples across all energies for each particle type (proton, helium, oxygen, and iron);
however, due to the computational time involved in generating a full simulated data set ($\sim$60000 CPU years), the alternate models include 1/10th the number of events, in proton and iron only.   
As a result of the limited size of the alternate simulated data sets, a full analysis is \textit{not} repeated using those datasets for the simulated templates. 
Rather, the difference in the number of high-energy muons and in \s~in the alternative sample with respect to Sibyll~2.1 is calculated for each particle type and applied to the experimental data, which is then passed through the full analysis chain.  (Neither the stochastics nor the zenith angle are taken into account in this estimate due to their low impact on the results.)  
The results of this process are then weighted using the reconstructed elemental fractions for each particle type from the baseline Sibyll~2.1 analysis in order to obtain an estimate of what the final result would have been had the alternative interaction model been used as the nominal Monte Carlo simulation instead of Sibyll~2.1. 
It is probable that the alternate results reported here differ slightly from the actual fractions that would be reconstructed with a full alternative simulated dataset.  

\begin{figure}[ht]\begin{center}
\includegraphics[width=0.98\linewidth]{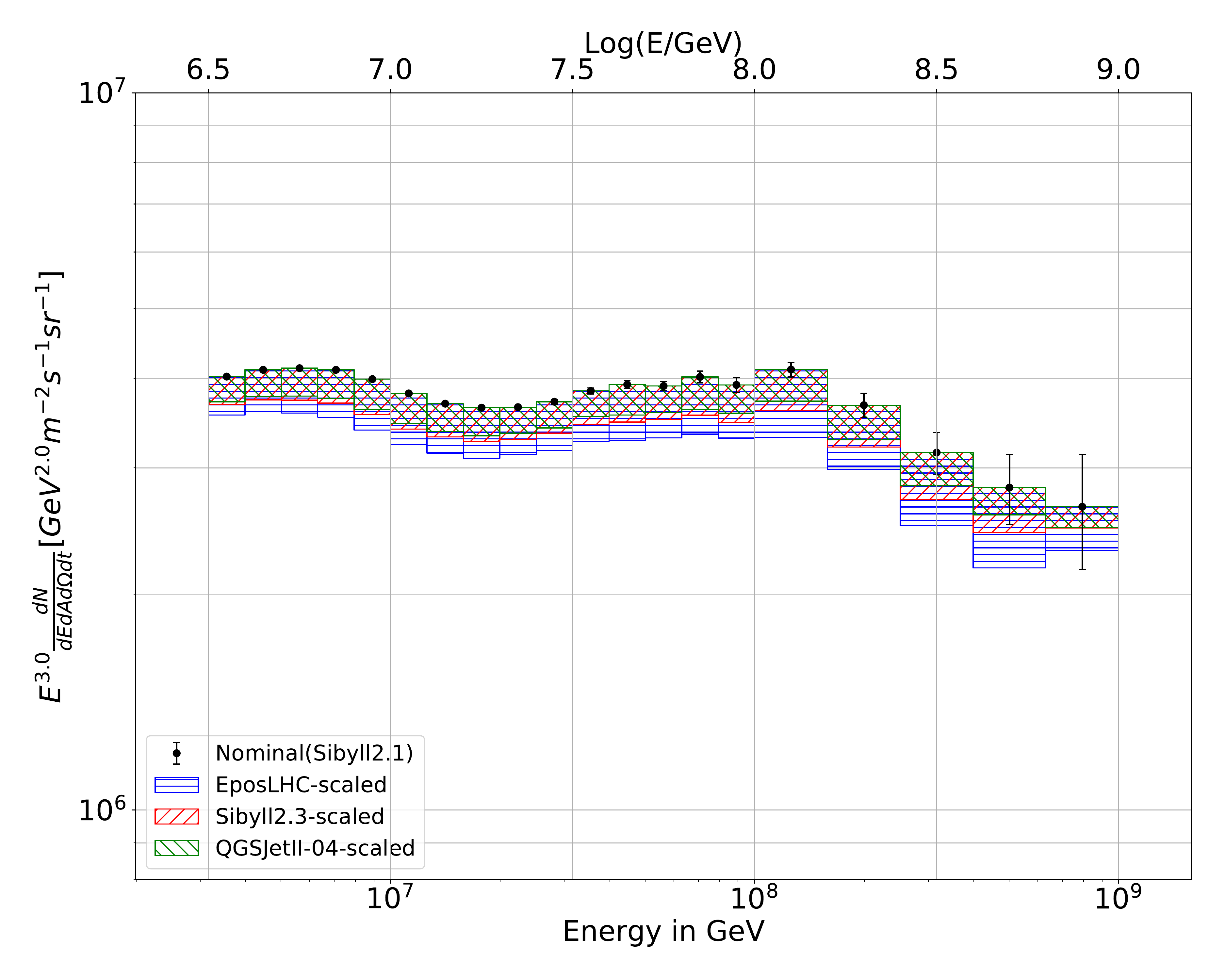}
\caption{Hadronic interaction model uncertainty range on all-particle energy spectrum based on EposLHC (blue), Sibyll2.3 (red) and QGSJetII-04 (green).}\fix{Caption needs polishing...}
\label{f:sys_had_single_spec}
\end{center}\end{figure}

\begin{figure}[ht]\begin{center}
\includegraphics[width=0.98\linewidth]{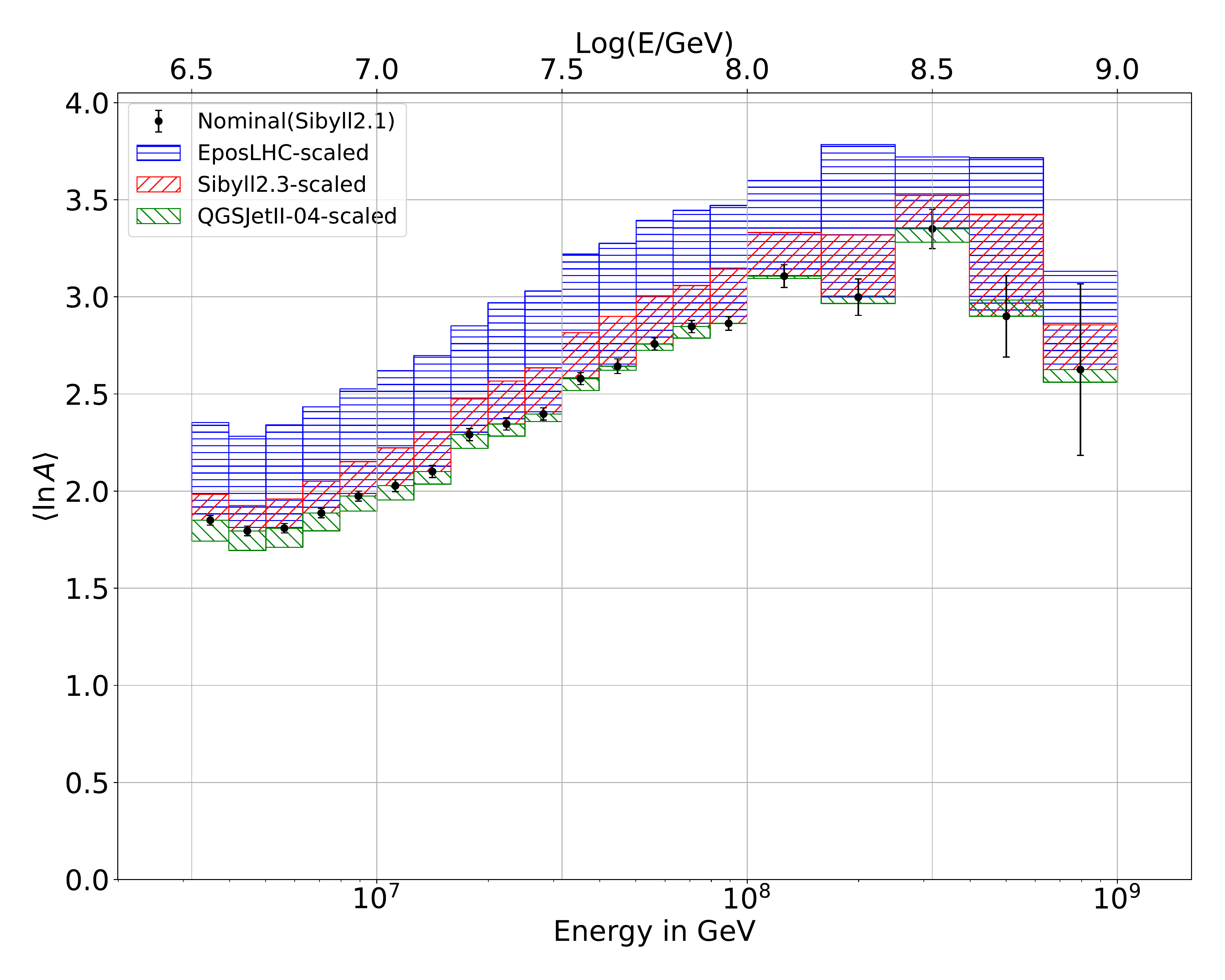}
\caption{Hadronic interaction model uncertainty range on \lna\, based on EposLHC (blue), Sibyll2.3 (red) and QGSJetII-04 (green).}\fix{Caption needs polishing...}
\label{f:sys_had_lnA}
\end{center}\end{figure}

\begin{figure}[t]\begin{center}
\includegraphics[width=1.0\linewidth]{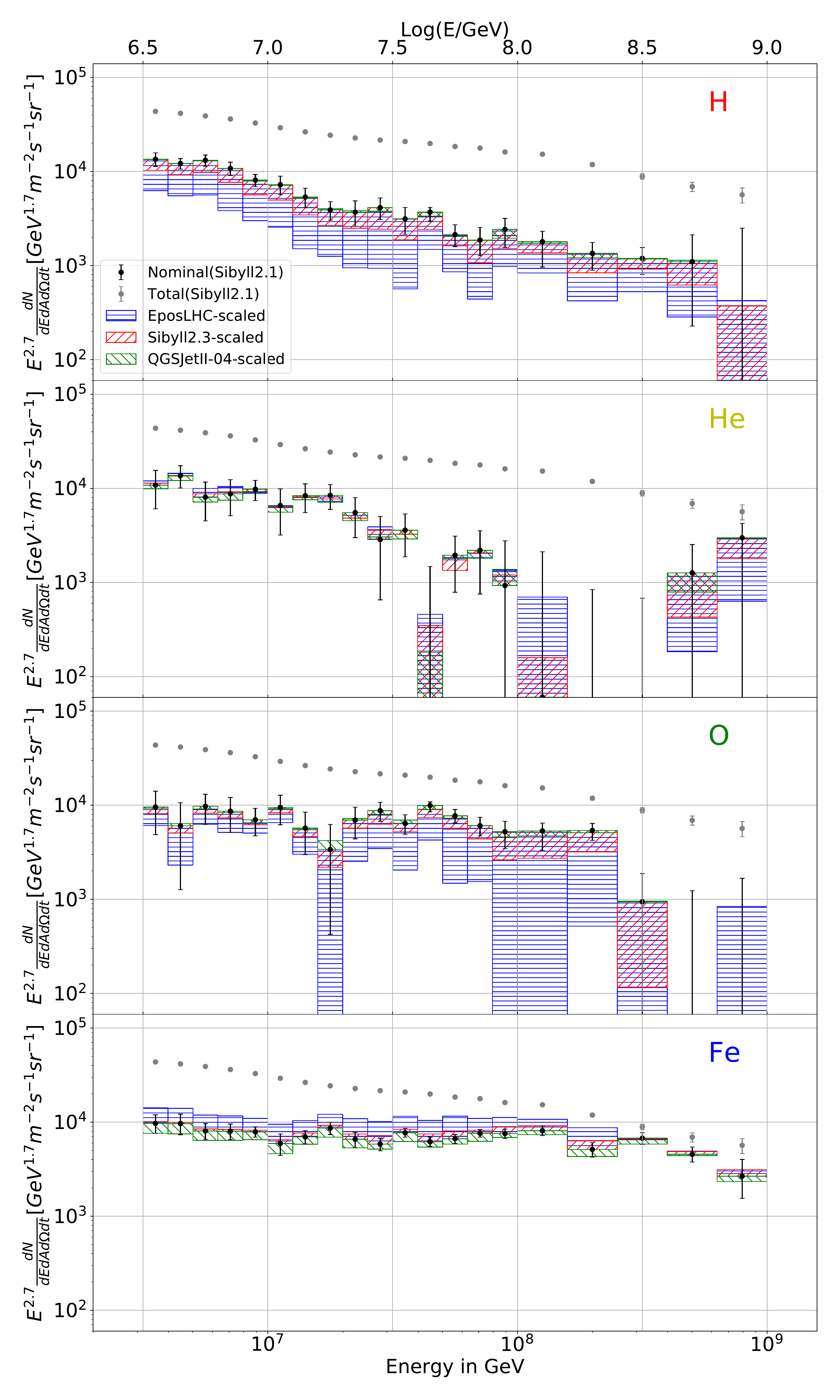}
\caption{Hadronic interaction model uncertainty range on elementary energy spectra based on EposLHC (blue), Sibyll2.3 (red) and QGSJetII-04 (green).}\fix{Caption needs polishing...}
\label{f:sys_had_elem_spec}
\end{center}\end{figure}

The estimated uncertainty due to each alternative model is shown in Figures \ref{f:sys_had_single_spec}, \ref{f:sys_had_lnA}, and \ref{f:sys_had_elem_spec}, which depict the effect of the choice of hadronic interaction model in the all-particle energy spectra, mean log mass, and the individual elemental spectra.  Although the uncertainties due to the hadronic interaction models is large, the shapes of the distributions with respect to energy do not change significantly between the models.  

\section{Results, Discussion and Outlook}
\label{sec:results}

\subsection{Energy Spectra}

\begin{figure}[t]
\begin{centering}
\includegraphics[width=.5\textwidth]{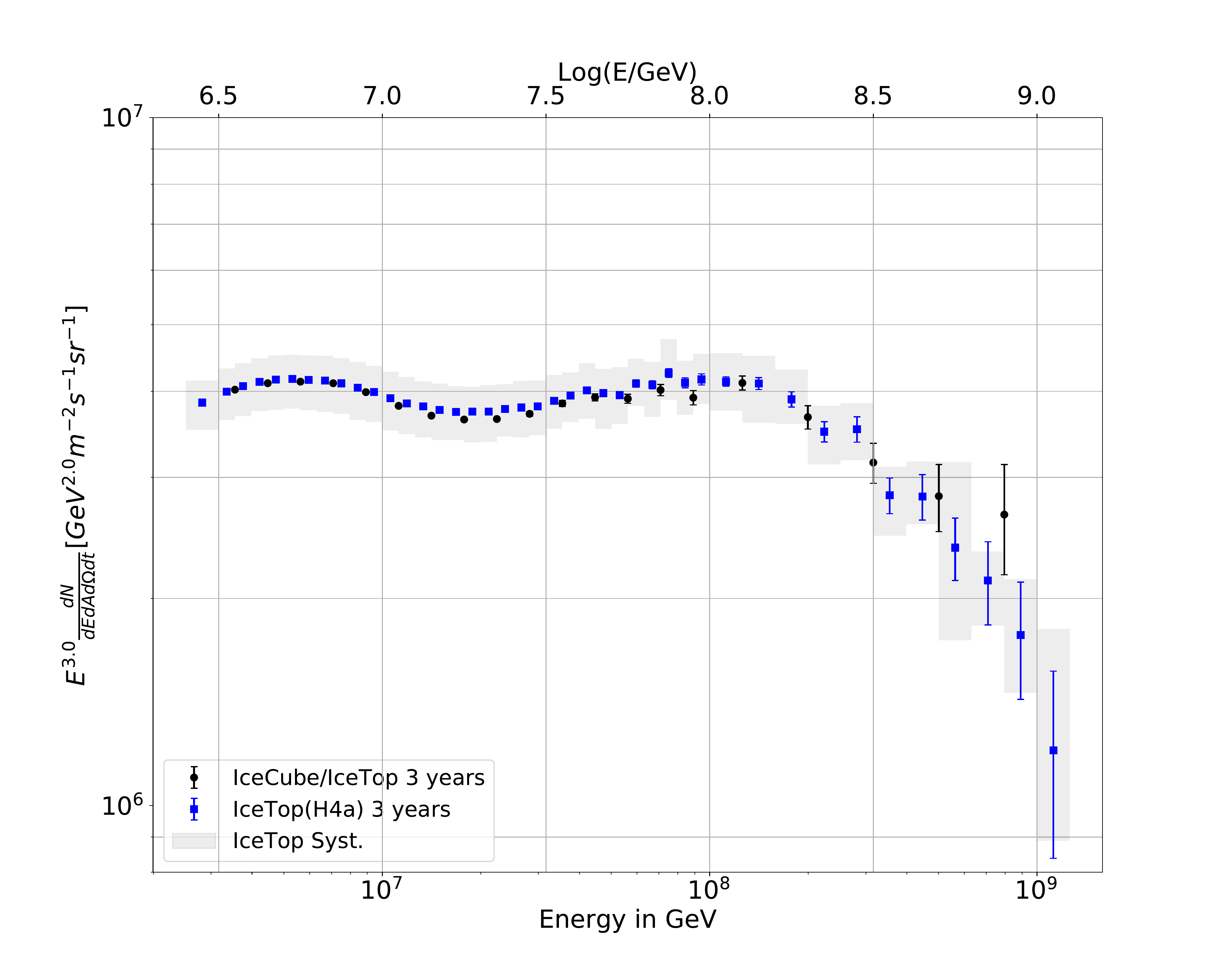}
\caption{A comparison of the combined three-year spectra from the two analyses in this paper: the IceTop-alone analysis, and the coincident analysis. The gray band represents the total systematic uncertaintity of the IceTop detector, as described in \cite{it73_icetopalone_spectrum}.  Sibyll~2.1 was used for the hadronic interaction model in the simulated dataset.
\internalcite{IceTop\_3year\_update}}
\label{f:espec_comparison}
\end{centering}
\end{figure}

Figure~\ref{f:espec_comparison} compares the energy spectra resulting from the IceTop-alone analysis and the coincident analysis as described herein.  The analyses are consistent with each other within the statistical and systematic uncertainties (only the smaller IceTop-alone systematic uncertainties are shown, for clarity). This good agreement indicates that the dependence of the IceTop-alone analysis on composition model has been effectively mitigated through the analysis technique, since the IceTop/InIce coincident analysis doesn't require prior knowledge of the composition.

\subsection{Mass composition}

\begin{figure*}[t]
\begin{centering}
\includegraphics[width=.8\textwidth]{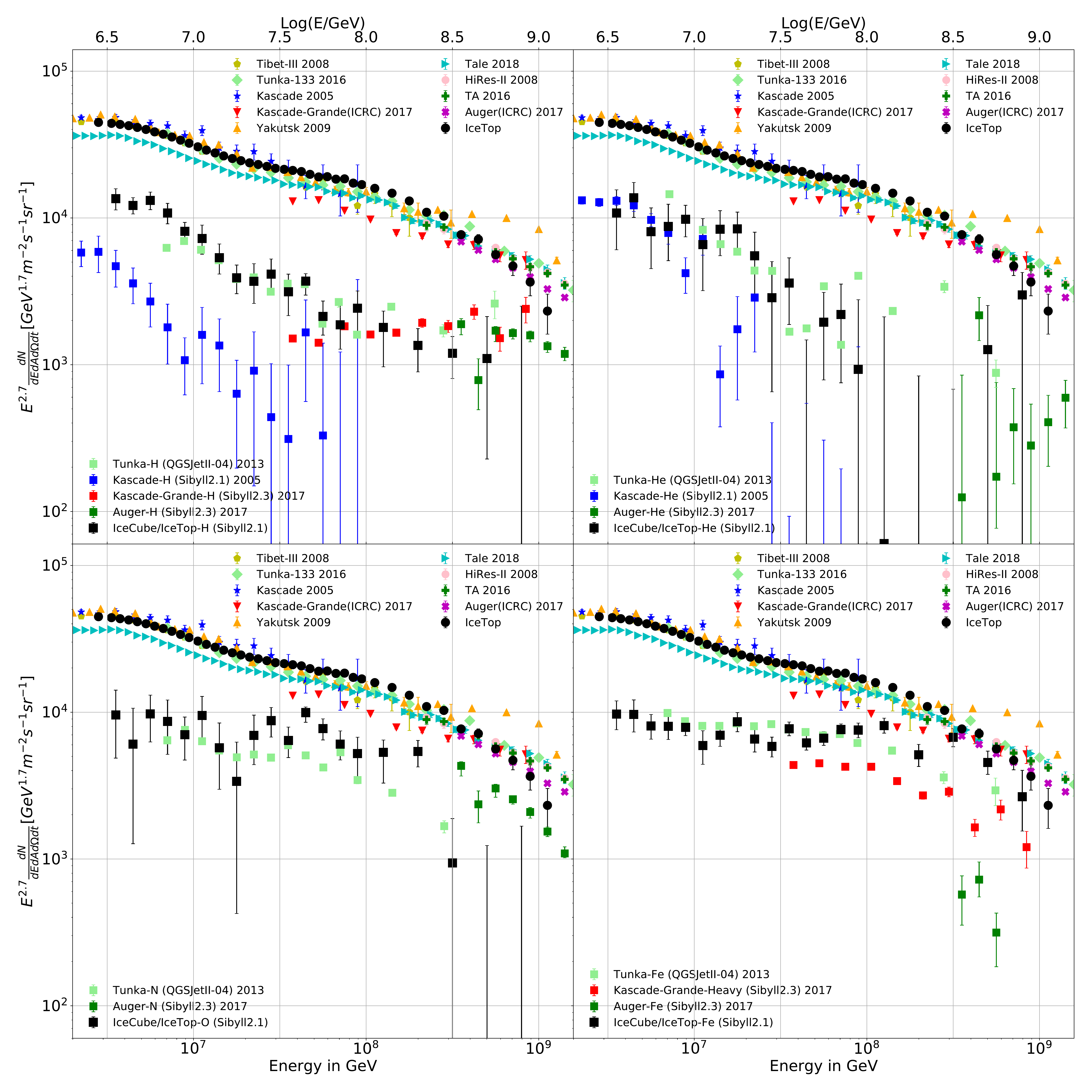}

\caption{Comparison of the all-particle and composition spectra of the four elemental groups H, He, O and Fe from this analysis using Sibyll~2.1 (black) with other experiments. The data set for the all-particle spectra are taken from Tibet \cite{TibetIII_2008}, Tunka \cite{Tunka_2016}, Yakutsk \cite{Yakutsk_2009}, Tale \cite{Tale_2018}, Hires \cite{Hires_2008} and Telescope Array \cite{TA_2016}.
The Kascade \cite{Kascade_2005} results are using a 5 component fit of H, He, CNO, MgSi and Fe groups using Sibyll~2.1. Therefore only the H and He spectra are compared directly as the other groups are strongly correlated.
Kascade-Grande \cite{Kascade-Grande_2017} results are using a 3 component fit of H, HeCNO and Heavy groups using Sibyll~2.3. Therefore only the H and Heavy spectra are compared, as the HeCNO group cannot be deconvoluted into a He and CNO part.
The Tunka \cite{Tunka_2013} results are using a 4 component fit of H, He, N and Fe groups using QGSJET~II-04.
The Pierre Auger Observatory \cite{Auger_2017} results are calculated by using their published elementary group fraction for H, He, N and Fe using Sibyll~2.3 convoluted with their most recent energy spectrum.  Note that differences in how different experiments handle intermediate elements (not one of the four groups used here) may lead to some small systematic differences in flux measurements between different experiments.
%\internalcite{IceTop\_3year\_update}
}
\label{f:individual_spectra_comparison}
\end{centering}
\end{figure*}

The elemental spectra results, shown in Figure~\ref{f:individual_spectra}, agree well with the recent H3a and H4a phenomenological models of the transition region between galactic and extragalactic cosmic rays (\cite{Gaisser_H4a}), in which heavier elements retain a harder spectral index to higher energies.  Within the statistical and systematical uncertainties the elemental spectra are also compatible with the phenomenological data fits, Gaisser-Stanev-Tilav (GST) \cite{GST_2013} fit and Global Spline Fit (GSF) \cite{Dembinski:2017zsh}.
These elemental spectra correlate with an increase in the mean-log-mass as a function of energy until about 100--200~PeV, as shown in Figure~\ref{f:lnA} (which is also derived from the individual fractions). 
Beyond this energy, 
given the statistical and systematic uncertainties 
the data are consistent with a composition that is either unchanging or decreasing. 
In Figure~\ref{f:individual_spectra_comparison}, our results are compared with those from other recent experiments: the results reported here indicate a higher flux in the iron group at high energies.  As shown in Section~\ref{sec:syst_hadintmodel}, the absolute scale of the composition is strongly dependent on which hadronic interaction model is used for the simulations.  In fact, the uncertainty due to the choice of hadronic interaction model is the biggest limitation on our analysis.

\subsection{Outlook}
The all-particle energy spectrum results presented here are consistent with each other and with previously published IceCube results \cite{it73_icetopalone_spectrum}.  The limiting systematic effect is the uncertainty of the snow coverage over the tanks.
The composition analysis results presented here are significantly improved from previously published results, which included only one month of data taken with a partly completed array \cite{Andeen:thesis,ic40_coincidence};
however, the present results are still limited by the amount of data on hand, the systematic uncertainty due to detector effects (particularly the light yield in the ice), and the dependence on the choice of hadronic interaction model used for the simulations.  For future analyses, we plan to include more years of experimental data, to simulate more intermediate elements, to investigate new composition-sensitive parameters currently under development, and to incorporate results from new internal studies to reduce the detector systematic uncertainties.  These updates will improve the precision of both analyses, and enable the extension of the analyses to higher and lower energies.  
Furthermore, the analyses presented here are well-suited to capitalize on future extensions to the IceCube Neutrino Observatory \cite{IceCubeGen2}.

\begin{acknowledgements}

The IceCube collaboration acknowledges the significant contributions to this manuscript from Marquette University, the University of Alaska Anchorage, and the University of Gent.  This paper also profited enormously from the contributions of our departed colleague Stefan Westerhoff (1967-2018).

USA {\textendash} U.S. National Science Foundation-Office of Polar Programs,
U.S. National Science Foundation-Physics Division,
Wisconsin Alumni Research Foundation,
Center for High Throughput Computing (CHTC) at the University of Wisconsin-Madison,
Open Science Grid (OSG),
Extreme Science and Engineering Discovery Environment (XSEDE),
U.S. Department of Energy-National Energy Research Scientific Computing Center,
Particle astrophysics research computing center at the University of Maryland,
Institute for Cyber-Enabled Research at Michigan State University,
and Astroparticle physics computational facility at Marquette University;
Belgium {\textendash} Funds for Scientific Research (FRS-FNRS and FWO),
FWO Odysseus and Big Science programmes,
and Belgian Federal Science Policy Office (Belspo);
Germany {\textendash} Bundesministerium f{\"u}r Bildung und Forschung (BMBF),
Deutsche Forschungsgemeinschaft (DFG),
Helmholtz Alliance for Astroparticle Physics (HAP),
Initiative and Networking Fund of the Helmholtz Association,
Deutsches Elektronen Synchrotron (DESY),
and High Performance Computing cluster of the RWTH Aachen;
Sweden {\textendash} Swedish Research Council,
Swedish Polar Research Secretariat,
Swedish National Infrastructure for Computing (SNIC),
and Knut and Alice Wallenberg Foundation;
Australia {\textendash} Australian Research Council;
Canada {\textendash} Natural Sciences and Engineering Research Council of Canada,
Calcul Qu{\'e}bec, Compute Ontario, Canada Foundation for Innovation, WestGrid, and Compute Canada;
Denmark {\textendash} Villum Fonden, Danish National Research Foundation (DNRF), Carlsberg Foundation;
New Zealand {\textendash} Marsden Fund;
Japan {\textendash} Japan Society for Promotion of Science (JSPS)
and Institute for Global Prominent Research (IGPR) of Chiba University;
Korea {\textendash} National Research Foundation of Korea (NRF);
Switzerland {\textendash} Swiss National Science Foundation (SNSF);
United Kingdom {\textendash} Department of Physics, University of Oxford.
\end{acknowledgements}

\bibliography{largebib}

\newpage
\appendix

\subsection{Table of Results}
\label{appendix:tables}

The values of the all particle energy spectrum from the coincident analysis, and the corresponding statistical and detector-related systematic uncertainties, as shown in Figure \ref{f:espec_italone}, are listed in Table \ref{table:icetop_only_all_particle}.
The values of the all particle energy spectrum from the coincident analysis, and the corresponding statistical and detector-related systematic uncertainties, as shown in Figure \ref{f:espec_coinc}, are listed in Table \ref{table:coinc_all_particle}.  The values for the elemental fluxes, as shown in Figure \ref{f:individual_spectra}, are listed in Tables \ref{table:proton_helium} and \ref{table:oxygen_iron}.

\newcolumntype{C}[1]{>{\centering\arraybackslash}p{#1}}

\begin{table*}[htb!]
\caption{Total flux IceTop-only analysis.}
\label{table:icetop_only_all_particle}
\begin{center}
\begin{tabular}{C{1.75cm}|C{1.75cm}|C{3.5cm}|C{3.5cm}|C{1.725cm}|C{1.725cm}|C{1.5cm}}
\hline
\multicolumn{1}{c}{Energy}& Bin Width	&	Flux & $\pm$ Stat. & \multicolumn{2}{c|}{$+$ Det.Syst. $-$}&  \\ 
\multicolumn{2}{c|}{(log$_{10}$(E$_{\mathrm{reco}}$ /GeV))} & GeV$^{-1}$ m$^{-2}$ s$^{-1}$ sr$^{-1}$& GeV$^{-1}$ m$^{-2}$ s$^{-1}$ sr$^{-1}$& \multicolumn{2}{c|}{GeV$^{-1}$ m$^{-2}$ s$^{-1}$ sr$^{-1}$} &   \\\hline \hline
%new
6.450 & 0.10 & 1.730 & 0.001 & 0.131 & 0.151 & $\times 10^{-13}$ \\
6.525 & 0.05 & 1.063 & 0.001 & 0.085 & 0.097 & $\times 10^{-13}$ \\
6.575 & 0.05 & 7.667 & 0.006 & 0.612 & 0.734 & $\times 10^{-14}$ \\
6.625 & 0.05 & 5.508 & 0.005 & 0.448 & 0.512 & $\times 10^{-14}$ \\
6.675 & 0.05 & 3.927 & 0.004 & 0.330 & 0.380 & $\times 10^{-14}$ \\
6.725 & 0.05 & 2.787 & 0.003 & 0.235 & 0.264 & $\times 10^{-14}$ \\
6.775 & 0.05 & 1.967 & 0.002 & 0.168 & 0.190 & $\times 10^{-14}$ \\
6.825 & 0.05 & 1.390 & 0.002 & 0.120 & 0.139 & $\times 10^{-14}$ \\
6.875 & 0.05 & 9.742 & 0.015 & 0.859 & 0.943 & $\times 10^{-15}$ \\
6.925 & 0.05 & 6.797 & 0.012 & 0.614 & 0.694 & $\times 10^{-15}$ \\
6.975 & 0.05 & 4.745 & 0.009 & 0.433 & 0.455 & $\times 10^{-15}$ \\
7.025 & 0.05 & 3.288 & 0.007 & 0.305 & 0.338 & $\times 10^{-15}$ \\
7.075 & 0.05 & 2.289 & 0.006 & 0.212 & 0.225 & $\times 10^{-15}$ \\
7.125 & 0.05 & 1.604 & 0.004 & 0.142 & 0.158 & $\times 10^{-15}$ \\
7.175 & 0.05 & 1.122 & 0.004 & 0.104 & 0.108 & $\times 10^{-15}$ \\
7.225 & 0.05 & 7.893 & 0.028 & 0.709 & 0.709 & $\times 10^{-16}$ \\
7.275 & 0.05 & 5.591 & 0.022 & 0.488 & 0.547 & $\times 10^{-16}$ \\
7.325 & 0.05 & 3.960 & 0.018 & 0.368 & 0.385 & $\times 10^{-16}$ \\
7.375 & 0.05 & 2.828 & 0.014 & 0.244 & 0.248 & $\times 10^{-16}$ \\
7.425 & 0.05 & 2.011 & 0.011 & 0.188 & 0.192 & $\times 10^{-16}$ \\
7.475 & 0.05 & 1.430 & 0.009 & 0.128 & 0.132 & $\times 10^{-16}$ \\
7.525 & 0.05 & 1.031 & 0.007 & 0.093 & 0.091 & $\times 10^{-16}$ \\
7.575 & 0.05 & 7.431 & 0.057 & 0.599 & 0.631 & $\times 10^{-17}$ \\
7.625 & 0.05 & 5.350 & 0.045 & 0.507 & 0.486 & $\times 10^{-17}$ \\
7.675 & 0.05 & 3.754 & 0.036 & 0.317 & 0.423 & $\times 10^{-17}$ \\
7.725 & 0.05 & 2.640 & 0.028 & 0.258 & 0.245 & $\times 10^{-17}$ \\
7.775 & 0.05 & 1.943 & 0.023 & 0.168 & 0.141 & $\times 10^{-17}$ \\
7.825 & 0.05 & 1.369 & 0.018 & 0.109 & 0.140 & $\times 10^{-17}$ \\
7.875 & 0.05 & 1.008 & 0.015 & 0.121 & 0.086 & $\times 10^{-17}$ \\
7.925 & 0.05 & 6.912 & 0.115 & 0.529 & 0.704 & $\times 10^{-18}$ \\
7.975 & 0.05 & 4.948 & 0.092 & 0.441 & 0.393 & $\times 10^{-18}$ \\
8.050 & 0.10 & 2.934 & 0.046 & 0.291 & 0.273 & $\times 10^{-18}$ \\
8.150 & 0.10 & 1.468 & 0.029 & 0.142 & 0.179 & $\times 10^{-18}$ \\
8.250 & 0.10 & 6.991 & 0.178 & 0.725 & 0.551 & $\times 10^{-19}$ \\
8.350 & 0.10 & 3.146 & 0.106 & 0.282 & 0.324 & $\times 10^{-19}$ \\
8.450 & 0.10 & 1.602 & 0.068 & 0.144 & 0.155 & $\times 10^{-19}$ \\
8.550 & 0.10 & 6.468 & 0.382 & 0.637 & 0.798 & $\times 10^{-20}$ \\
8.650 & 0.10 & 3.169 & 0.239 & 0.390 & 0.280 & $\times 10^{-20}$ \\
8.750 & 0.10 & 1.350 & 0.139 & 0.443 & 0.355 & $\times 10^{-20}$ \\
8.850 & 0.10 & 6.091 & 0.837 & 0.612 & 0.842 & $\times 10^{-21}$ \\
8.950 & 0.10 & 2.492 & 0.480 & 0.514 & 0.440 & $\times 10^{-21}$ \\
9.050 & 0.10 & 8.199 & 2.472 & 4.278 & 2.228 & $\times 10^{-22}$ \\
\end{tabular}
\label{table:appendix_result_italone}
\end{center}
\end{table*}

\begin{table*}[htb!]
\caption{Total flux coincidence analysis.}
\label{table:coinc_all_particle}
\begin{center}
\begin{tabular}{C{1.75cm}|C{1.75cm}|C{3.5cm}|C{3.5cm}|C{1.725cm}|C{1.725cm}|C{1.5cm}}
\hline
 \multicolumn{1}{c}{Energy} & Bin Width	&	Flux & $\pm$ Stat. & \multicolumn{2}{c|}{$-$ Det.Syst. $+$} & \\ 
\multicolumn{2}{c|}{(log$_{10}$(E$_{\mathrm{reco}}$ /GeV))} & GeV$^{-1}$ m$^{-2}$ s$^{-1}$ sr$^{-1}$& GeV$^{-1}$ m$^{-2}$ s$^{-1}$ sr$^{-1}$& \multicolumn{2}{c|}{GeV$^{-1}$ m$^{-2}$ s$^{-1}$ sr$^{-1}$} & \\\hline \hline
%New
6.55 & 0.1 & 9.005 & 0.008 & 0.587 & 0.668 & $\times 10^{-14}$ \\
6.65 & 0.1 & 4.612 & 0.005 & 0.317 & 0.377 & $\times 10^{-14}$ \\
6.75 & 0.1 & 2.323 & 0.003 & 0.168 & 0.203 & $\times 10^{-14}$ \\
6.85 & 0.1 & 1.158 & 0.002 & 0.087 & 0.103 & $\times 10^{-14}$ \\
6.95 & 0.1 & 5.635 & 0.014 & 0.446 & 0.524 & $\times 10^{-15}$ \\
7.05 & 0.1 & 2.698 & 0.008 & 0.220 & 0.254 & $\times 10^{-15}$ \\
7.15 & 0.1 & 1.308 & 0.005 & 0.094 & 0.117 & $\times 10^{-15}$ \\
7.25 & 0.1 & 6.473 & 0.032 & 0.473 & 0.592 & $\times 10^{-16}$ \\
7.35 & 0.1 & 3.250 & 0.020 & 0.235 & 0.266 & $\times 10^{-16}$ \\
7.45 & 0.1 & 1.657 & 0.013 & 0.116 & 0.133 & $\times 10^{-16}$ \\
7.55 & 0.1 & 8.600 & 0.083 & 0.596 & 0.685 & $\times 10^{-17}$ \\
7.65 & 0.1 & 4.399 & 0.053 & 0.377 & 0.430 & $\times 10^{-17}$ \\
7.75 & 0.1 & 2.194 & 0.033 & 0.137 & 0.219 & $\times 10^{-17}$ \\
7.85 & 0.1 & 1.132 & 0.021 & 0.098 & 0.086 & $\times 10^{-17}$ \\
7.95 & 0.1 & 5.532 & 0.133 & 0.483 & 0.574 & $\times 10^{-18}$ \\
8.10 & 0.2 & 2.062 & 0.048 & 0.148 & 0.205 & $\times 10^{-18}$ \\
8.30 & 0.2 & 4.619 & 0.181 & 0.429 & 0.549 & $\times 10^{-19}$ \\
8.50 & 0.2 & 9.966 & 0.666 & 1.019 & 1.495 & $\times 10^{-20}$ \\
8.70 & 0.2 & 2.237 & 0.250 & 0.124 & 0.206 & $\times 10^{-20}$ \\
8.90 & 0.2 & 5.283 & 0.964 & 0.720 & 1.021 & $\times 10^{-21}$ \\
%Old
% 6.55 & 0.1 & 9.0049e-14 & 8.4237e-17 & 5.8746e-15 & 6.6752e-15\\
% 6.65 & 0.1 & 4.6120e-14 & 5.3830e-17 & 3.1719e-15 & 3.7672e-15\\
% 6.75 & 0.1 & 2.3234e-14 & 3.4246e-17 & 1.6812e-15 & 2.0252e-15\\
% 6.85 & 0.1 & 1.1584e-14 & 2.1716e-17 & 8.7407e-16 & 1.0305e-15\\
% 6.95 & 0.1 & 5.6351e-15 & 1.3548e-17 & 4.4583e-16 & 5.2388e-16\\
% 7.05 & 0.1 & 2.6977e-15 & 8.3460e-18 & 2.1975e-16 & 2.5376e-16\\
% 7.15 & 0.1 & 1.3083e-15 & 5.1748e-18 & 9.4493e-17 & 1.1737e-16\\
% 7.25 & 0.1 & 6.4728e-16 & 3.2409e-18 & 4.7337e-17 & 5.9231e-17\\
% 7.35 & 0.1 & 3.2496e-16 & 2.0445e-18 & 2.3462e-17 & 2.6625e-17\\
% 7.45 & 0.1 & 1.6572e-16 & 1.3000e-18 & 1.1554e-17 & 1.3341e-17\\
% 7.55 & 0.1 & 8.6003e-17 & 8.3382e-19 & 5.9614e-18 & 6.8521e-18\\
% 7.65 & 0.1 & 4.3990e-17 & 5.3096e-19 & 3.7708e-18 & 4.3033e-18\\
% 7.75 & 0.1 & 2.1944e-17 & 3.3389e-19 & 1.3697e-18 & 2.1877e-18\\
% 7.85 & 0.1 & 1.1324e-17 & 2.1356e-19 & 9.8267e-19 & 8.6129e-19\\
% 7.95 & 0.1 & 5.5318e-18 & 1.3290e-19 & 4.8330e-19 & 5.7379e-19\\
% 8.10 & 0.2 & 2.0619e-18 & 4.8202e-20 & 1.4835e-19 & 2.0505e-19\\
% 8.30 & 0.2 & 4.6190e-19 & 1.8086e-20 & 4.2901e-20 & 5.4912e-20\\
% 8.50 & 0.2 & 9.9658e-20 & 6.6602e-21 & 1.0195e-20 & 1.4951e-20\\
% 8.70 & 0.2 & 2.2371e-20 & 2.5016e-21 & 1.2386e-21 & 2.0616e-21\\
% 8.90 & 0.2 & 5.2831e-21 & 9.6379e-22 & 7.2014e-22 & 1.0212e-21\\
\end{tabular}
\label{table:appendix_result}
\end{center}
\end{table*}

%6.55 & 0.1 & 9.0049e-14 & 8.4237e-17 & 5.8687e-15 & 6.6752e-15\\
%6.65 & 0.1 & 4.6120e-14 & 5.3830e-17 & 3.1722e-15 & 3.7672e-15\\
%6.75 & 0.1 & 2.3234e-14 & 3.4246e-17 & 1.6810e-15 & 2.0252e-15\\
%6.85 & 0.1 & 1.1584e-14 & 2.1716e-17 & 8.7362e-16 & 1.0305e-15\\
%6.95 & 0.1 & 5.6351e-15 & 1.3548e-17 & 4.4578e-16 & 5.2388e-16\\
%7.05 & 0.1 & 2.6977e-15 & 8.3460e-18 & 2.1966e-16 & 2.5376e-16\\
%7.15 & 0.1 & 1.3083e-15 & 5.1748e-18 & 9.4404e-17 & 1.1737e-16\\
%7.25 & 0.1 & 6.4728e-16 & 3.2409e-18 & 4.7338e-17 & 5.9226e-17\\
%7.35 & 0.1 & 3.2496e-16 & 2.0445e-18 & 2.3457e-17 & 2.6609e-17\\
%7.45 & 0.1 & 1.6572e-16 & 1.3000e-18 & 1.1554e-17 & 1.3341e-17\\
%7.55 & 0.1 & 8.6003e-17 & 8.3382e-19 & 5.9613e-18 & 6.8521e-18\\
%7.65 & 0.1 & 4.3990e-17 & 5.3096e-19 & 3.7695e-18 & 4.3033e-18\\
%7.75 & 0.1 & 2.1944e-17 & 3.3389e-19 & 1.3697e-18 & 2.1877e-18\\
%7.85 & 0.1 & 1.1324e-17 & 2.1356e-19 & 9.8301e-19 & 8.6129e-19\\
%7.95 & 0.1 & 5.5318e-18 & 1.3290e-19 & 4.8324e-19 & 5.7379e-19\\
%8.10 & 0.2 & 2.0619e-18 & 4.8202e-20 & 1.4835e-19 & 2.0509e-19\\
%8.30 & 0.2 & 4.6190e-19 & 1.8086e-20 & 4.2901e-20 & 5.4912e-20\\
%8.50 & 0.2 & 9.9658e-20 & 6.6602e-21 & 1.0195e-20 & 1.4952e-20\\
%8.70 & 0.2 & 2.2371e-20 & 2.5016e-21 & 1.2386e-21 & 2.0616e-21\\
%8.90 & 0.2 & 5.2831e-21 & 9.6379e-22 & 7.2014e-22 & 1.0212e-21\\

\begin{table*}[htb!]
\caption{Proton and helium group flux.}
\label{table:proton_helium}
\begin{center}
\begin{tabular}{C{1.75cm}|C{1.75cm}|C{3.5cm}||C{1.725cm}|C{1.725cm}|C{1.725cm}|C{1.725cm}|C{1.5cm}}
\hline
\multicolumn{1}{c}{Energy} & Bin Width & Flux & \multicolumn{2}{c|}{$-$ Stat. $+$} & \multicolumn{2}{c|}{$-$ Det.Syst. $+$} & \\ 
\multicolumn{2}{c|}{(log$_{10}$(E$_{\mathrm{reco}}$ /GeV))} & GeV$^{-1}$ m$^{-2}$ s$^{-1}$ sr$^{-1}$& \multicolumn{2}{c|}{GeV$^{-1}$ m$^{-2}$ s$^{-1}$ sr$^{-1}$}& \multicolumn{2}{c|}{GeV$^{-1}$ m$^{-2}$ s$^{-1}$ sr$^{-1}$} & \\\hline \hline
Proton & & & & & & & \\
\hline
6.55 & 0.1 & 2.791 & 0.448 & 0.471 & 1.253 & 1.289 & $\times 10^{-14}$ \\
6.65 & 0.1 & 1.350 & 0.170 & 0.173 & 0.637 & 0.629 & $\times 10^{-14}$ \\
6.75 & 0.1 & 7.838 & 1.055 & 1.092 & 3.834 & 3.790 & $\times 10^{-15}$ \\
6.85 & 0.1 & 3.456 & 0.542 & 0.563 & 2.013 & 2.045 & $\times 10^{-15}$ \\
6.95 & 0.1 & 1.393 & 0.201 & 0.207 & 0.785 & 0.816 & $\times 10^{-15}$ \\
7.05 & 0.1 & 6.684 & 1.462 & 1.550 & 4.106 & 4.335 & $\times 10^{-16}$ \\
7.15 & 0.1 & 2.656 & 0.601 & 0.635 & 1.771 & 1.943 & $\times 10^{-16}$ \\
7.25 & 0.1 & 1.040 & 0.230 & 0.231 & 0.643 & 0.727 & $\times 10^{-16}$ \\
7.35 & 0.1 & 5.291 & 1.556 & 1.671 & 3.677 & 4.417 & $\times 10^{-17}$ \\
7.45 & 0.1 & 3.172 & 0.807 & 0.851 & 2.218 & 2.168 & $\times 10^{-17}$ \\
7.55 & 0.1 & 1.294 & 0.405 & 0.412 & 1.082 & 1.037 & $\times 10^{-17}$ \\
7.65 & 0.1 & 8.205 & 1.609 & 1.022 & 4.838 & 4.461 & $\times 10^{-18}$ \\
7.75 & 0.1 & 2.536 & 0.646 & 0.691 & 1.167 & 1.866 & $\times 10^{-18}$ \\
7.85 & 0.1 & 1.194 & 0.379 & 0.431 & 0.796 & 1.246 & $\times 10^{-18}$ \\
7.95 & 0.1 & 8.318 & 2.957 & 2.574 & 6.256 & 2.819 & $\times 10^{-19}$ \\
8.10 & 0.2 & 2.422 & 1.118 & 0.703 & 1.441 & 1.101 & $\times 10^{-19}$ \\
8.30 & 0.2 & 5.257 & 1.781 & 1.589 & 3.663 & 1.860 & $\times 10^{-20}$ \\
8.50 & 0.2 & 1.340 & 0.437 & 0.406 & 0.631 & 0.845 & $\times 10^{-20}$ \\
8.70 & 0.2 & 3.562 & 2.825 & 3.316 & 2.908 & 5.462 & $\times 10^{-21}$ \\
8.90 & 0.2 & 0.016 & 0.016 & 2.319 & 0.000 & 2.115 & $\times 10^{-21}$ \\
% 6.55 & 0.1 & 2.7906e-14 & 4.4786e-15 & 4.7067e-15 & 1.2534e-14 & 1.2888e-14\\
% 6.65 & 0.1 & 1.3496e-14 & 1.6966e-15 & 1.7348e-15 & 6.3740e-15 & 6.2854e-15\\
% 6.75 & 0.1 & 7.8379e-15 & 1.0555e-15 & 1.0919e-15 & 3.8338e-15 & 3.7944e-15\\
% 6.85 & 0.1 & 3.4563e-15 & 5.4215e-16 & 5.6256e-16 & 2.0130e-15 & 2.0453e-15\\
% 6.95 & 0.1 & 1.3932e-15 & 2.0142e-16 & 2.0666e-16 & 7.8527e-16 & 8.1643e-16\\
% 7.05 & 0.1 & 6.6844e-16 & 1.4620e-16 & 1.5496e-16 & 4.1056e-16 & 4.3345e-16\\
% 7.15 & 0.1 & 2.6561e-16 & 6.0132e-17 & 6.3519e-17 & 1.7713e-16 & 1.9418e-16\\
% 7.25 & 0.1 & 1.0401e-16 & 2.2997e-17 & 2.3054e-17 & 6.4190e-17 & 7.2726e-17\\
% 7.35 & 0.1 & 5.2912e-17 & 1.5563e-17 & 1.6705e-17 & 3.6871e-17 & 4.3973e-17\\
% 7.45 & 0.1 & 3.1722e-17 & 8.0669e-18 & 8.5088e-18 & 2.2176e-17 & 2.1677e-17\\
% 7.55 & 0.1 & 1.2937e-17 & 4.0482e-18 & 4.1211e-18 & 1.0820e-17 & 1.0349e-17\\
% 7.65 & 0.1 & 8.2046e-18 & 1.6089e-18 & 1.0219e-18 & 4.8378e-18 & 4.4545e-18\\
% 7.75 & 0.1 & 2.5361e-18 & 6.4621e-19 & 6.9071e-19 & 1.1671e-18 & 1.8659e-18\\
% 7.85 & 0.1 & 1.1937e-18 & 3.7941e-19 & 4.3092e-19 & 7.9555e-19 & 1.2367e-18\\
% 7.95 & 0.1 & 8.3181e-19 & 2.9567e-19 & 2.5739e-19 & 6.2558e-19 & 2.8144e-19\\
% 8.10 & 0.2 & 2.4217e-19 & 1.1176e-19 & 7.0320e-20 & 1.4412e-19 & 1.1033e-19\\
% 8.30 & 0.2 & 5.2570e-20 & 1.7814e-20 & 1.5887e-20 & 3.6627e-20 & 1.8596e-20\\
% 8.50 & 0.2 & 1.3401e-20 & 4.3691e-21 & 4.0624e-21 & 6.3068e-21 & 8.5046e-21\\
% 8.70 & 0.2 & 3.5617e-21 & 2.8248e-21 & 3.3161e-21 & 2.9081e-21 & 5.4622e-21\\
% 8.90 & 0.2 & 1.6361e-23 & 1.6361e-23 & 2.3193e-21 & 0.0000e+00 & 2.1155e-21\\

\hline
Helium & & & & & & & \\
\hline
6.55 & 0.1 & 2.232 & 0.981 & 0.974 & 0.761 & 0.514 & $\times 10^{-14}$ \\
6.65 & 0.1 & 1.520 & 0.401 & 0.412 & 0.373 & 0.367 & $\times 10^{-14}$ \\
6.75 & 0.1 & 4.798 & 2.109 & 2.168 & 2.146 & 1.984 & $\times 10^{-15}$ \\
6.85 & 0.1 & 2.801 & 1.166 & 1.161 & 1.360 & 1.174 & $\times 10^{-15}$ \\
6.95 & 0.1 & 1.682 & 0.406 & 0.412 & 0.301 & 0.219 & $\times 10^{-15}$ \\
7.05 & 0.1 & 6.067 & 3.120 & 3.049 & 2.295 & 1.549 & $\times 10^{-16}$ \\
7.15 & 0.1 & 4.146 & 1.407 & 1.399 & 0.804 & 0.592 & $\times 10^{-16}$ \\
7.25 & 0.1 & 2.241 & 0.649 & 0.674 & 0.290 & 0.108 & $\times 10^{-16}$ \\
7.35 & 0.1 & 7.893 & 3.603 & 3.526 & 2.571 & 1.885 & $\times 10^{-17}$ \\
7.45 & 0.1 & 2.194 & 1.693 & 1.668 & 0.738 & 1.121 & $\times 10^{-17}$ \\
7.55 & 0.1 & 1.482 & 0.708 & 0.720 & 0.521 & 0.718 & $\times 10^{-17}$ \\
7.65 & 0.1 & 0.118 & 0.118 & 3.143 & 0.118 & 2.046 & $\times 10^{-18}$ \\
7.75 & 0.1 & 2.316 & 1.380 & 1.379 & 0.948 & 0.449 & $\times 10^{-18}$ \\
7.85 & 0.1 & 1.402 & 0.922 & 0.858 & 1.067 & 0.110 & $\times 10^{-18}$ \\
7.95 & 0.1 & 3.182 & 3.182 & 6.307 & 0.000 & 8.377 & $\times 10^{-19}$ \\
8.10 & 0.2 & 0.081 & 0.081 & 2.774 & 0.081 & 2.159 & $\times 10^{-19}$ \\
8.30 & 0.2 & 0.000 & 0.000 & 3.264 & 0.000 & 7.476 & $\times 10^{-20}$ \\
8.50 & 0.2 & 0.001 & 0.001 & 7.633 & 0.001 & 0.070 & $\times 10^{-21}$ \\
8.70 & 0.2 & 4.096 & 4.096 & 4.094 & 4.096 & 2.171 & $\times 10^{-21}$ \\
8.90 & 0.2 & 2.785 & 2.785 & 1.181 & 2.785 & 0.165 & $\times 10^{-21}$ \\
% 6.55 & 0.1 & 2.2320e-14 & 9.8057e-15 & 9.7445e-15 & 7.5798e-15 & 5.1364e-15\\
% 6.65 & 0.1 & 1.5202e-14 & 4.0093e-15 & 4.1164e-15 & 3.7262e-15 & 3.6669e-15\\
% 6.75 & 0.1 & 4.7981e-15 & 2.1095e-15 & 2.1679e-15 & 2.1518e-15 & 1.9837e-15\\
% 6.85 & 0.1 & 2.8014e-15 & 1.1658e-15 & 1.1614e-15 & 1.3592e-15 & 1.1742e-15\\
% 6.95 & 0.1 & 1.6819e-15 & 4.0626e-16 & 4.1180e-16 & 3.0194e-16 & 2.1932e-16\\
% 7.05 & 0.1 & 6.0673e-16 & 3.1203e-16 & 3.0493e-16 & 2.2788e-16 & 1.5485e-16\\
% 7.15 & 0.1 & 4.1458e-16 & 1.4075e-16 & 1.3990e-16 & 7.9148e-17 & 5.9167e-17\\
% 7.25 & 0.1 & 2.2406e-16 & 6.4867e-17 & 6.7408e-17 & 2.8992e-17 & 1.0523e-17\\
% 7.35 & 0.1 & 7.8931e-17 & 3.6026e-17 & 3.5255e-17 & 2.5378e-17 & 1.9071e-17\\
% 7.45 & 0.1 & 2.1944e-17 & 1.6932e-17 & 1.6684e-17 & 7.3765e-18 & 1.1212e-17\\
% 7.55 & 0.1 & 1.4821e-17 & 7.0807e-18 & 7.1988e-18 & 5.1506e-18 & 7.1775e-18\\
% 7.65 & 0.1 & 1.1842e-19 & 1.1842e-19 & 3.1426e-18 & 1.1842e-19 & 2.0605e-18\\
% 7.75 & 0.1 & 2.3156e-18 & 1.3796e-18 & 1.3792e-18 & 9.4788e-19 & 4.4872e-19\\
% 7.85 & 0.1 & 1.4020e-18 & 9.2161e-19 & 8.5754e-19 & 1.0573e-18 & 1.0964e-19\\
% 7.95 & 0.1 & 3.1820e-19 & 3.1820e-19 & 6.3070e-19 & 0.0000e+00 & 8.3918e-19\\
% 8.10 & 0.2 & 8.1497e-21 & 8.1497e-21 & 2.7743e-19 & 8.1497e-21 & 2.1588e-19\\
% 8.30 & 0.2 & 5.0014e-25 & 5.0014e-25 & 3.2636e-20 & 0.0000e+00 & 7.4760e-20\\
% 8.50 & 0.2 & 1.1816e-24 & 1.1816e-24 & 7.6326e-21 & 1.1816e-24 & 6.9811e-23\\
% 8.70 & 0.2 & 4.0957e-21 & 4.0957e-21 & 4.0937e-21 & 4.0957e-21 & 2.1715e-21\\
% 8.90 & 0.2 & 2.7852e-21 & 2.7852e-21 & 1.1812e-21 & 2.7852e-21 & 1.6489e-22\\
\end{tabular}
\label{table:appendix_result_H_He}
\end{center}
\end{table*}

\begin{table*}[htb!]
\caption{Oxygen and iron group fluxes.}
\label{table:oxygen_iron}
\begin{center}
\begin{tabular}{C{1.75cm}|C{1.75cm}|C{3.5cm}||C{1.725cm}|C{1.725cm}|C{1.725cm}|C{1.725cm}|C{1.5cm}}
\hline
\multicolumn{1}{c}{Energy} & Bin Width & Flux & \multicolumn{2}{c|}{$-$ Stat. $+$} & \multicolumn{2}{c|}{$-$ Det.Syst. $+$} & \\ 
\multicolumn{2}{c|}{(log$_{10}$(E$_{\mathrm{reco}}$ /GeV))} & GeV$^{-1}$ m$^{-2}$ s$^{-1}$ sr$^{-1}$& \multicolumn{2}{c|}{GeV$^{-1}$ m$^{-2}$ s$^{-1}$ sr$^{-1}$}& \multicolumn{2}{c|}{GeV$^{-1}$ m$^{-2}$ s$^{-1}$ sr$^{-1}$} & \\\hline \hline
Oxygen & & & & & & &\\
\hline
6.55 & 0.1 & 1.975 & 0.972 & 0.942 & 0.857 & 0.479 & $\times 10^{-14}$ \\
6.65 & 0.1 & 6.714 & 5.305 & 5.097 & 5.909 & 2.597 & $\times 10^{-15}$ \\
6.75 & 0.1 & 5.803 & 2.068 & 1.978 & 2.040 & 1.016 & $\times 10^{-15}$ \\
6.85 & 0.1 & 2.766 & 1.115 & 1.108 & 1.137 & 0.690 & $\times 10^{-15}$ \\
6.95 & 0.1 & 1.206 & 0.395 & 0.386 & 0.355 & 0.112 & $\times 10^{-15}$ \\
7.05 & 0.1 & 8.757 & 3.037 & 3.055 & 2.852 & 1.583 & $\times 10^{-16}$ \\
7.15 & 0.1 & 2.831 & 1.354 & 1.352 & 1.531 & 0.594 & $\times 10^{-16}$ \\
7.25 & 0.1 & 9.002 & 7.871 & 7.627 & 9.002 & 6.196 & $\times 10^{-17}$ \\
7.35 & 0.1 & 9.934 & 3.663 & 3.712 & 6.989 & 3.545 & $\times 10^{-17}$ \\
7.45 & 0.1 & 6.723 & 1.528 & 1.538 & 3.490 & 1.441 & $\times 10^{-17}$ \\
7.55 & 0.1 & 2.647 & 0.625 & 0.614 & 2.095 & 1.069 & $\times 10^{-17}$ \\
7.65 & 0.1 & 2.200 & 0.311 & 0.203 & 1.149 & 0.340 & $\times 10^{-17}$ \\
7.75 & 0.1 & 9.188 & 1.493 & 1.517 & 4.586 & 3.184 & $\times 10^{-18}$ \\
7.85 & 0.1 & 3.866 & 0.849 & 0.889 & 2.925 & 2.045 & $\times 10^{-18}$ \\
7.95 & 0.1 & 1.792 & 0.601 & 0.522 & 1.792 & 0.719 & $\times 10^{-18}$ \\
8.10 & 0.2 & 7.187 & 2.751 & 1.536 & 7.187 & 4.681 & $\times 10^{-19}$ \\
8.30 & 0.2 & 2.101 & 0.452 & 0.396 & 1.786 & 0.731 & $\times 10^{-19}$ \\
8.50 & 0.2 & 1.055 & 1.055 & 1.058 & 1.055 & 3.217 & $\times 10^{-20}$ \\
8.70 & 0.2 & 0.001 & 0.001 & 3.993 & 0.001 & 0.000 & $\times 10^{-21}$ \\
8.90 & 0.2 & 0.001 & 0.001 & 1.561 & 0.001 & 2.915 & $\times 10^{-21}$ \\
% 6.55 & 0.1 & 1.9747e-14 & 9.7185e-15 & 9.4212e-15 & 8.5661e-15 & 4.7599e-15\\
% 6.65 & 0.1 & 6.7138e-15 & 5.3052e-15 & 5.0970e-15 & 5.9089e-15 & 2.5963e-15\\
% 6.75 & 0.1 & 5.8033e-15 & 2.0678e-15 & 1.9783e-15 & 2.0396e-15 & 1.0165e-15\\
% 6.85 & 0.1 & 2.7657e-15 & 1.1149e-15 & 1.1081e-15 & 1.1372e-15 & 6.8890e-16\\
% 6.95 & 0.1 & 1.2059e-15 & 3.9459e-16 & 3.8636e-16 & 3.5492e-16 & 1.1224e-16\\
% 7.05 & 0.1 & 8.7568e-16 & 3.0373e-16 & 3.0551e-16 & 2.8525e-16 & 1.5737e-16\\
% 7.15 & 0.1 & 2.8307e-16 & 1.3536e-16 & 1.3522e-16 & 1.5314e-16 & 5.8205e-17\\
% 7.25 & 0.1 & 9.0022e-17 & 7.8710e-17 & 7.6269e-17 & 9.0022e-17 & 6.1959e-17\\
% 7.35 & 0.1 & 9.9338e-17 & 3.6632e-17 & 3.7124e-17 & 7.0244e-17 & 3.5200e-17\\
% 7.45 & 0.1 & 6.7229e-17 & 1.5277e-17 & 1.5375e-17 & 3.4903e-17 & 1.4413e-17\\
% 7.55 & 0.1 & 2.6472e-17 & 6.2532e-18 & 6.1411e-18 & 2.0952e-17 & 1.0616e-17\\
% 7.65 & 0.1 & 2.1996e-17 & 3.1135e-18 & 2.0254e-18 & 1.1487e-17 & 3.4055e-18\\
% 7.75 & 0.1 & 9.1884e-18 & 1.4926e-18 & 1.5172e-18 & 4.5858e-18 & 3.1836e-18\\
% 7.85 & 0.1 & 3.8662e-18 & 8.4914e-19 & 8.8871e-19 & 2.9245e-18 & 2.0327e-18\\
% 7.95 & 0.1 & 1.7920e-18 & 6.0110e-19 & 5.2246e-19 & 1.7920e-18 & 7.1089e-19\\
% 8.10 & 0.2 & 7.1872e-19 & 2.5524e-19 & 1.5363e-19 & 7.1872e-19 & 4.6924e-19\\
% 8.30 & 0.2 & 2.1006e-19 & 4.5208e-20 & 3.9567e-20 & 1.7857e-19 & 7.3093e-20\\
% 8.50 & 0.2 & 1.0553e-20 & 1.0553e-20 & 1.0585e-20 & 1.0553e-20 & 3.2064e-20\\
% 8.70 & 0.2 & 7.6933e-25 & 7.6933e-25 & 3.9934e-21 & 7.6933e-25 & 0.0000e+00\\
% 8.90 & 0.2 & 6.3612e-25 & 6.3612e-25 & 1.5606e-21 & 6.2234e-25 & 2.9150e-21\\
\hline
Iron & & & & & & & \\
\hline
6.55 & 0.1 & 2.008 & 0.436 & 0.460 & 0.897 & 1.733 & $\times 10^{-14}$ \\
6.65 & 0.1 & 1.071 & 0.257 & 0.274 & 0.475 & 0.945 & $\times 10^{-14}$ \\
6.75 & 0.1 & 4.795 & 0.894 & 0.949 & 2.073 & 4.133 & $\times 10^{-15}$ \\
6.85 & 0.1 & 2.561 & 0.477 & 0.489 & 1.107 & 2.061 & $\times 10^{-15}$ \\
6.95 & 0.1 & 1.354 & 0.165 & 0.176 & 0.496 & 0.989 & $\times 10^{-15}$ \\
7.05 & 0.1 & 5.468 & 1.395 & 1.428 & 2.907 & 5.366 & $\times 10^{-16}$ \\
7.15 & 0.1 & 3.450 & 0.548 & 0.567 & 1.447 & 2.791 & $\times 10^{-16}$ \\
7.25 & 0.1 & 2.292 & 0.332 & 0.352 & 0.953 & 1.529 & $\times 10^{-16}$ \\
7.35 & 0.1 & 9.378 & 1.726 & 1.759 & 5.213 & 9.010 & $\times 10^{-17}$ \\
7.45 & 0.1 & 4.482 & 0.679 & 0.703 & 2.183 & 4.418 & $\times 10^{-17}$ \\
7.55 & 0.1 & 3.177 & 0.328 & 0.353 & 1.539 & 2.486 & $\times 10^{-17}$ \\
7.65 & 0.1 & 1.367 & 0.145 & 0.162 & 0.630 & 1.353 & $\times 10^{-17}$ \\
7.75 & 0.1 & 7.904 & 0.842 & 0.853 & 4.190 & 6.535 & $\times 10^{-18}$ \\
7.85 & 0.1 & 4.862 & 0.441 & 0.445 & 2.142 & 3.552 & $\times 10^{-18}$ \\
7.95 & 0.1 & 2.590 & 0.293 & 0.299 & 1.135 & 1.809 & $\times 10^{-18}$ \\
8.10 & 0.2 & 1.093 & 0.121 & 0.135 & 0.480 & 0.683 & $\times 10^{-18}$ \\
8.30 & 0.2 & 1.993 & 0.334 & 0.354 & 0.815 & 1.818 & $\times 10^{-19}$ \\
8.50 & 0.2 & 7.570 & 1.043 & 1.092 & 3.116 & 2.111 & $\times 10^{-20}$ \\
8.70 & 0.2 & 1.471 & 0.251 & 0.286 & 0.133 & 0.419 & $\times 10^{-20}$ \\
8.90 & 0.2 & 2.481 & 1.036 & 1.266 & 1.043 & 1.693 & $\times 10^{-21}$ \\
% 6.55 & 0.1 & 2.0076e-14 & 4.3579e-15 & 4.6002e-15 & 8.9445e-15 & 1.7332e-14\\
% 6.65 & 0.1 & 1.0708e-14 & 2.5729e-15 & 2.7444e-15 & 4.7453e-15 & 9.4494e-15\\
% 6.75 & 0.1 & 4.7949e-15 & 8.9370e-16 & 9.4942e-16 & 2.0718e-15 & 4.1329e-15\\
% 6.85 & 0.1 & 2.5608e-15 & 4.7747e-16 & 4.8946e-16 & 1.1053e-15 & 2.0607e-15\\
% 6.95 & 0.1 & 1.3541e-15 & 1.6494e-16 & 1.7640e-16 & 4.9651e-16 & 9.8859e-16\\
% 7.05 & 0.1 & 5.4684e-16 & 1.3947e-16 & 1.4277e-16 & 2.9046e-16 & 5.3660e-16\\
% 7.15 & 0.1 & 3.4501e-16 & 5.4846e-17 & 5.6698e-17 & 1.4362e-16 & 2.7911e-16\\
% 7.25 & 0.1 & 2.2919e-16 & 3.3237e-17 & 3.5190e-17 & 9.5254e-17 & 1.5320e-16\\
% 7.35 & 0.1 & 9.3776e-17 & 1.7256e-17 & 1.7585e-17 & 5.1983e-17 & 9.0199e-17\\
% 7.45 & 0.1 & 4.4824e-17 & 6.7859e-18 & 7.0274e-18 & 2.1827e-17 & 4.4183e-17\\
% 7.55 & 0.1 & 3.1772e-17 & 3.2806e-18 & 3.5274e-18 & 1.5352e-17 & 2.4856e-17\\
% 7.65 & 0.1 & 1.3672e-17 & 1.4549e-18 & 1.6190e-18 & 6.3086e-18 & 1.3534e-17\\
% 7.75 & 0.1 & 7.9038e-18 & 8.4219e-19 & 8.5326e-19 & 4.1903e-18 & 6.5355e-18\\
% 7.85 & 0.1 & 4.8619e-18 & 4.4056e-19 & 4.4546e-19 & 2.1417e-18 & 3.5520e-18\\
% 7.95 & 0.1 & 2.5898e-18 & 2.9273e-19 & 2.9924e-19 & 1.1319e-18 & 1.8093e-18\\
% 8.10 & 0.2 & 1.0928e-18 & 1.2142e-19 & 1.3512e-19 & 4.8093e-19 & 6.8341e-19\\
% 8.30 & 0.2 & 1.9928e-19 & 3.3365e-20 & 3.5448e-20 & 8.1524e-20 & 1.8182e-19\\
% 8.50 & 0.2 & 7.5703e-20 & 1.0430e-20 & 1.0916e-20 & 3.1102e-20 & 2.1113e-20\\
% 8.70 & 0.2 & 1.4712e-20 & 2.5145e-21 & 2.8635e-21 & 1.3300e-21 & 4.1937e-21\\
% 8.90 & 0.2 & 2.4809e-21 & 1.0360e-21 & 1.2659e-21 & 1.0434e-21 & 1.6930e-21\\
\end{tabular}
\label{table:appendix_result_O_Fe}
\end{center}
\end{table*}

\clearpage

\newpage

\subsection{K.D.E.  templates and fit results}
\label{appendix:templates}
\begin{figure*}[h]
\begin{center}
\includegraphics[width=0.24\textwidth]{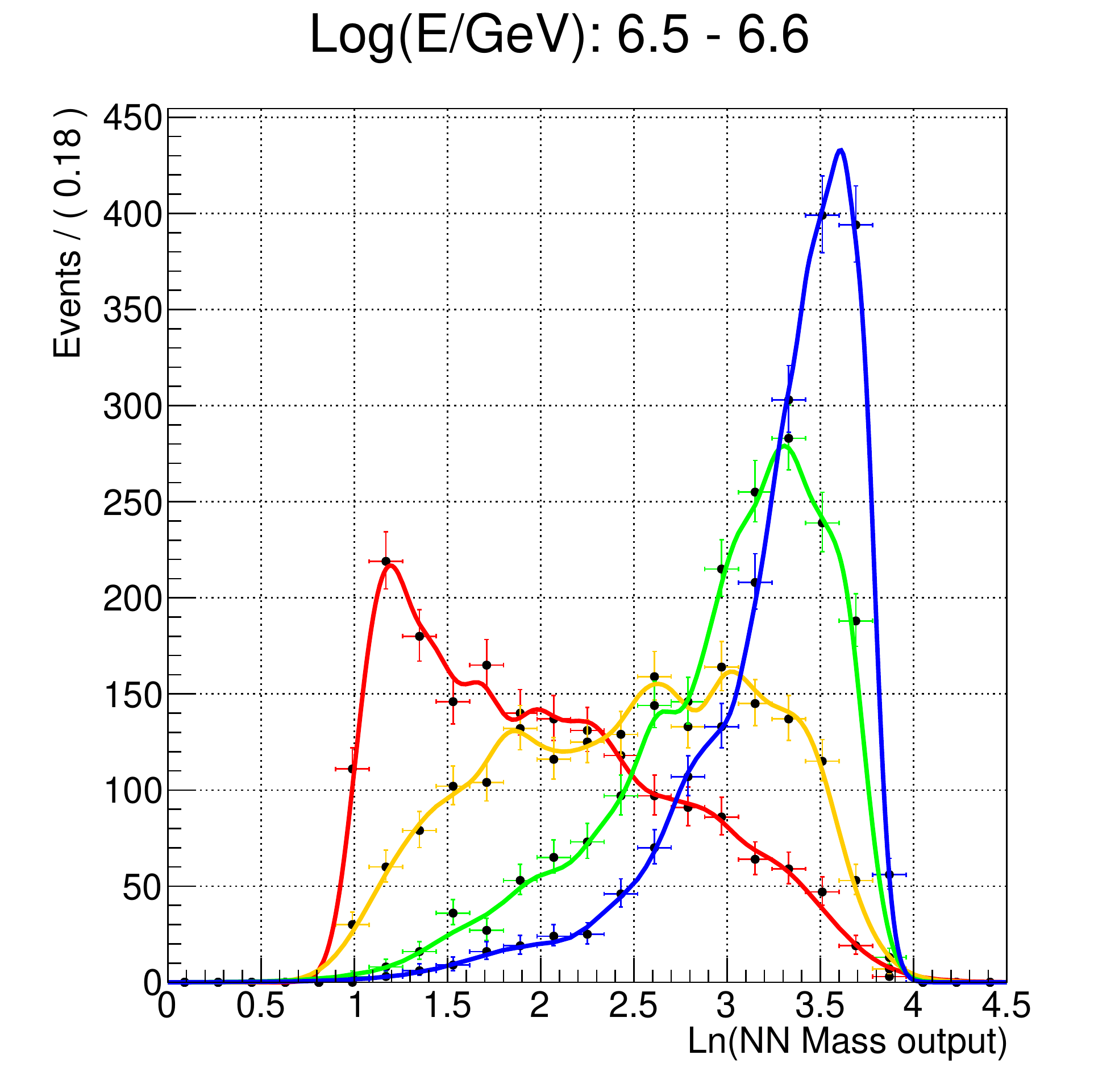}
\includegraphics[width=0.24\textwidth]{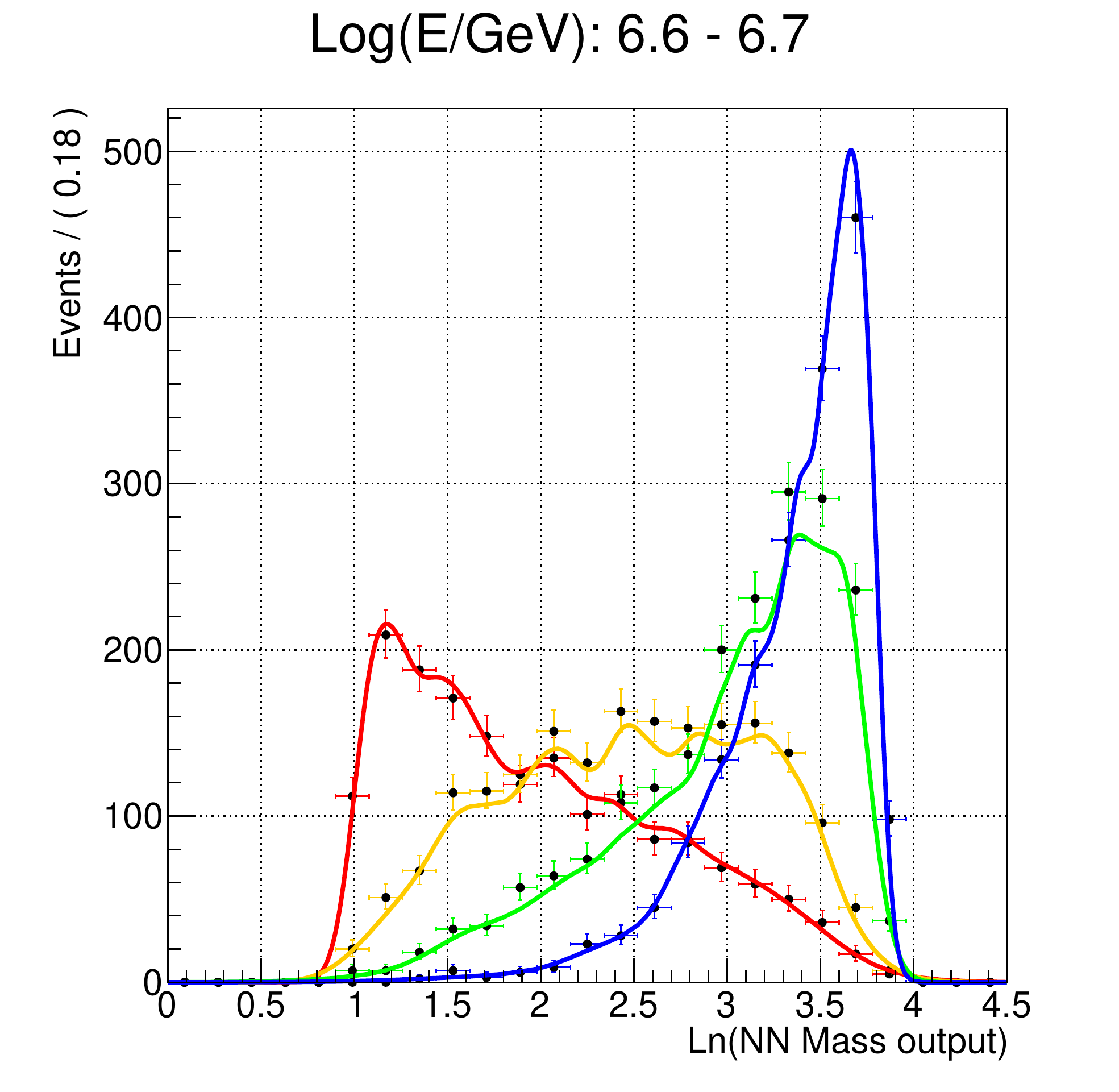}
\includegraphics[width=0.24\textwidth]{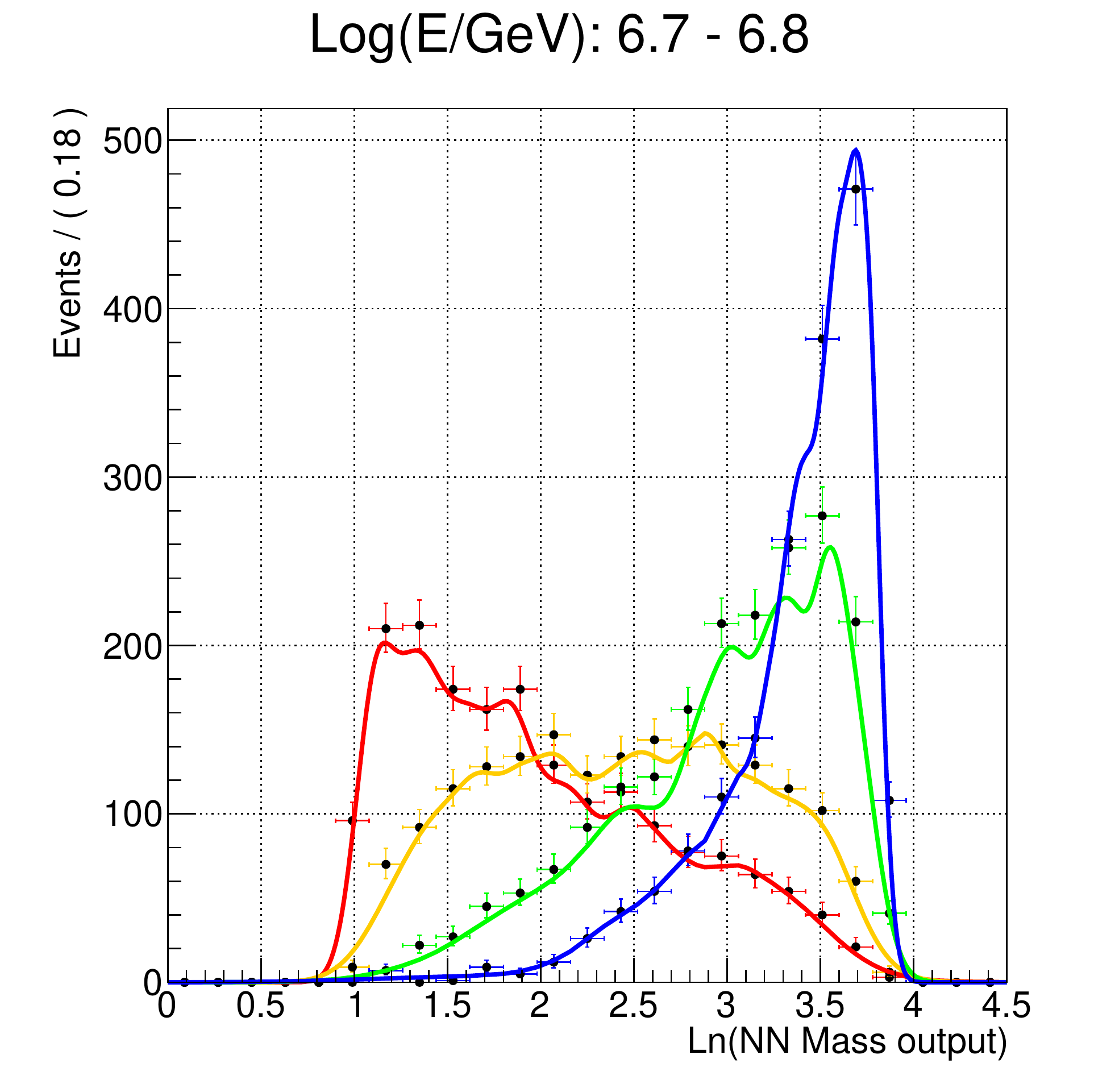}
\includegraphics[width=0.24\textwidth]{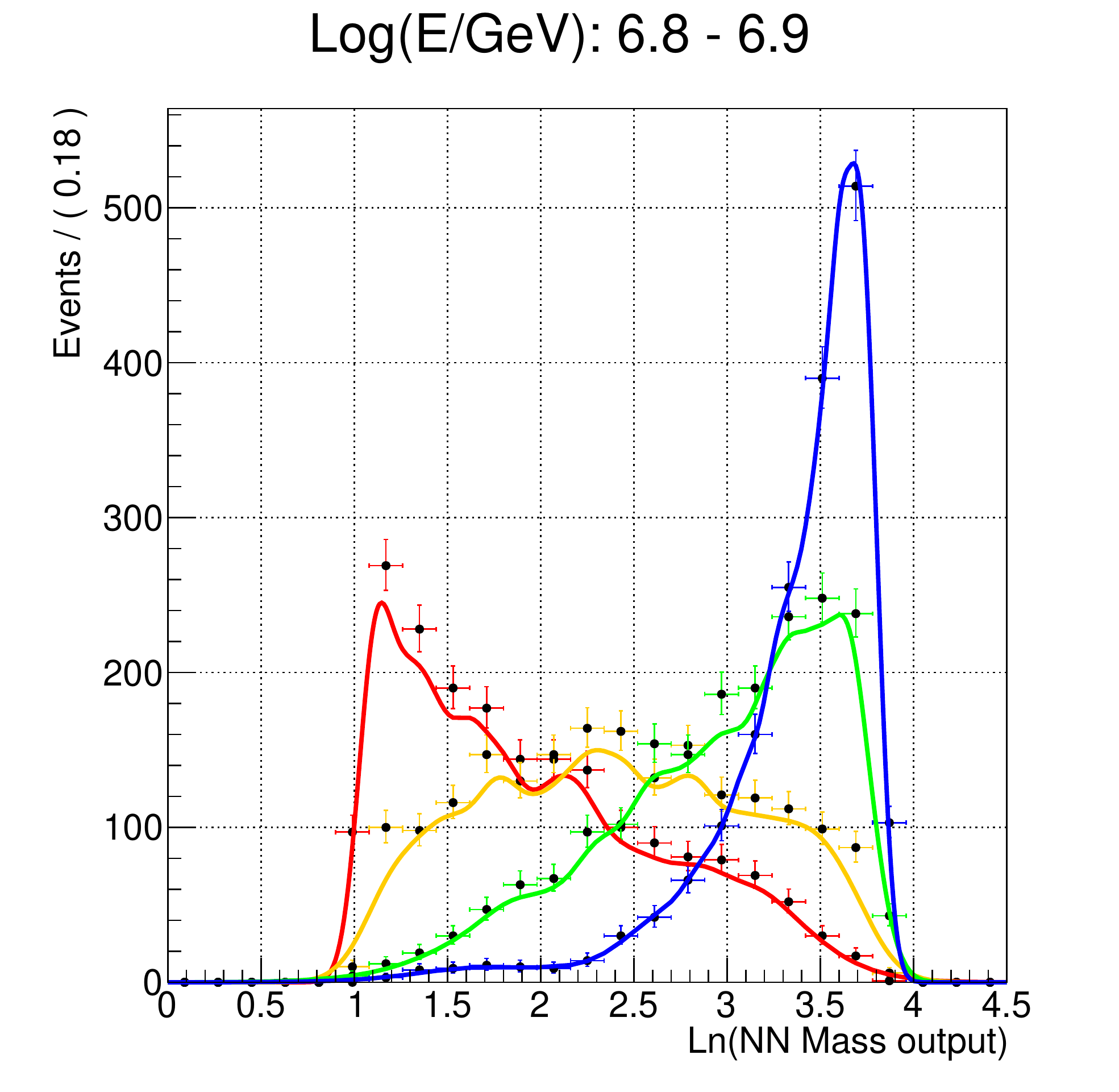}
\includegraphics[width=0.24\textwidth]{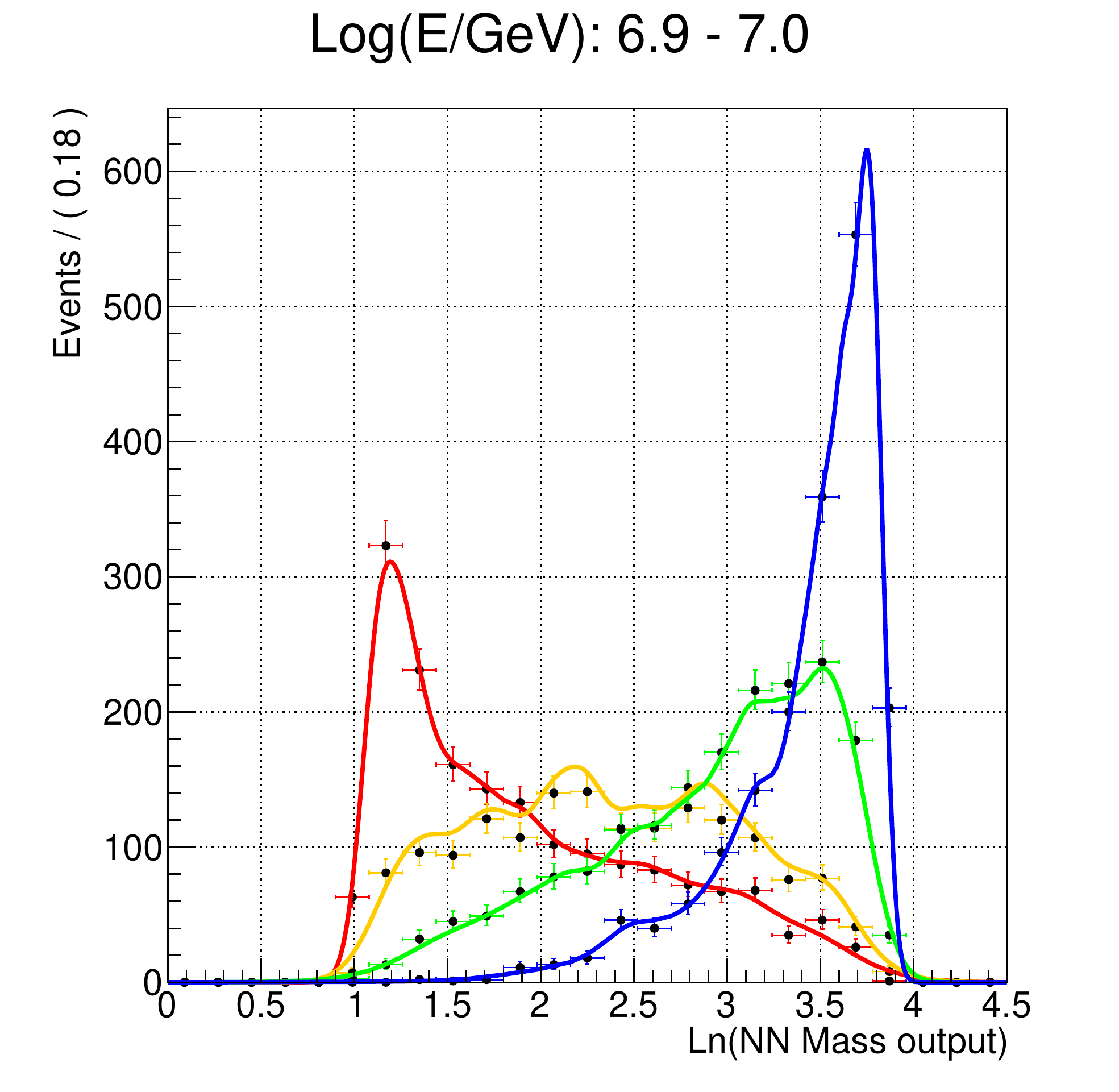}
\includegraphics[width=0.24\textwidth]{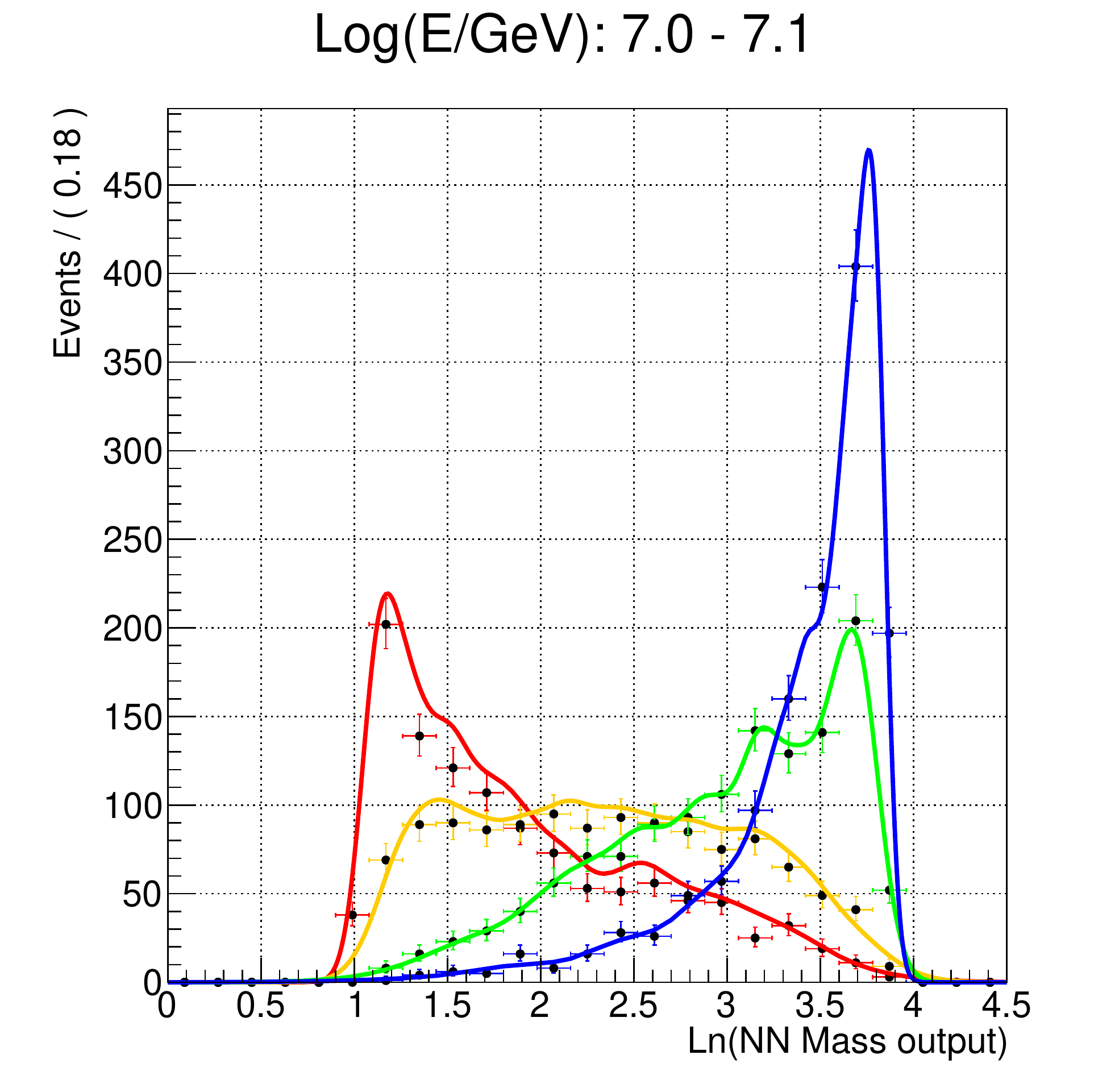}
\includegraphics[width=0.24\textwidth]{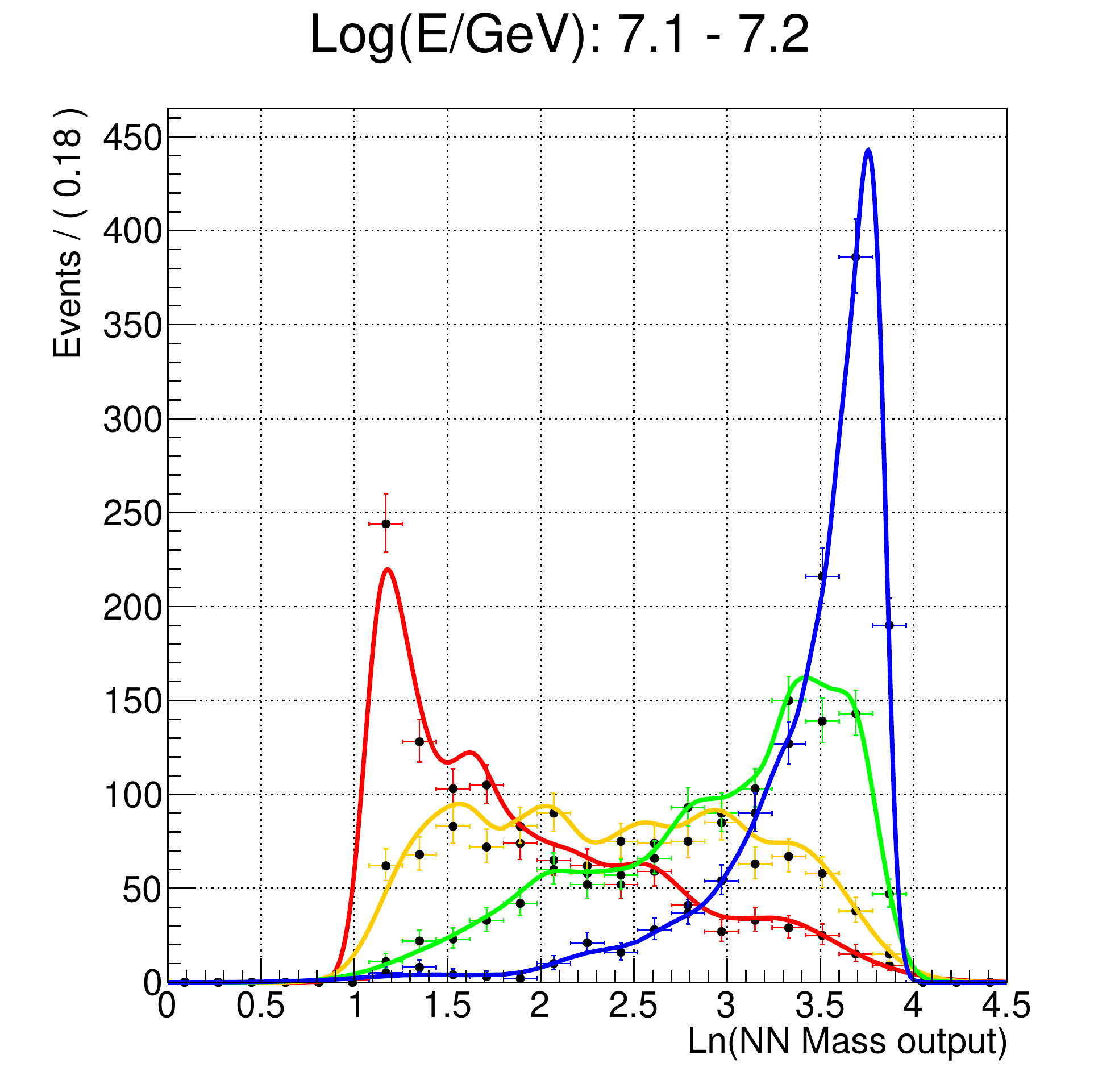}
\includegraphics[width=0.24\textwidth]{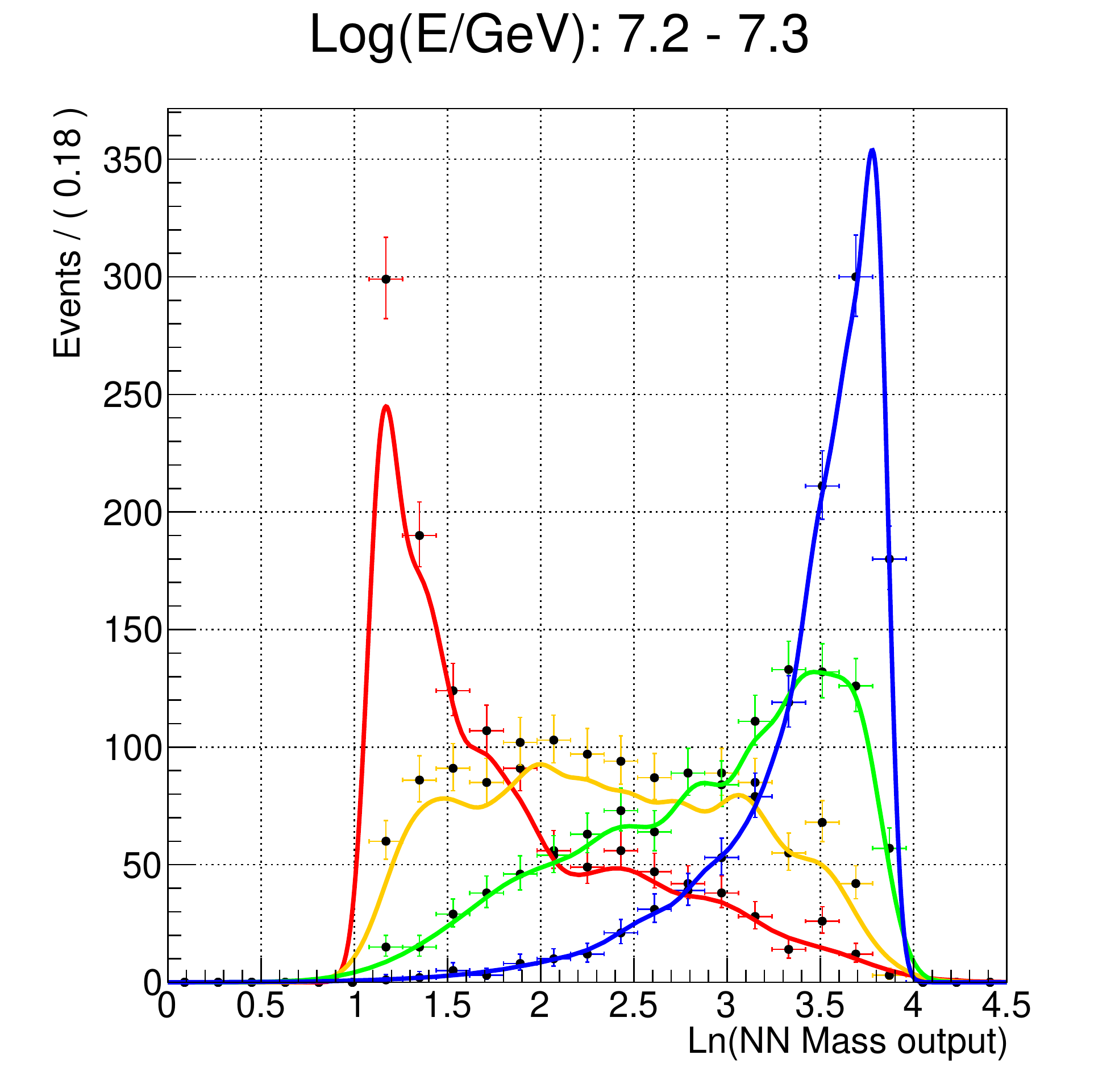}
\includegraphics[width=0.24\textwidth]{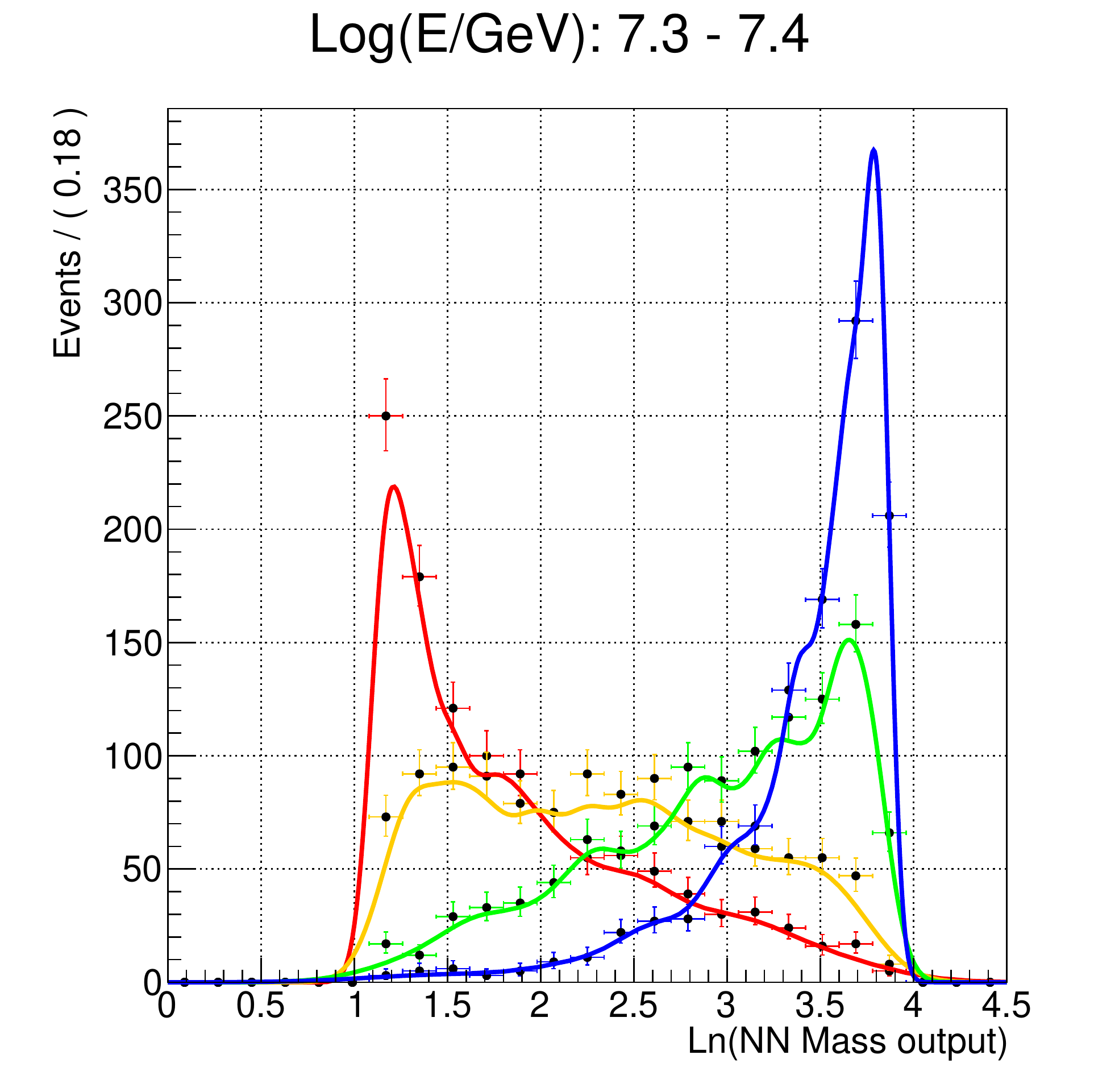}
\includegraphics[width=0.24\textwidth]{plots_actually_in_paper/coinc/{templates/template_all_7.4_7.5}.pdf}
\includegraphics[width=0.24\textwidth]{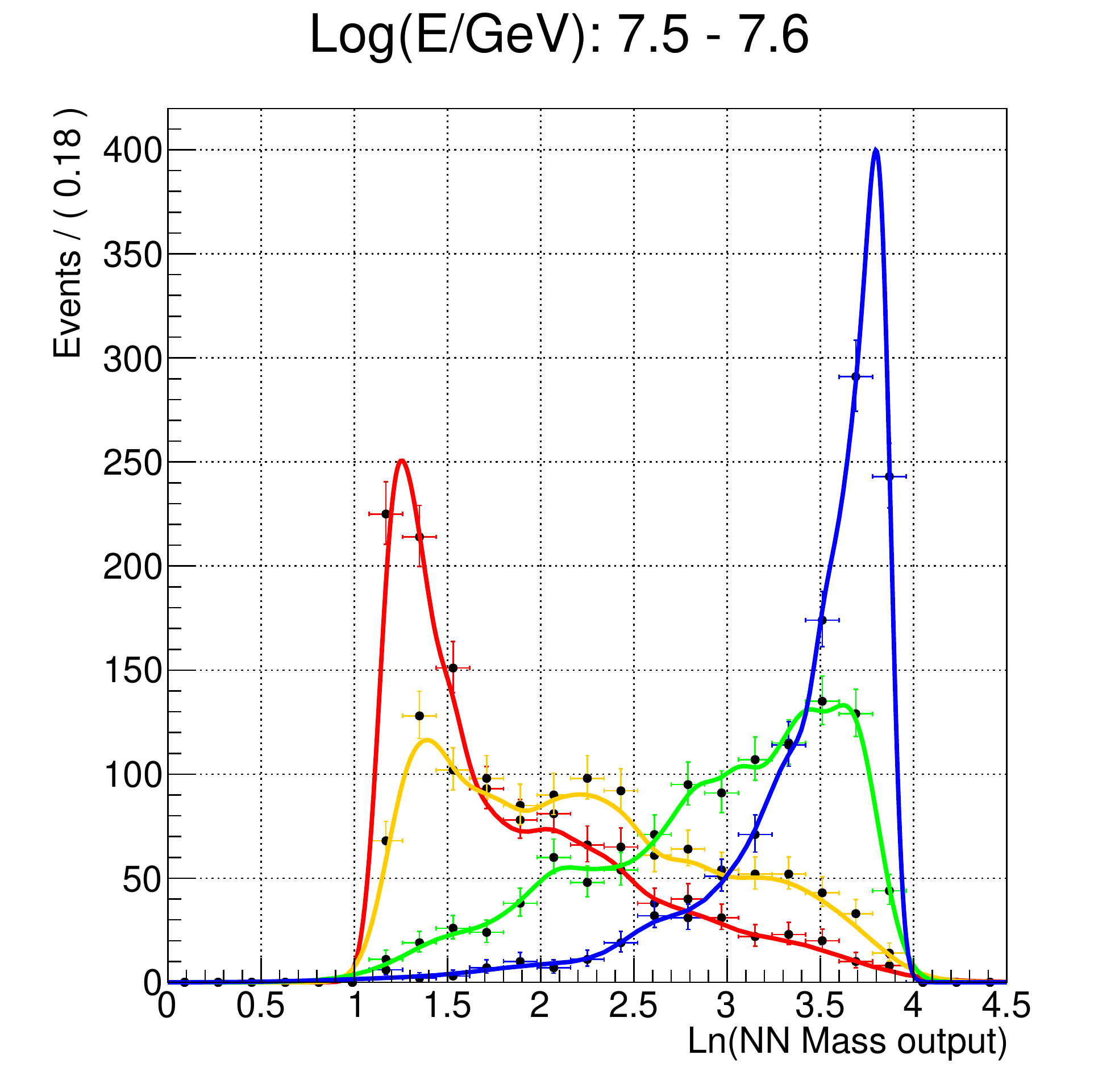}
\includegraphics[width=0.24\textwidth]{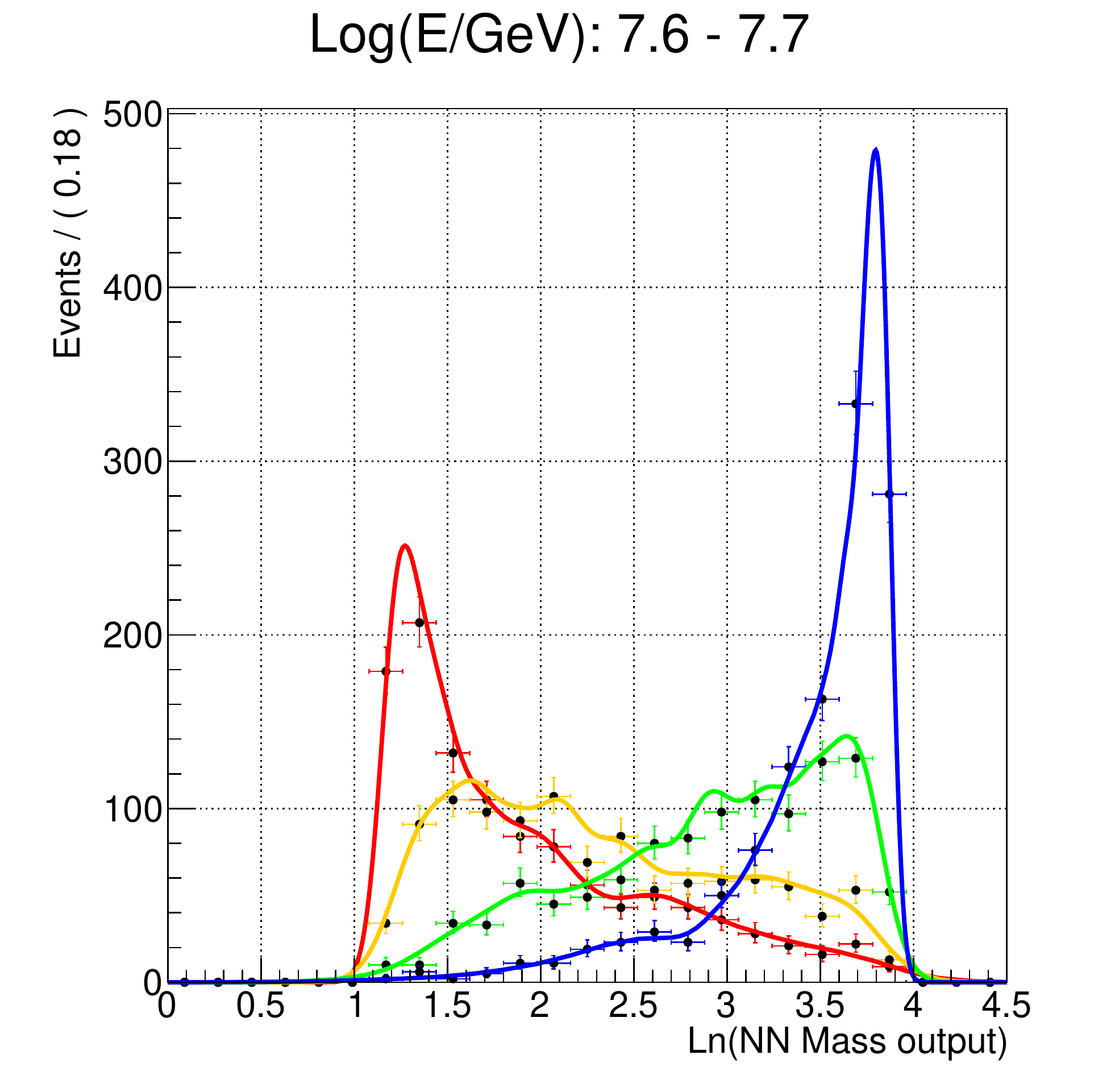}
\includegraphics[width=0.24\textwidth]{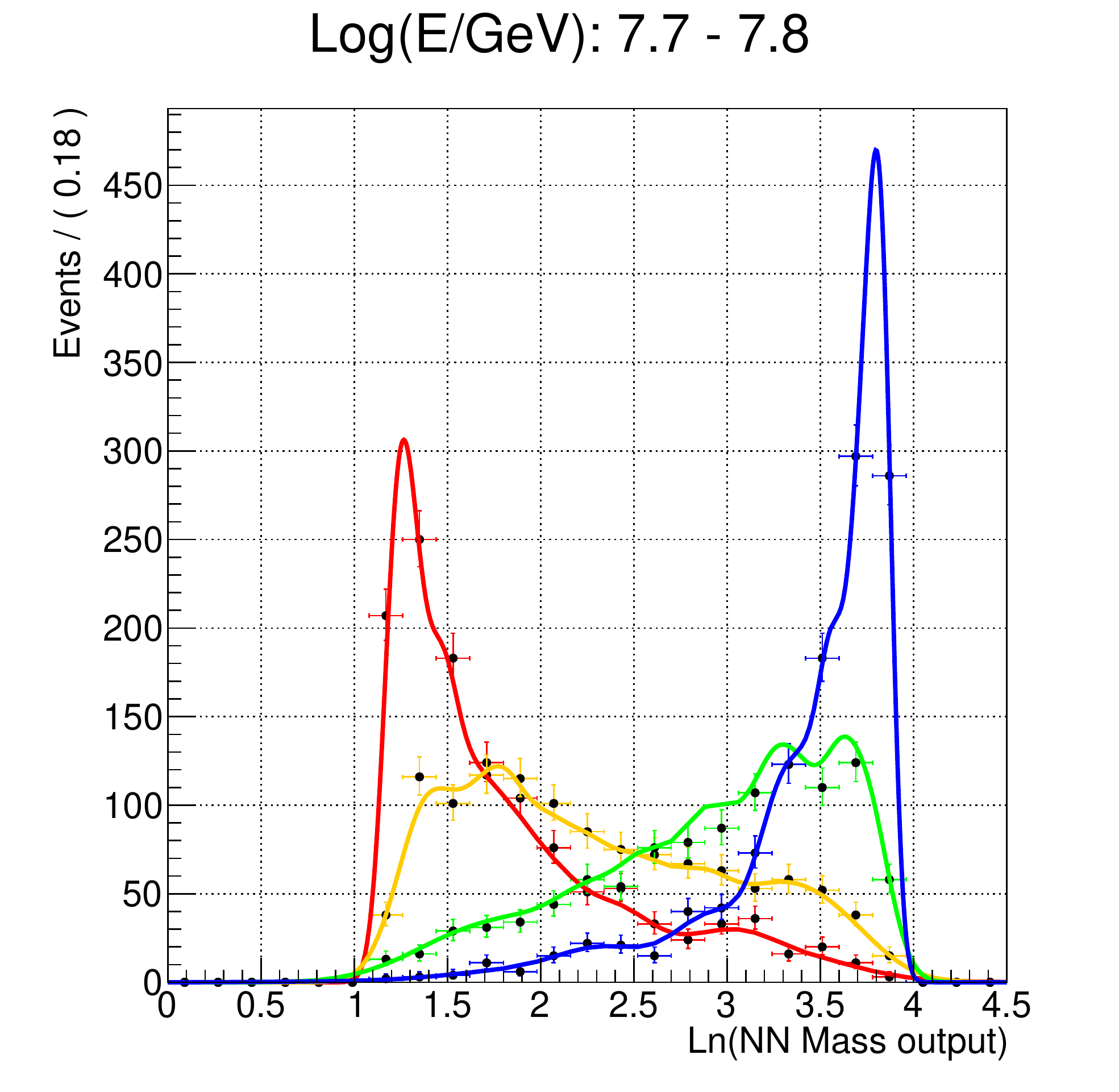}
\includegraphics[width=0.24\textwidth]{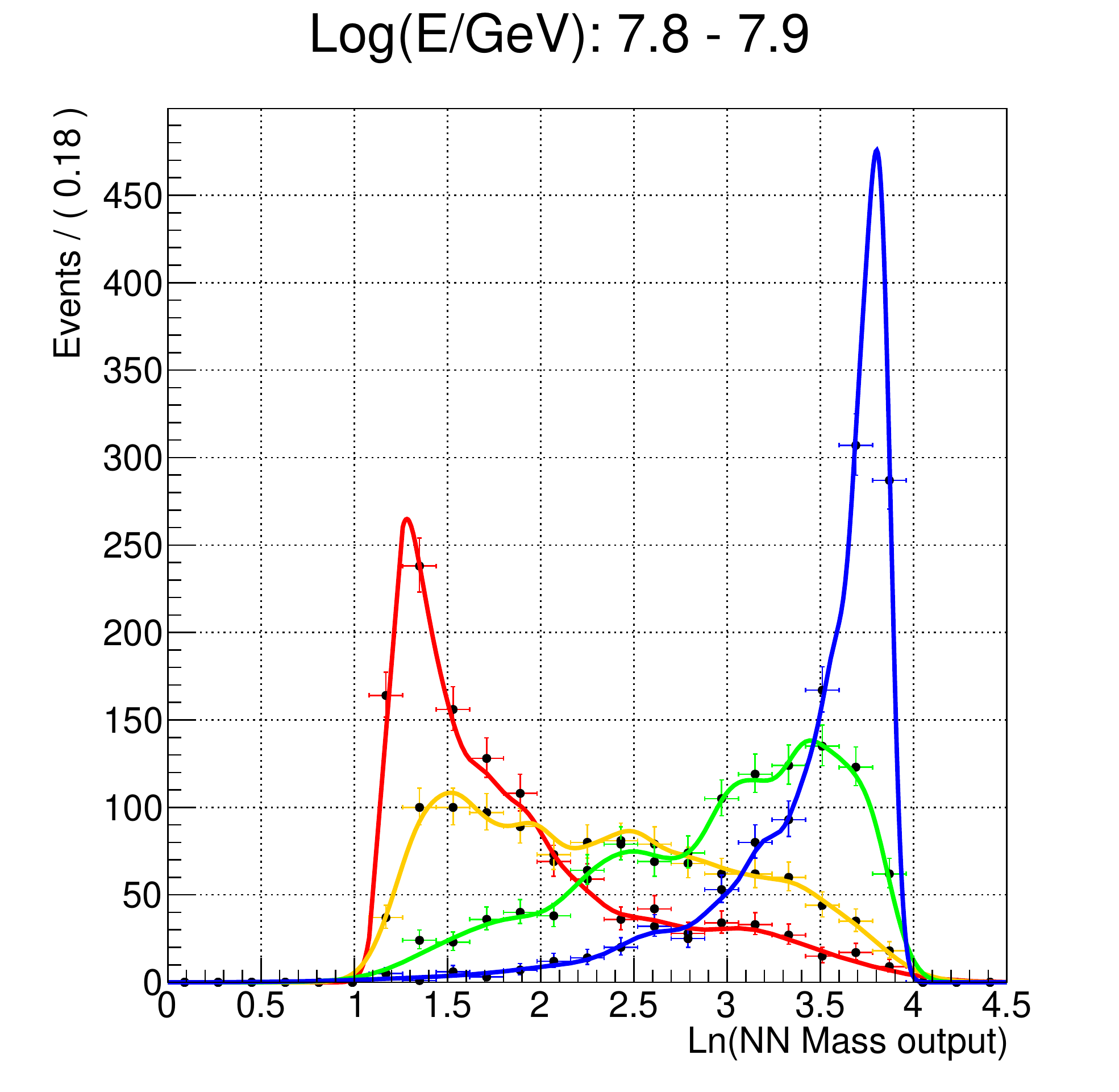}
\includegraphics[width=0.24\textwidth]{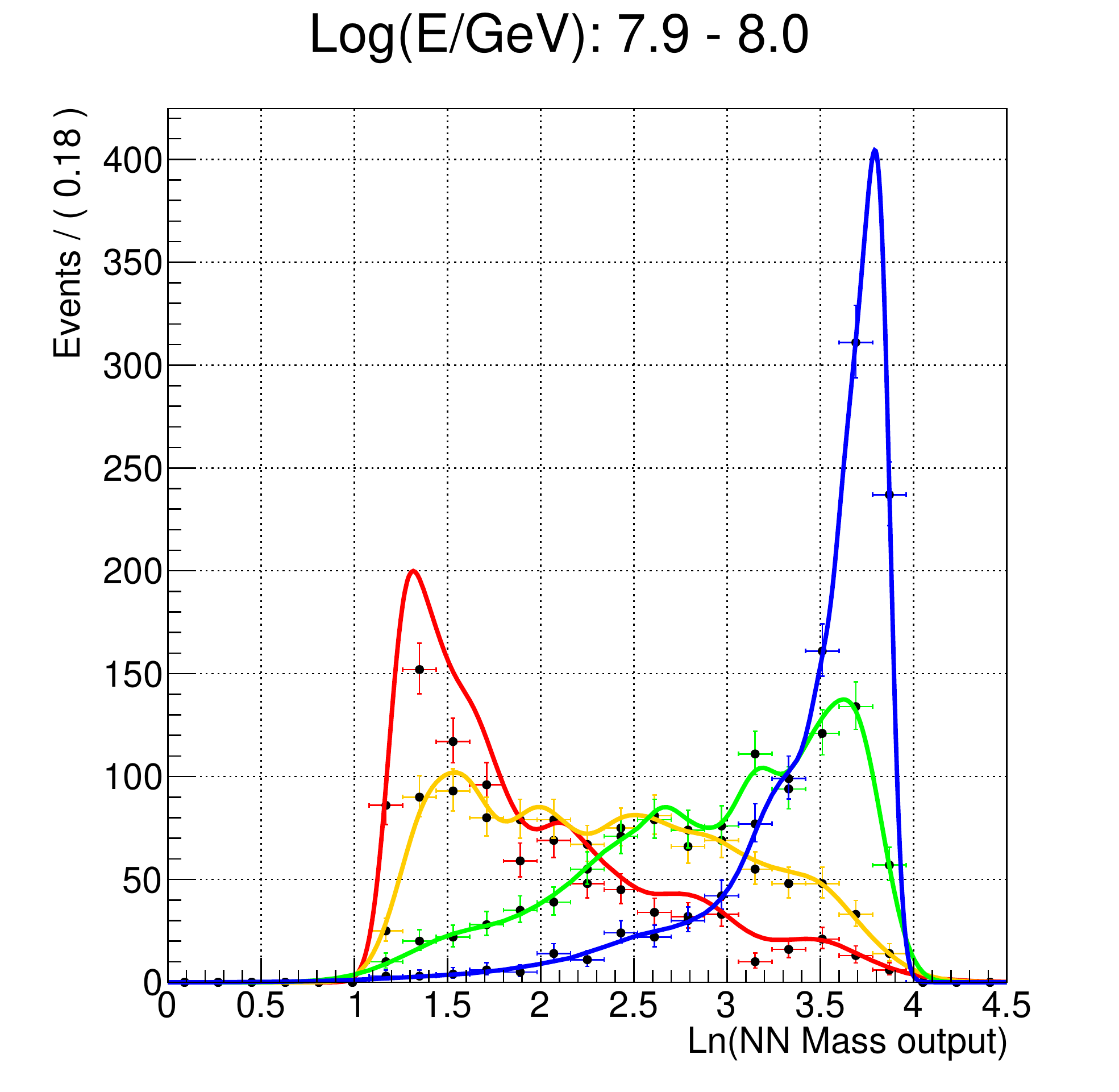}
\includegraphics[width=0.24\textwidth]{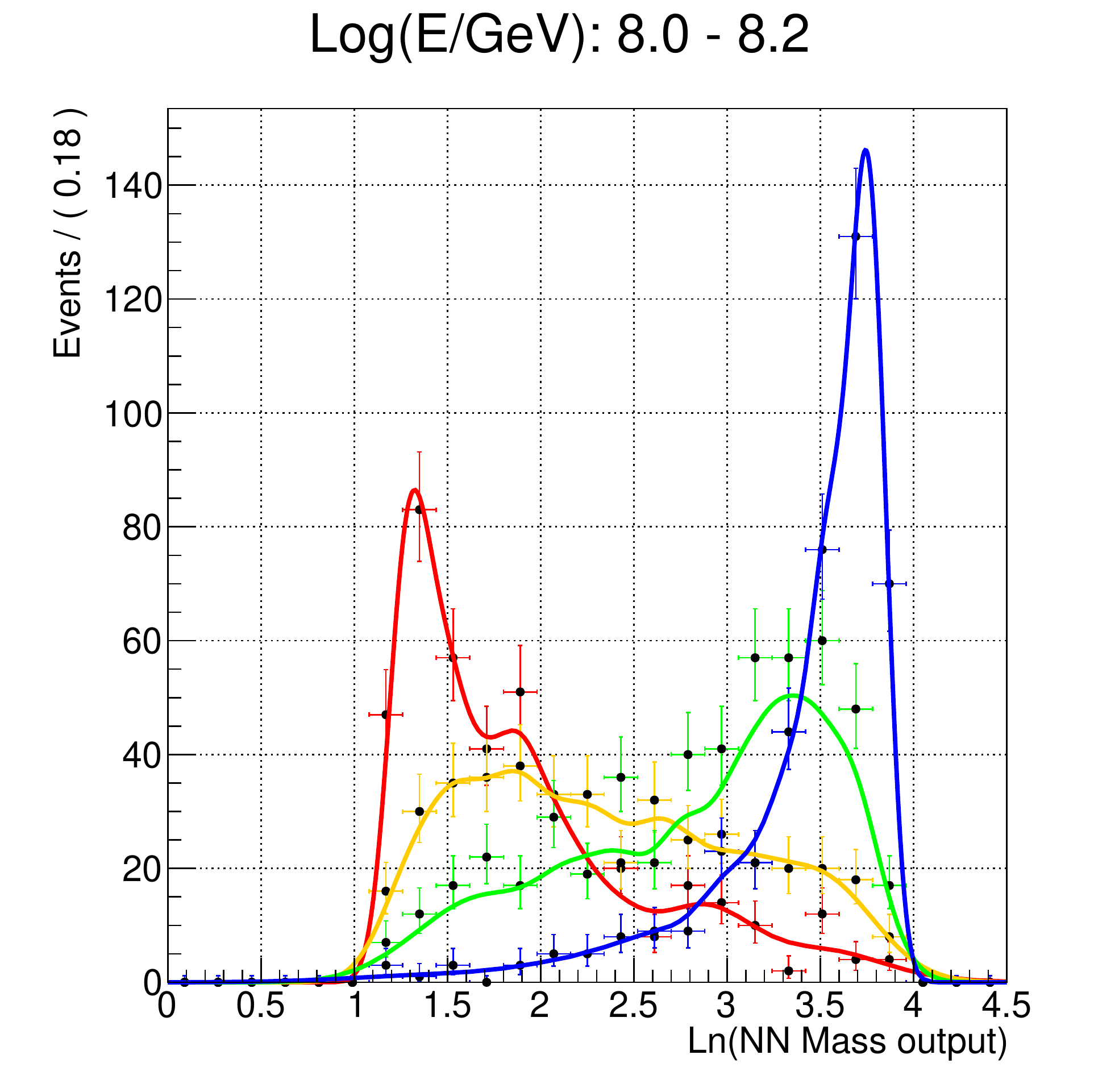}
\includegraphics[width=0.24\textwidth]{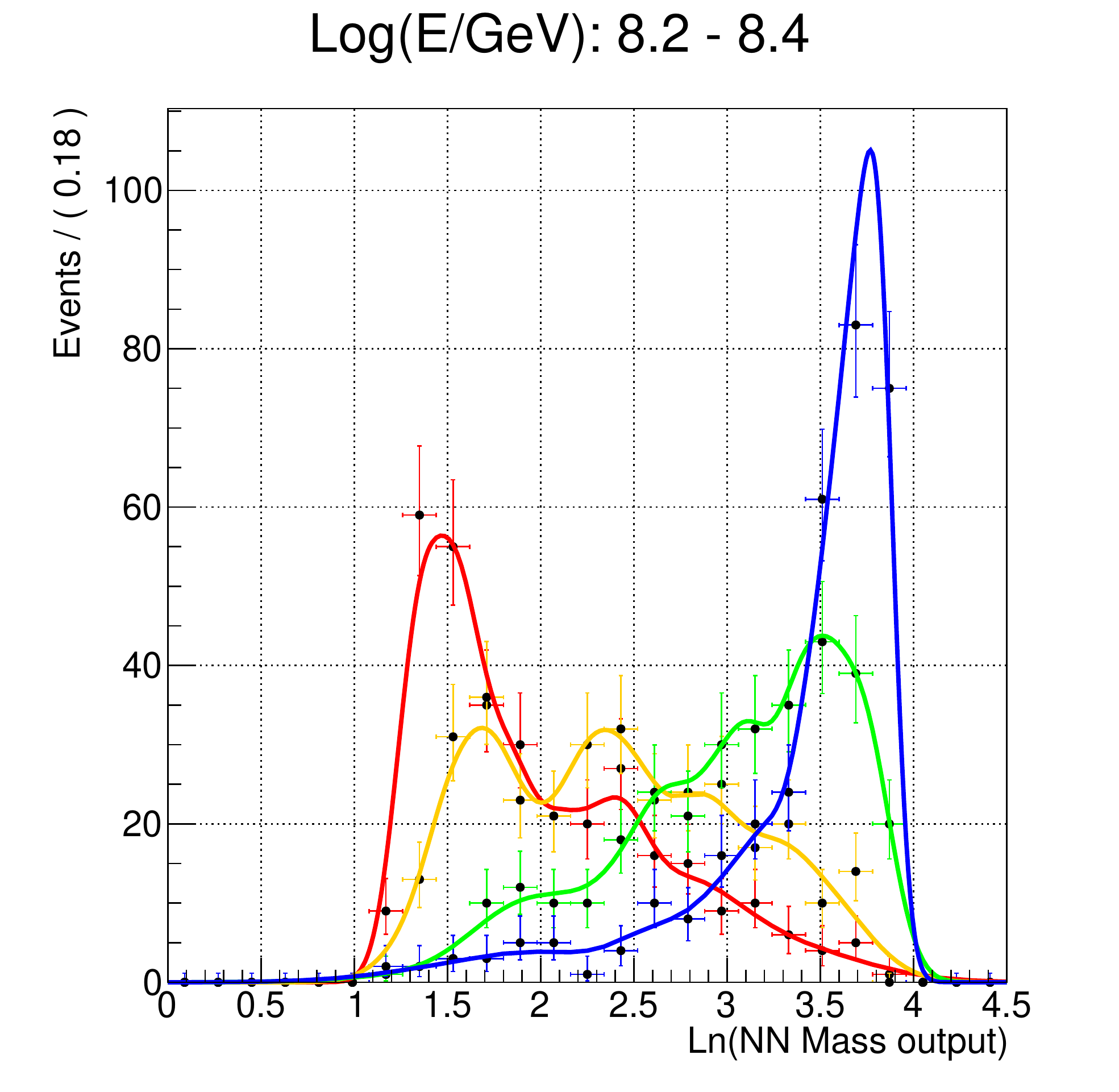}
\includegraphics[width=0.24\textwidth]{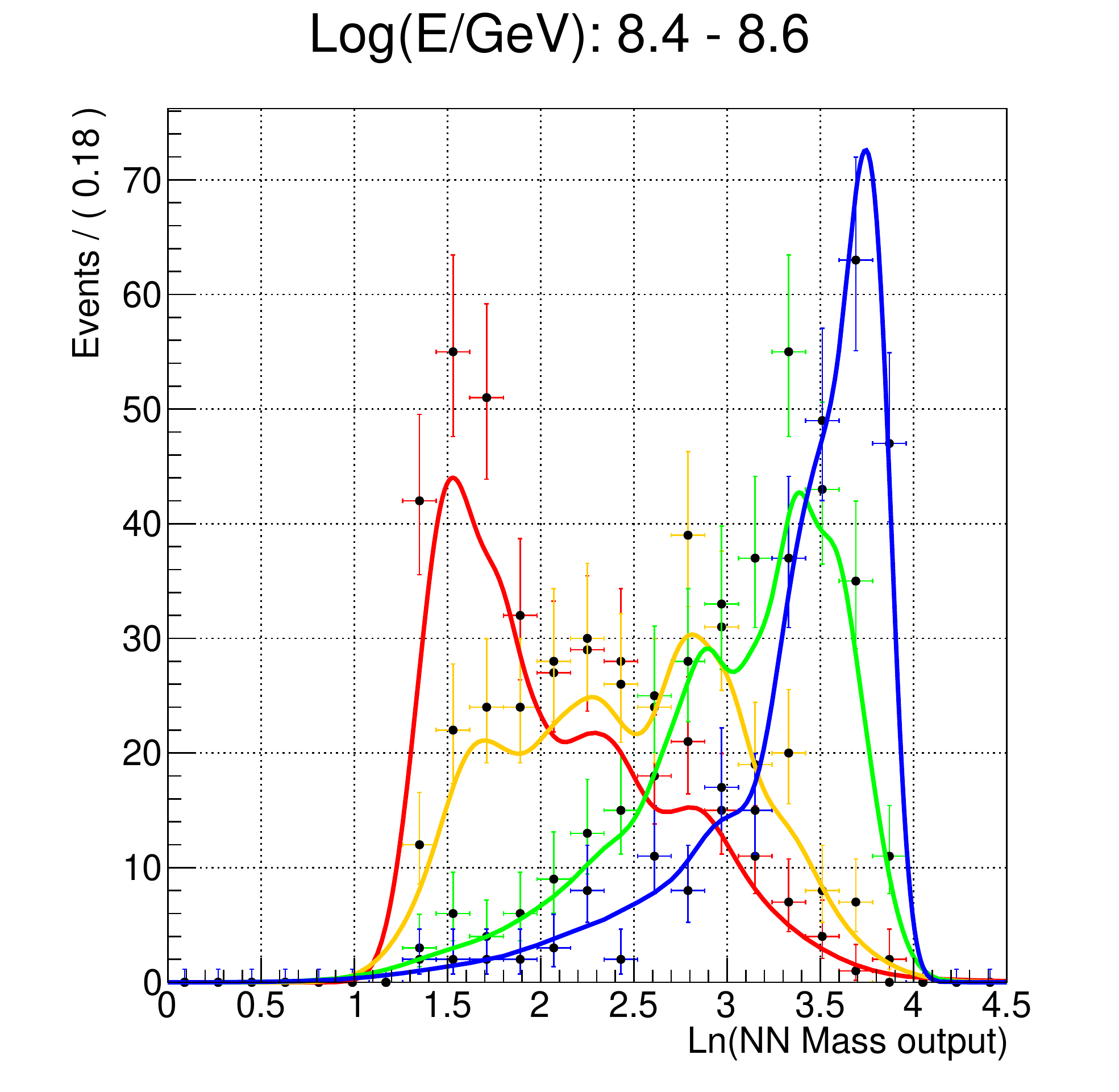}
\includegraphics[width=0.24\textwidth]{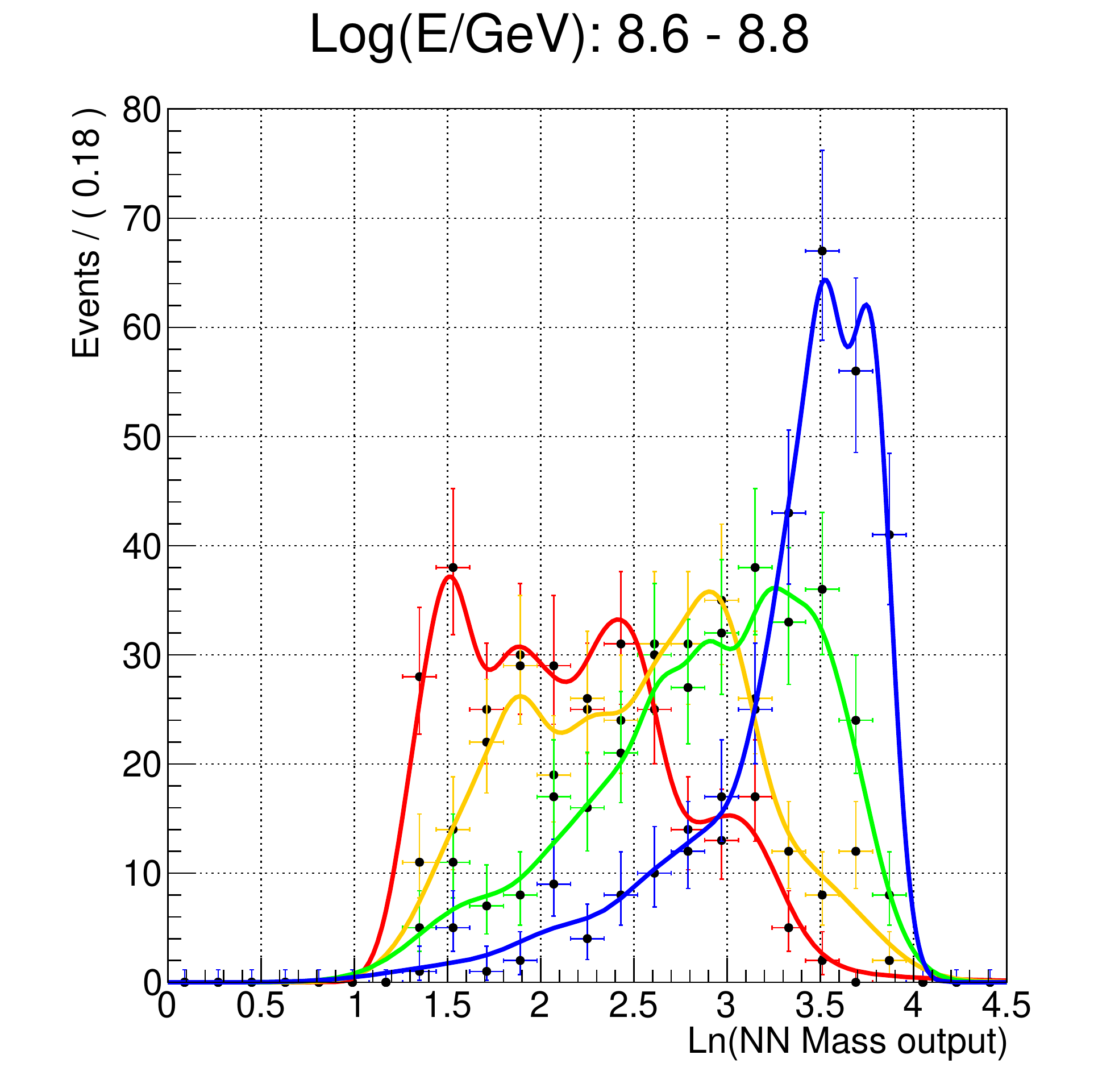}
\includegraphics[width=0.24\textwidth]{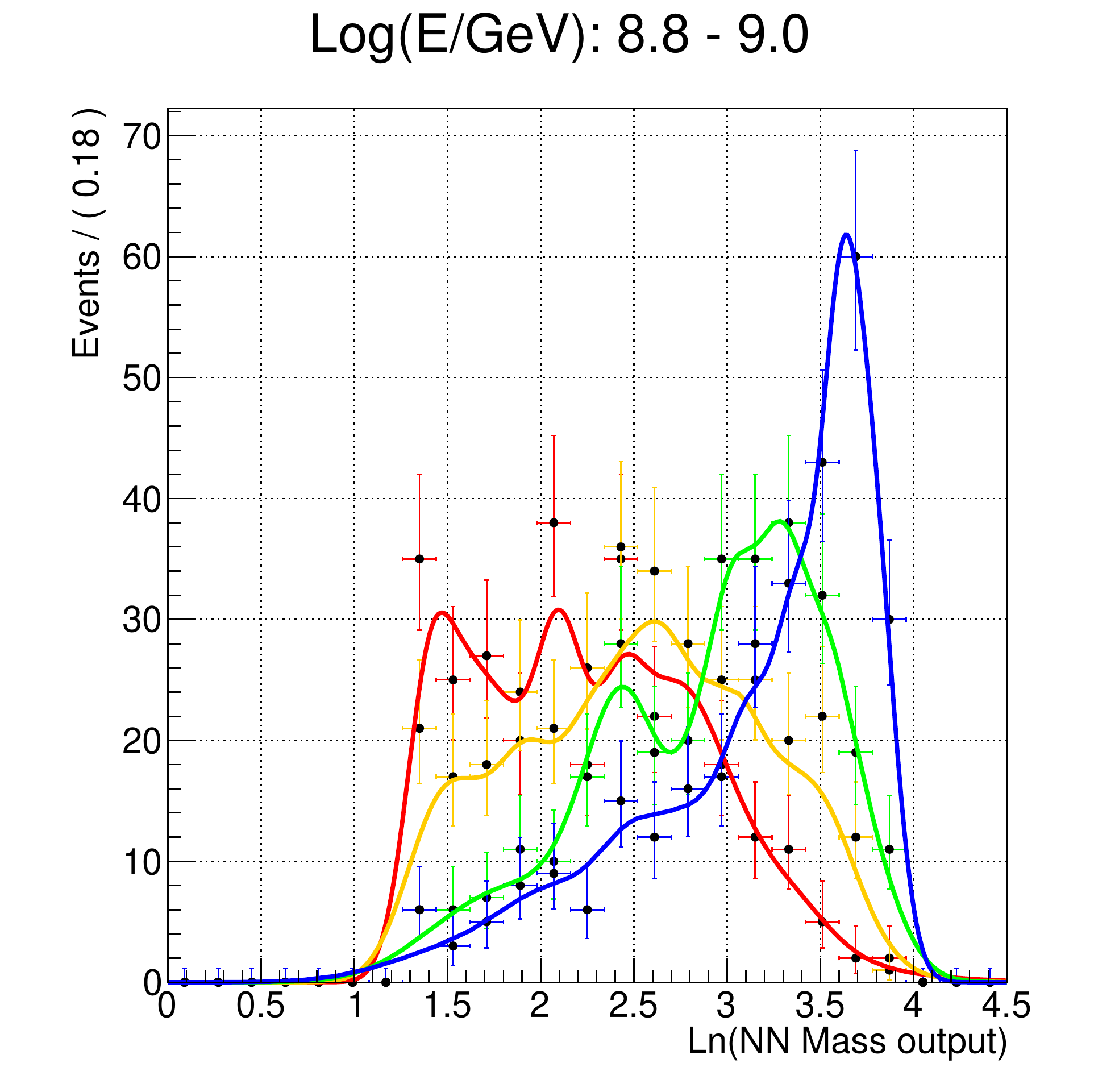}

\caption{Distributions of the natural logarithm of the neural network mass output for each slice in energy used in the coincident analysis.  Energy ranges are labeled in log$_{10}$(E$_{0,reco}$/GeV) in the titles of each figure.  The y-axis represents the the number of simulated events for proton (red), helium (orange), oxygen (green) and iron (blue). The solid line represents the template probability density functions found by the adaptive KDE fitting method.  }
\label{f:nn_templatehisto}
\end{center}
\end{figure*}

\begin{figure*}[h]
\begin{center}
\includegraphics[width=0.24\textwidth]{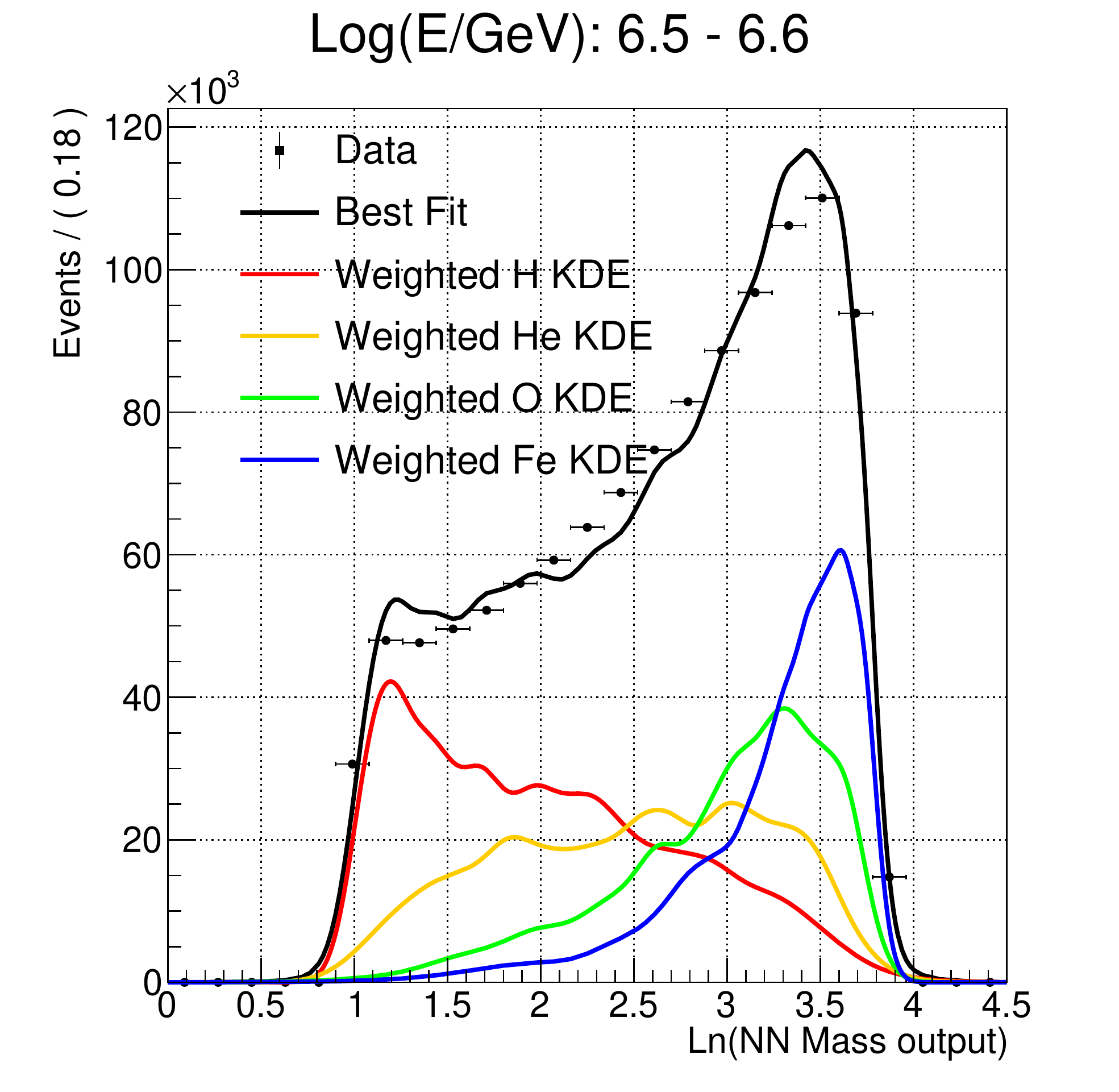}
\includegraphics[width=0.24\textwidth]{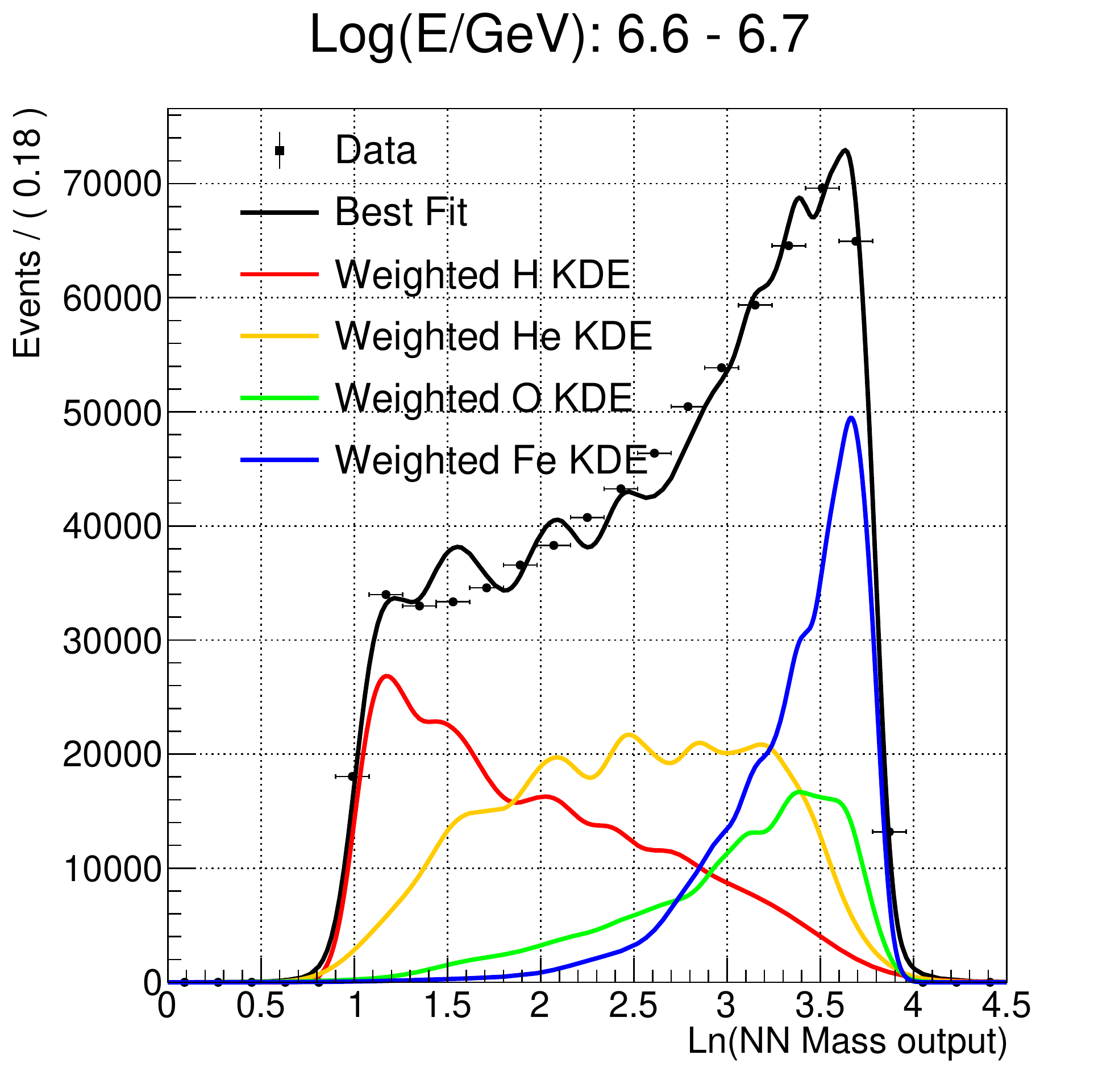}
\includegraphics[width=0.24\textwidth]{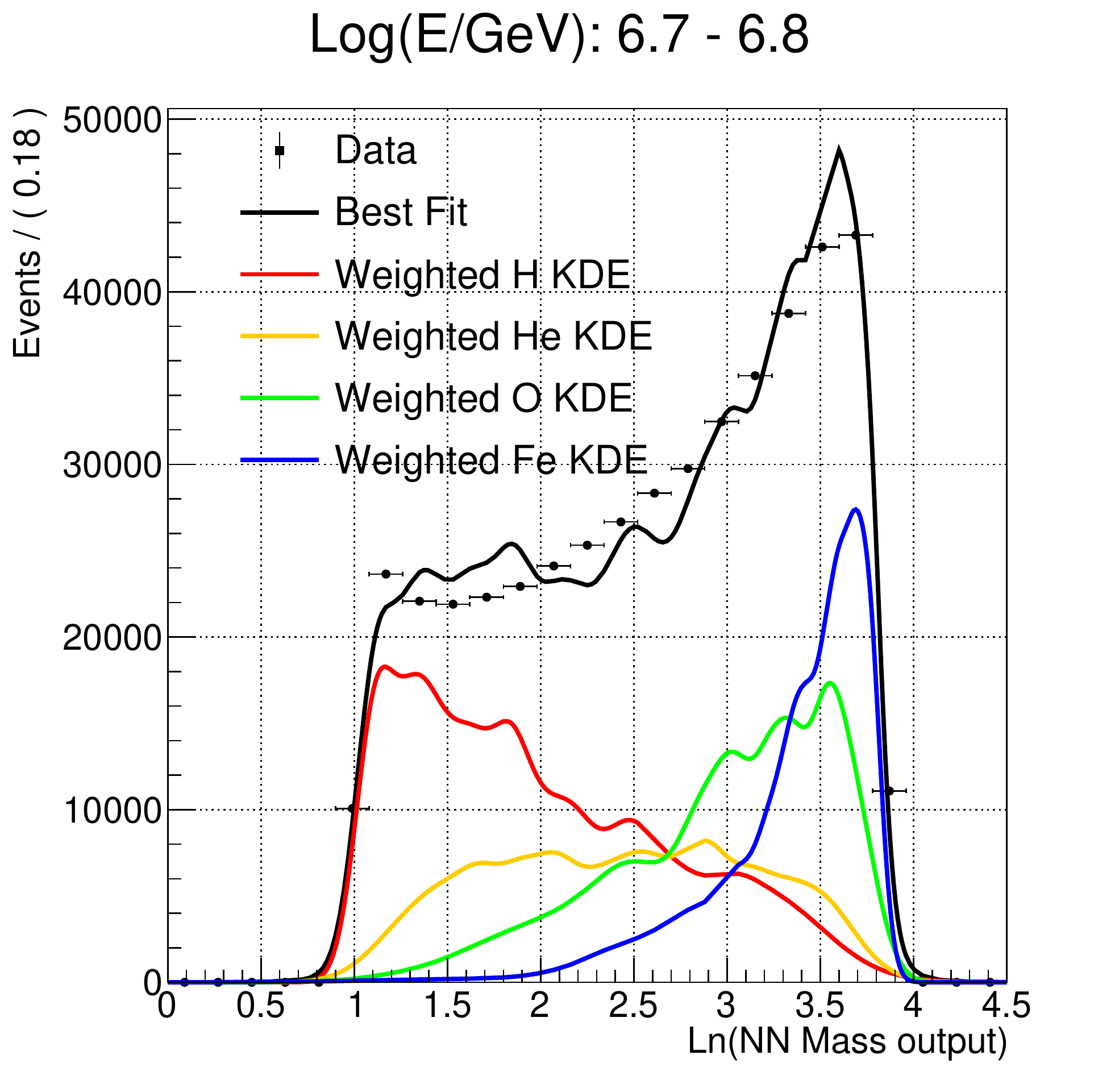}
\includegraphics[width=0.24\textwidth]{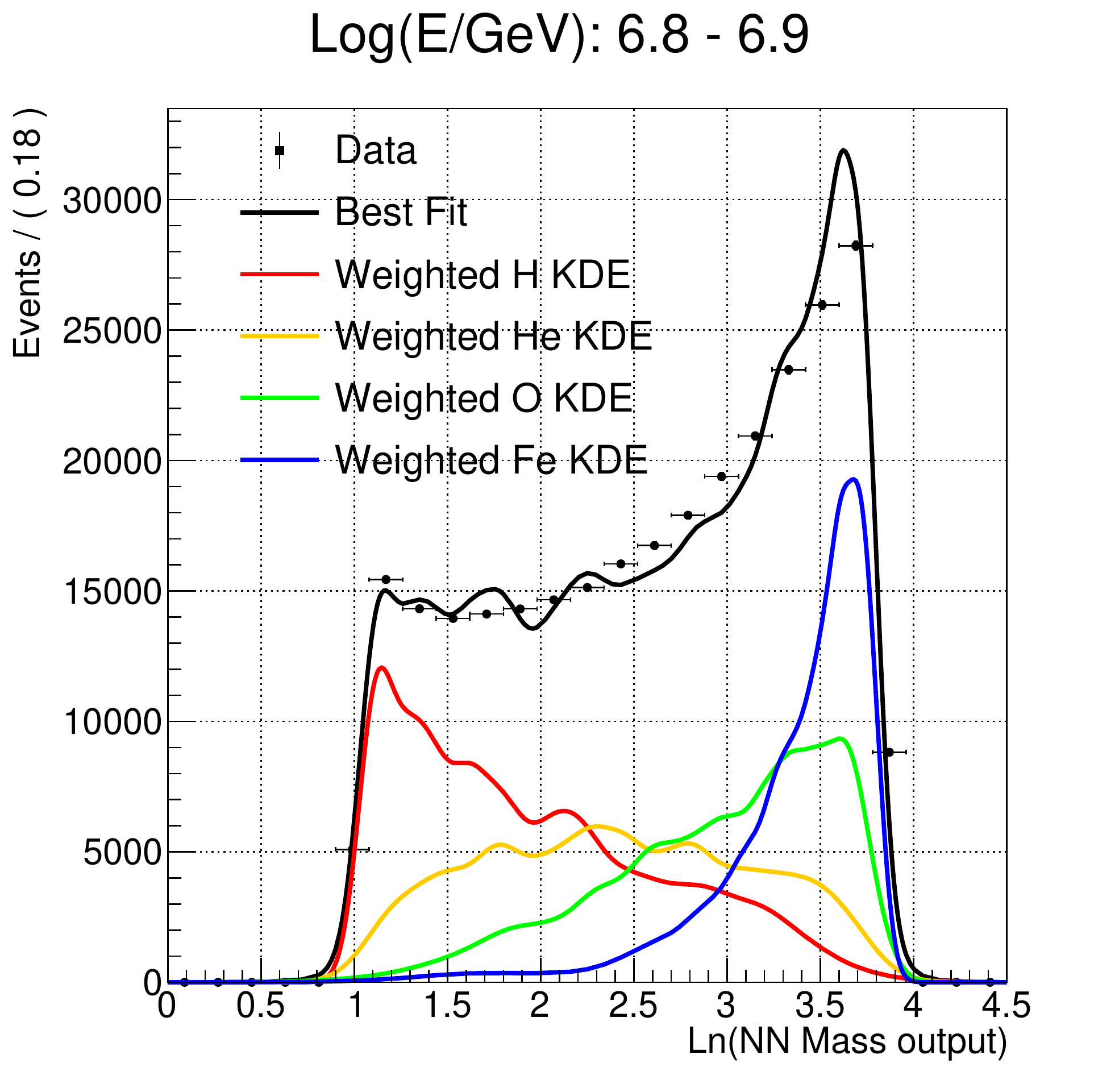}
\includegraphics[width=0.24\textwidth]{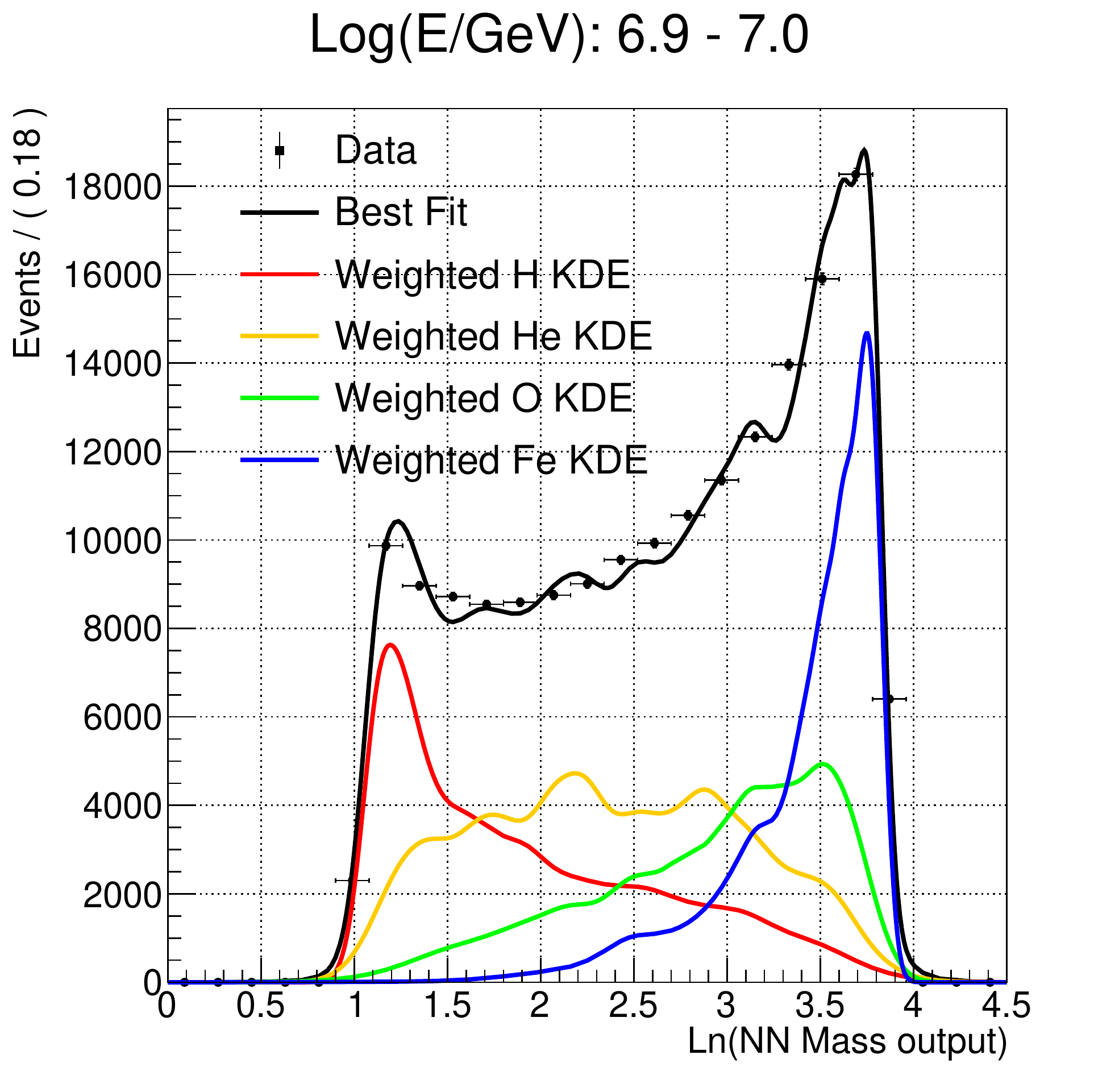}
\includegraphics[width=0.24\textwidth]{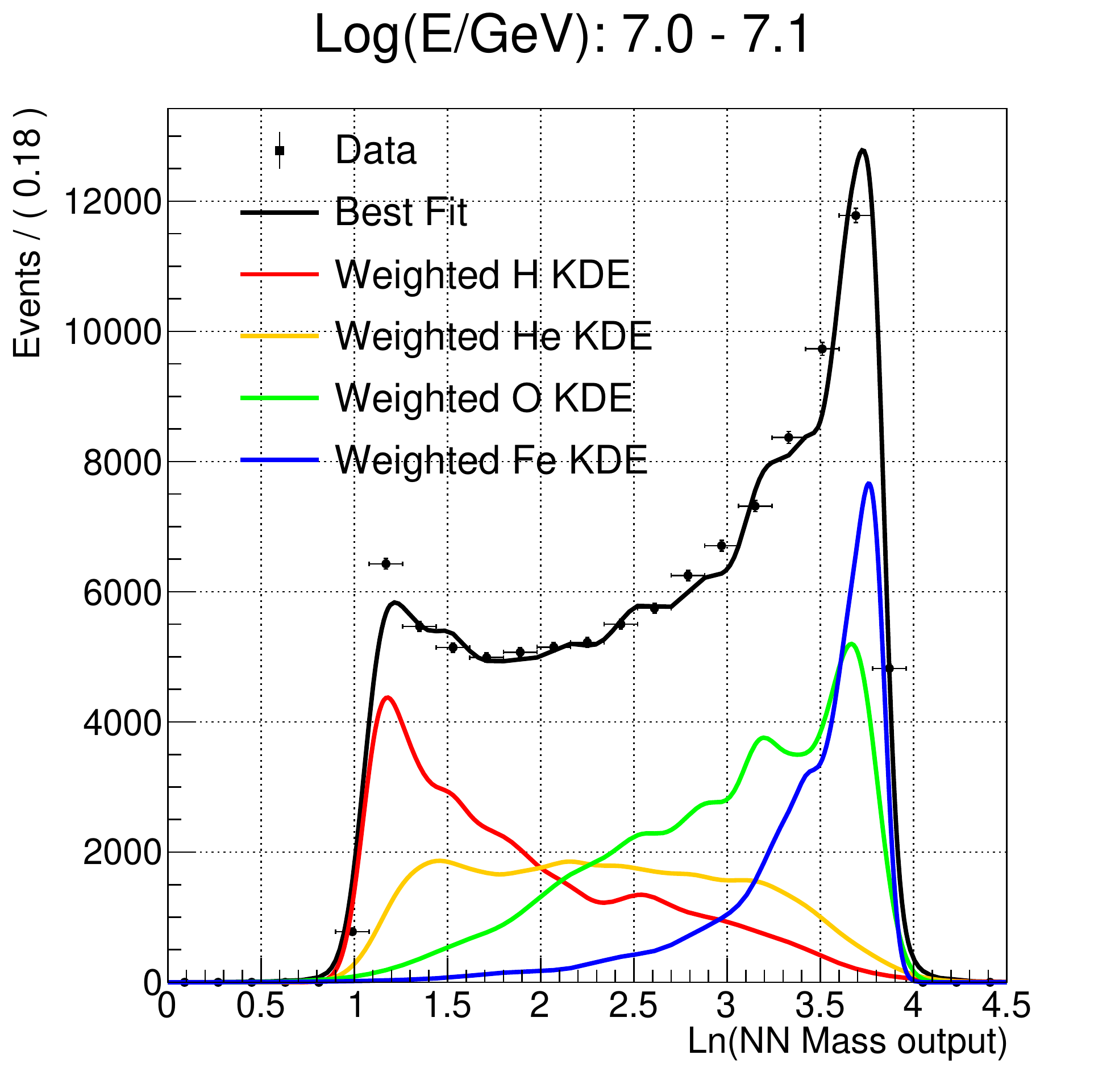}
\includegraphics[width=0.24\textwidth]{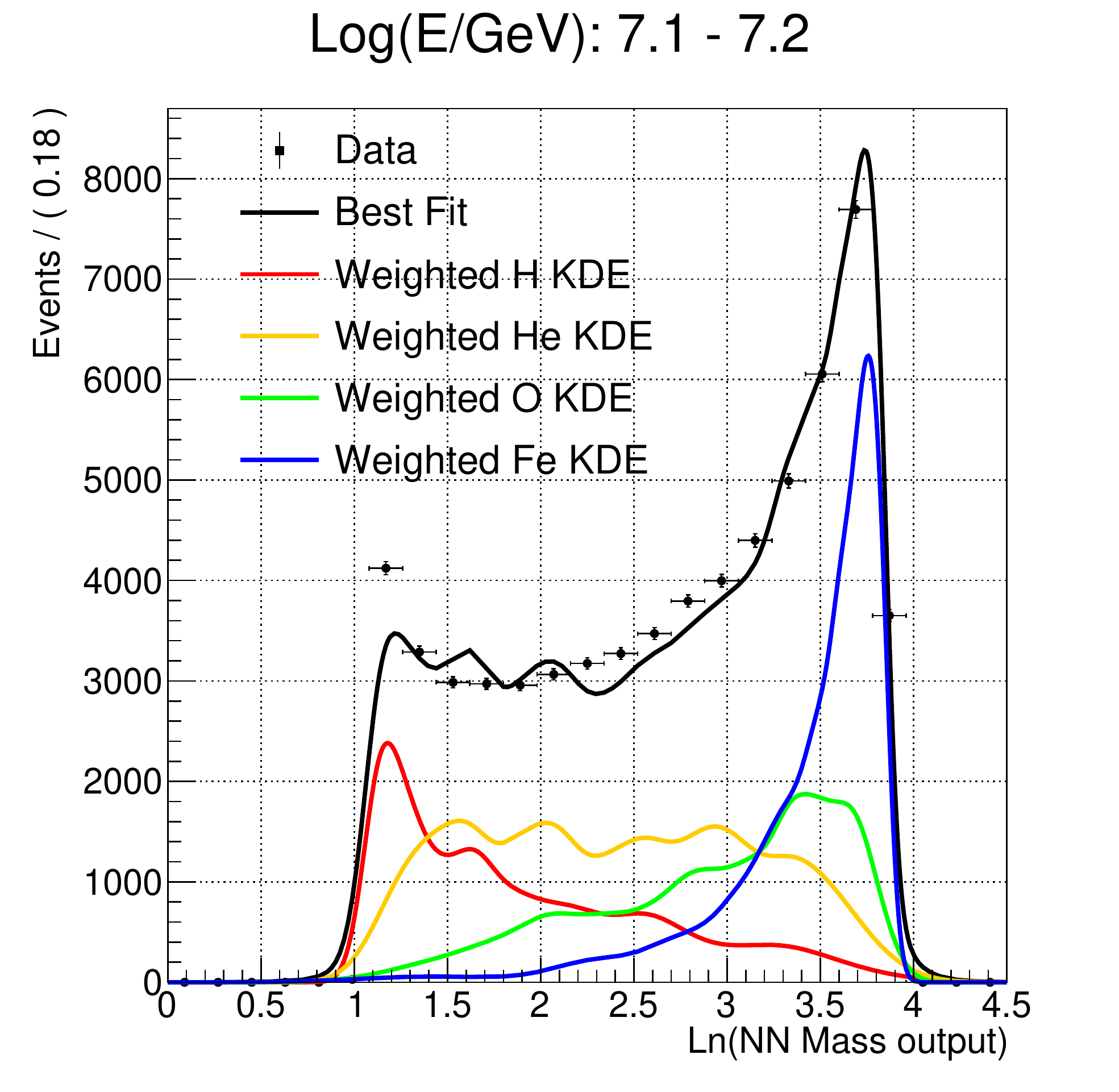}
\includegraphics[width=0.24\textwidth]{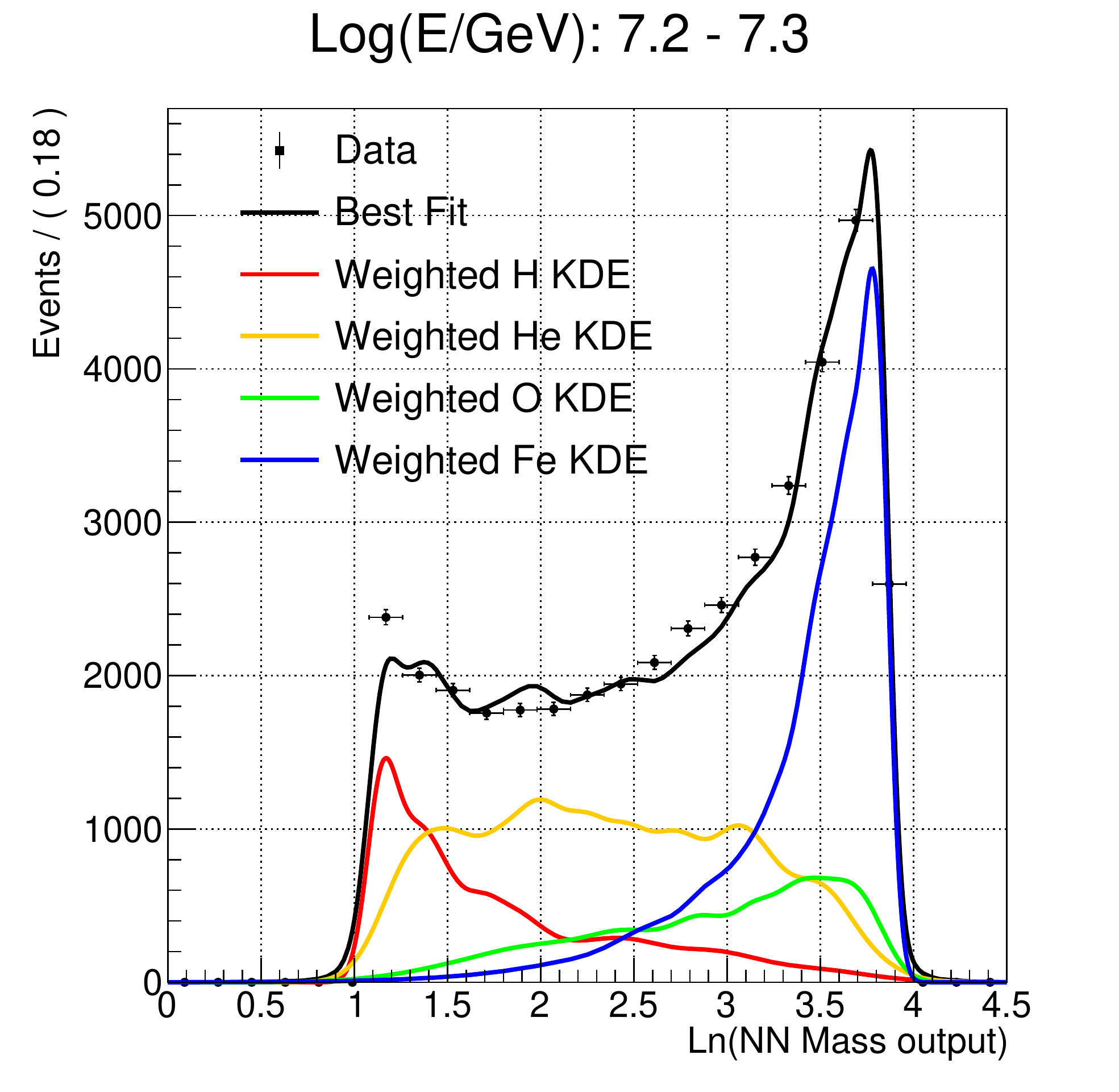}
\includegraphics[width=0.24\textwidth]{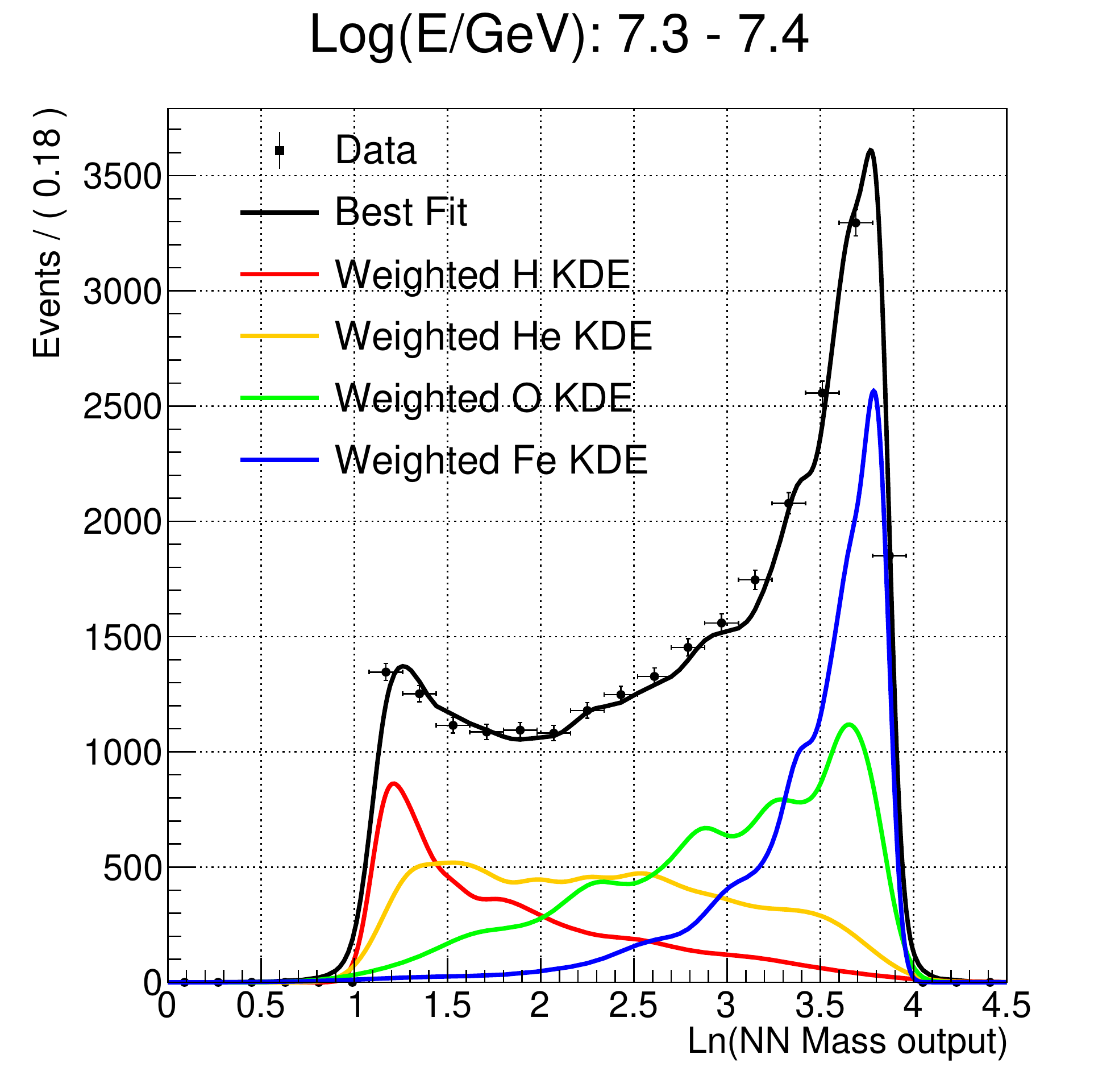}
\includegraphics[width=0.24\textwidth]{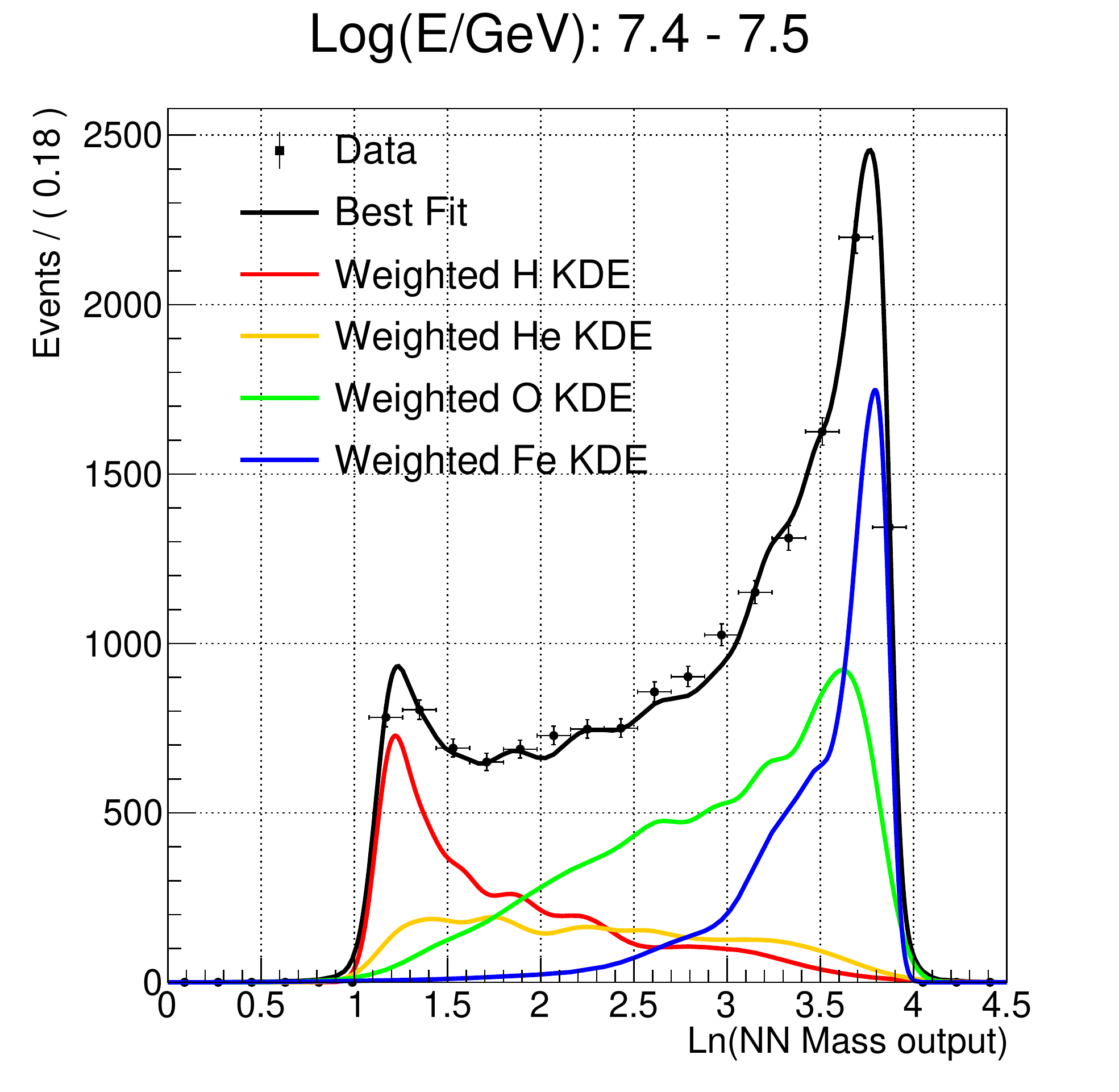}
\includegraphics[width=0.24\textwidth]{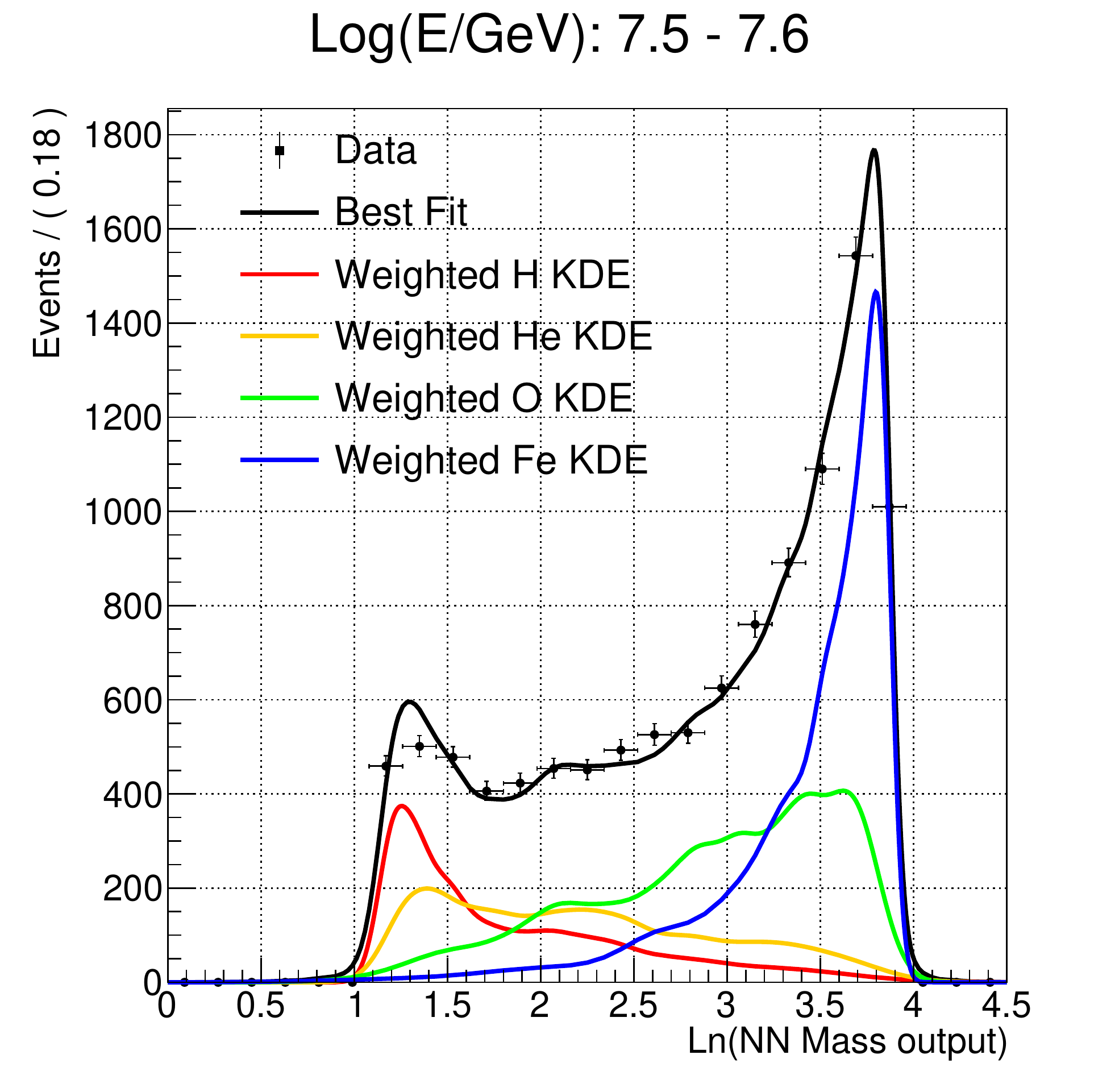}
\includegraphics[width=0.24\textwidth]{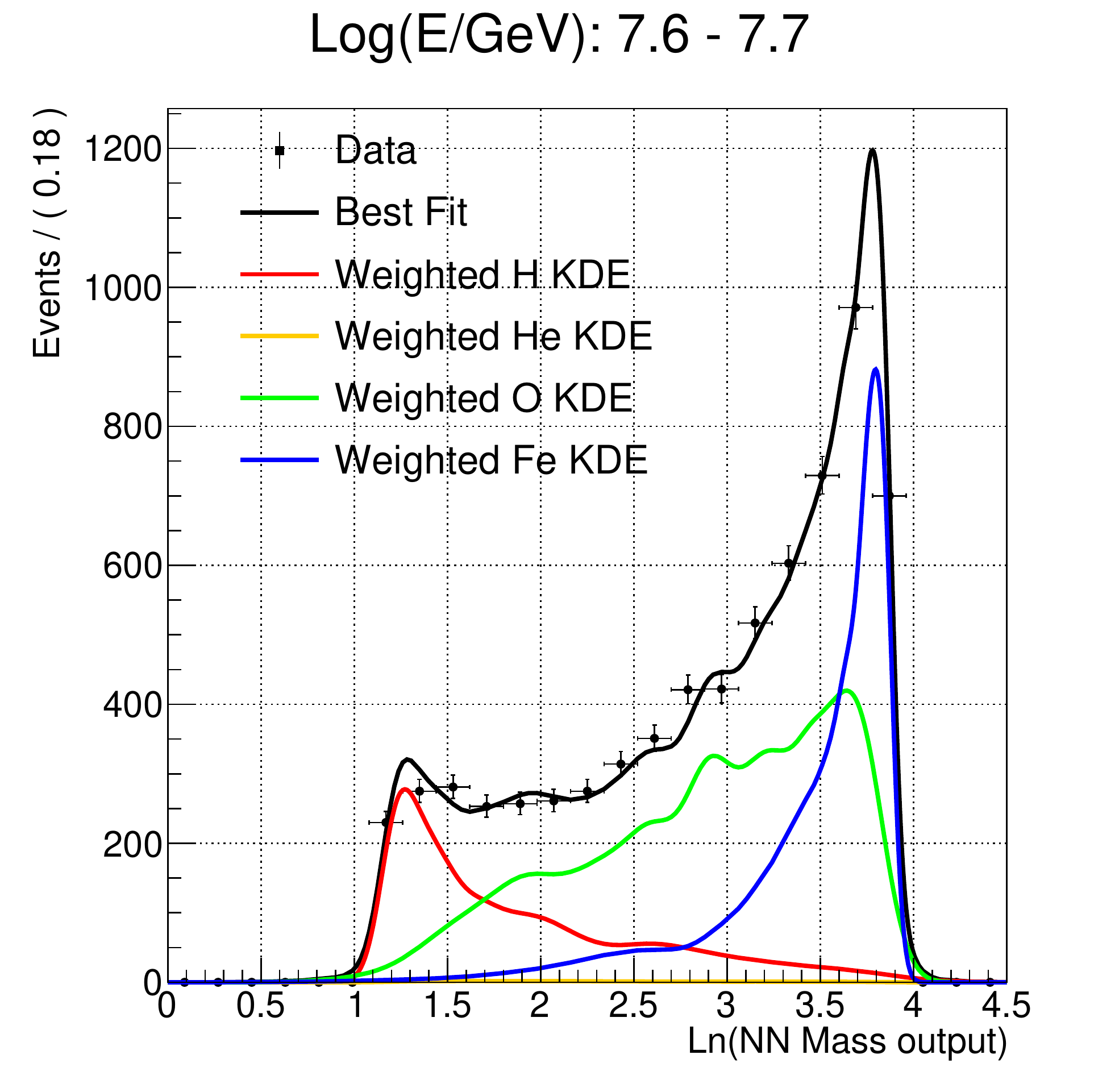}
\includegraphics[width=0.24\textwidth]{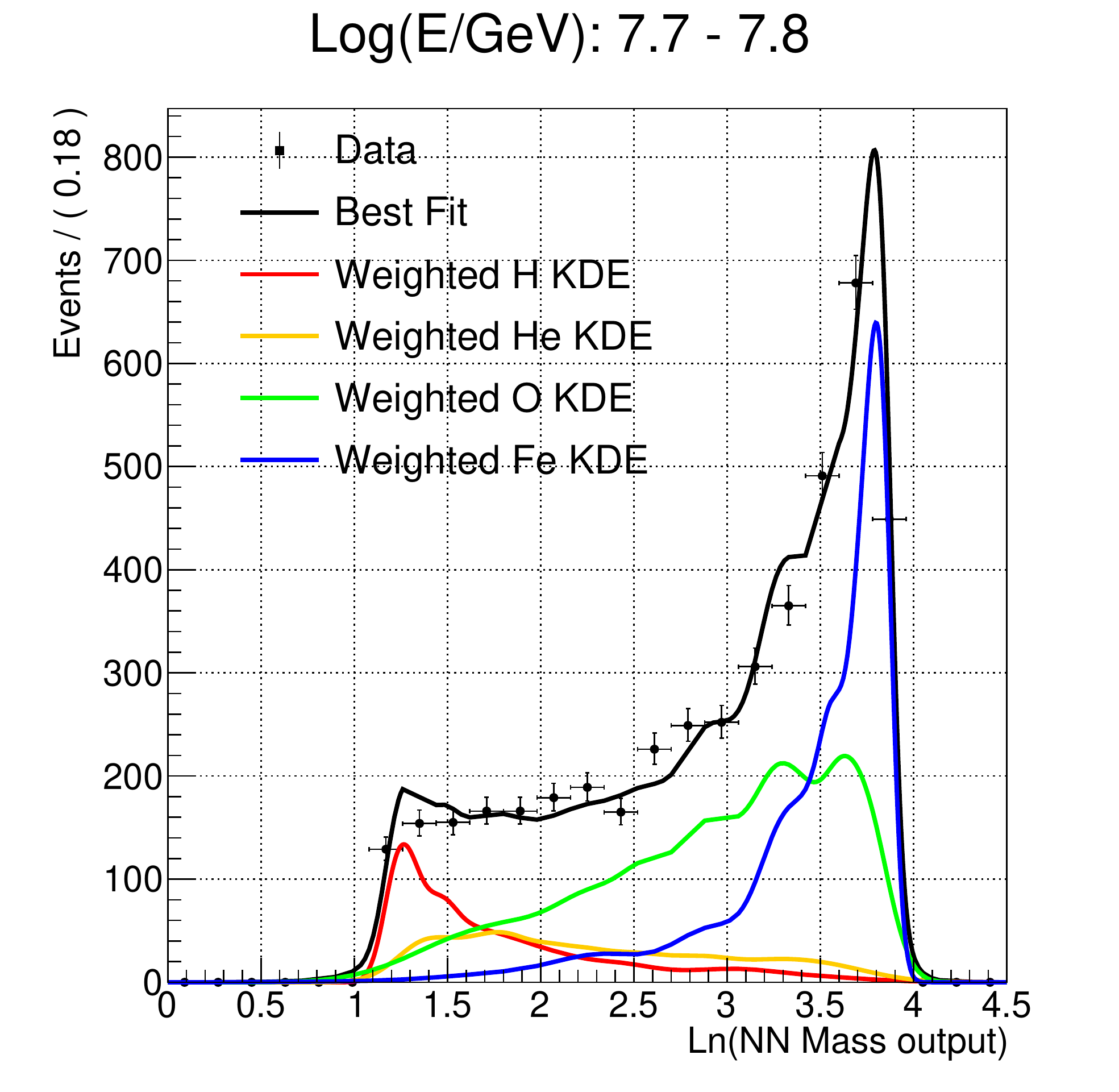}
\includegraphics[width=0.24\textwidth]{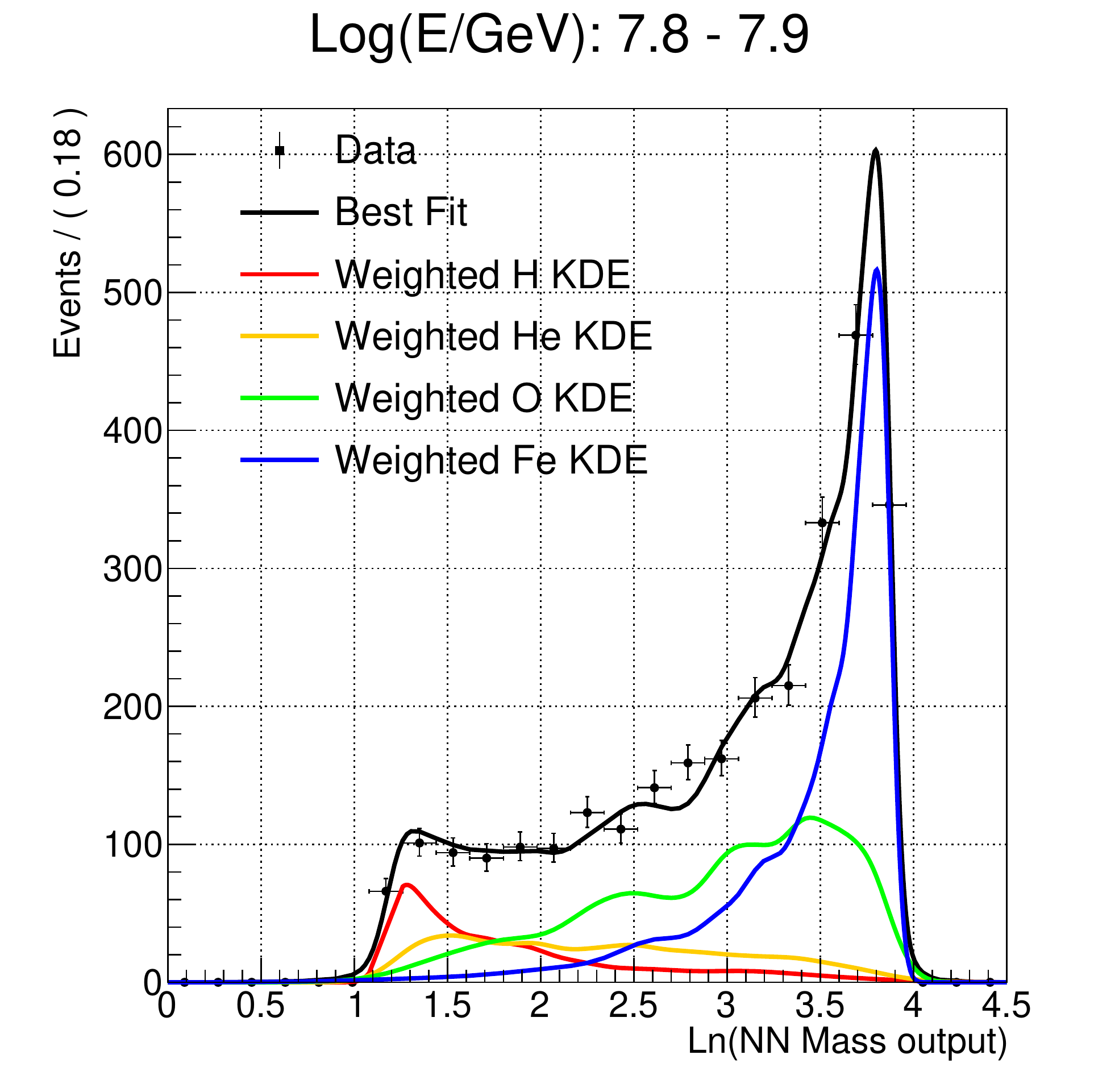}
\includegraphics[width=0.24\textwidth]{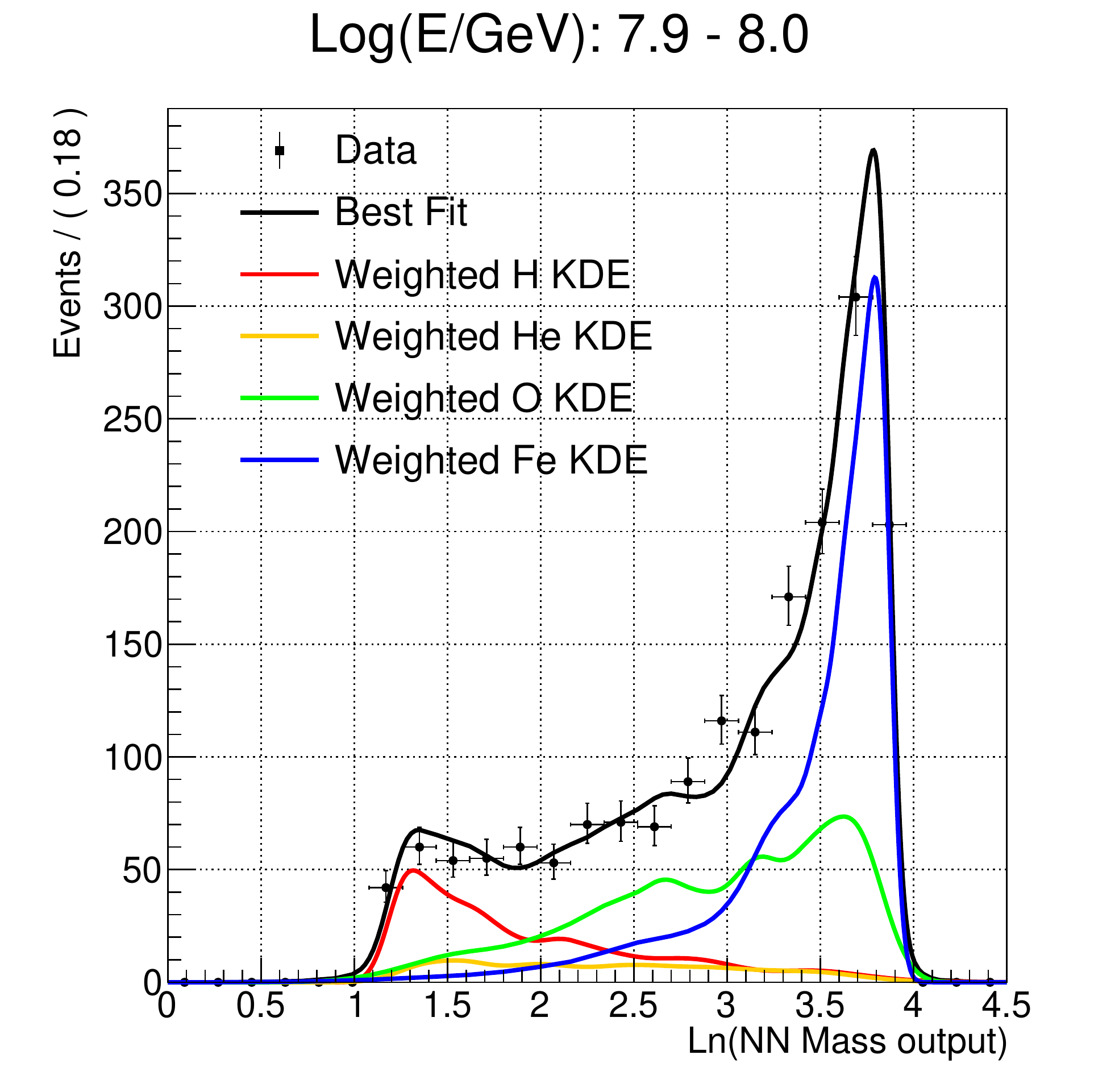}
\includegraphics[width=0.24\textwidth]{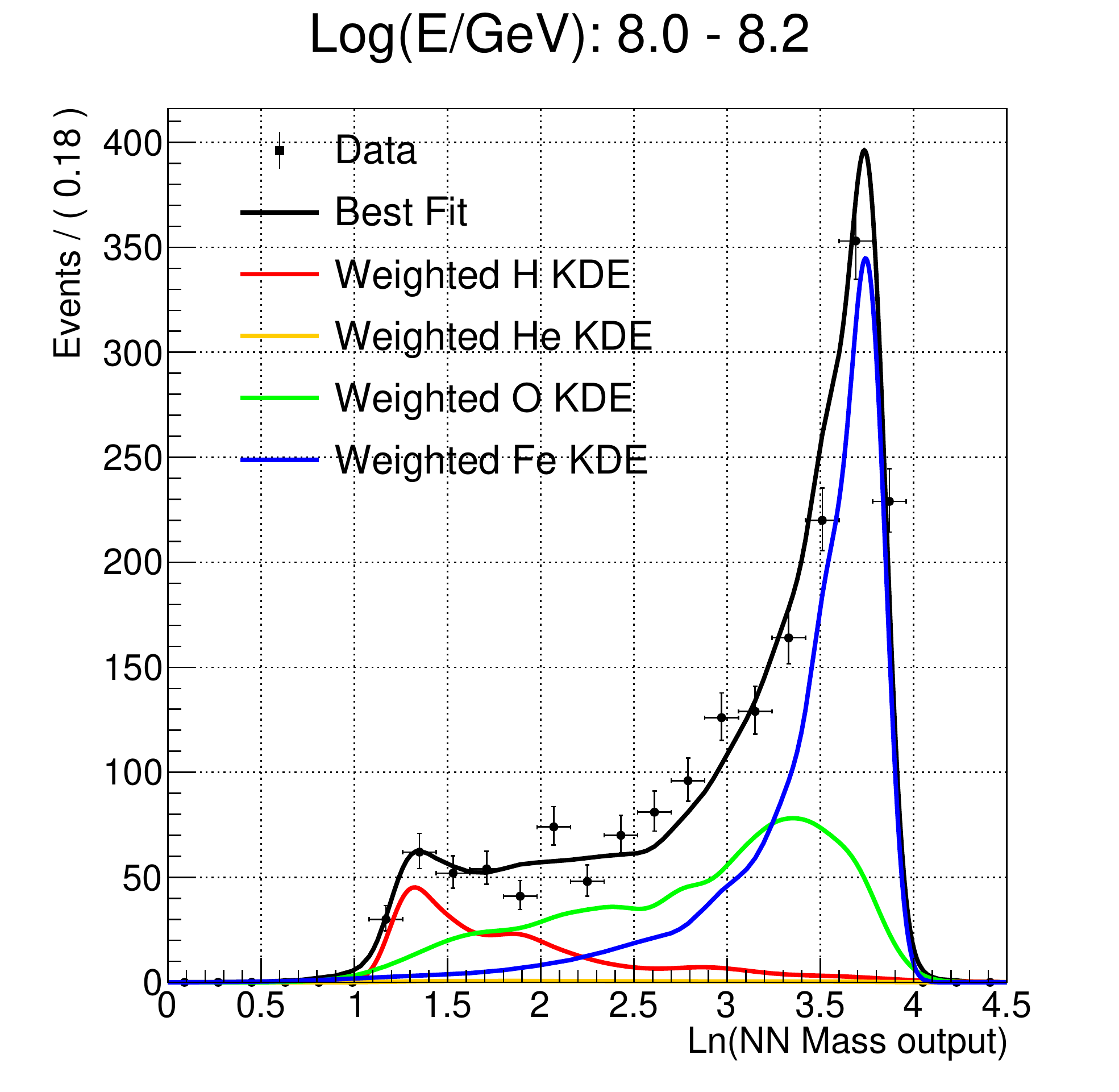}
\includegraphics[width=0.24\textwidth]{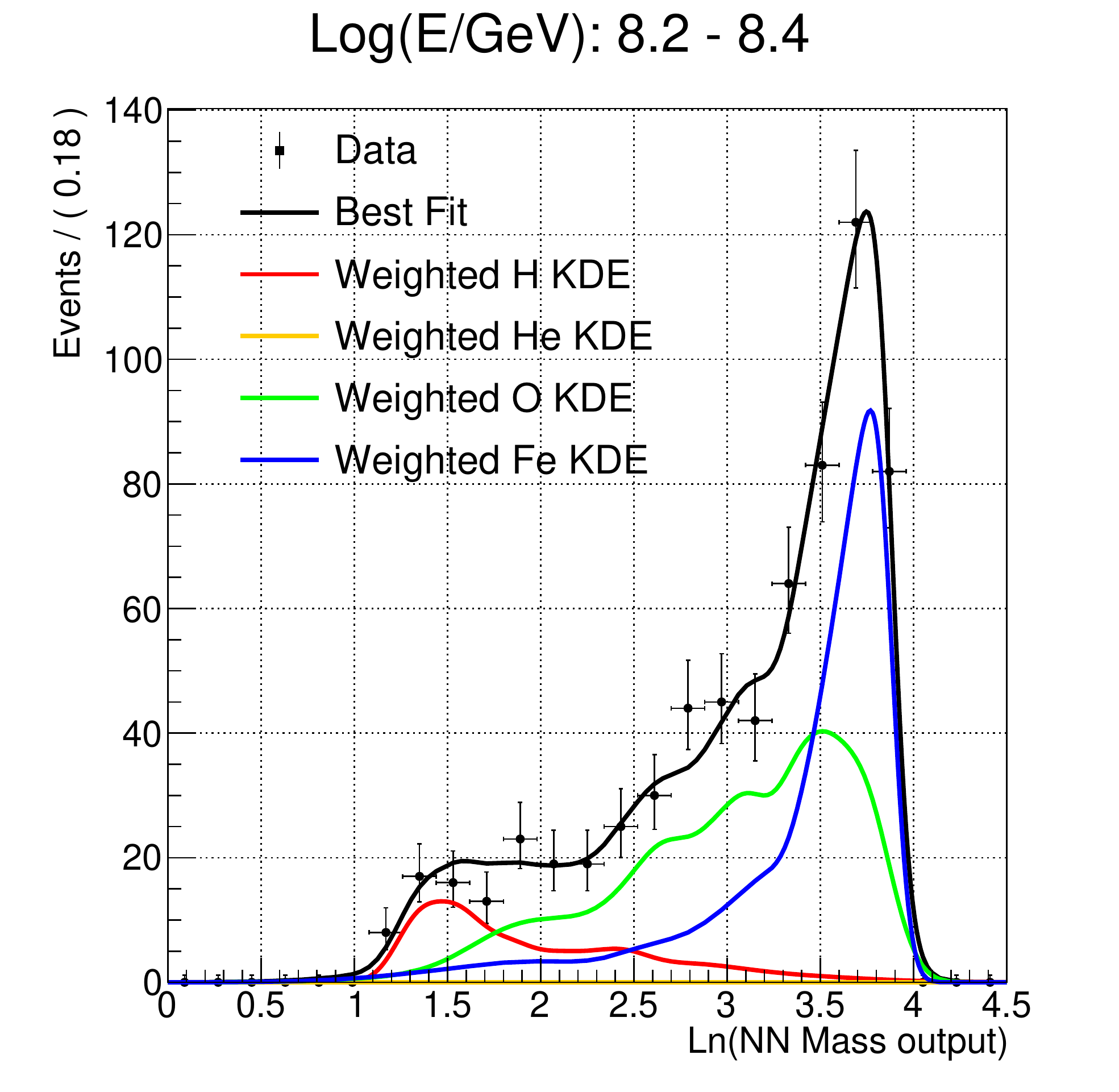}
\includegraphics[width=0.24\textwidth]{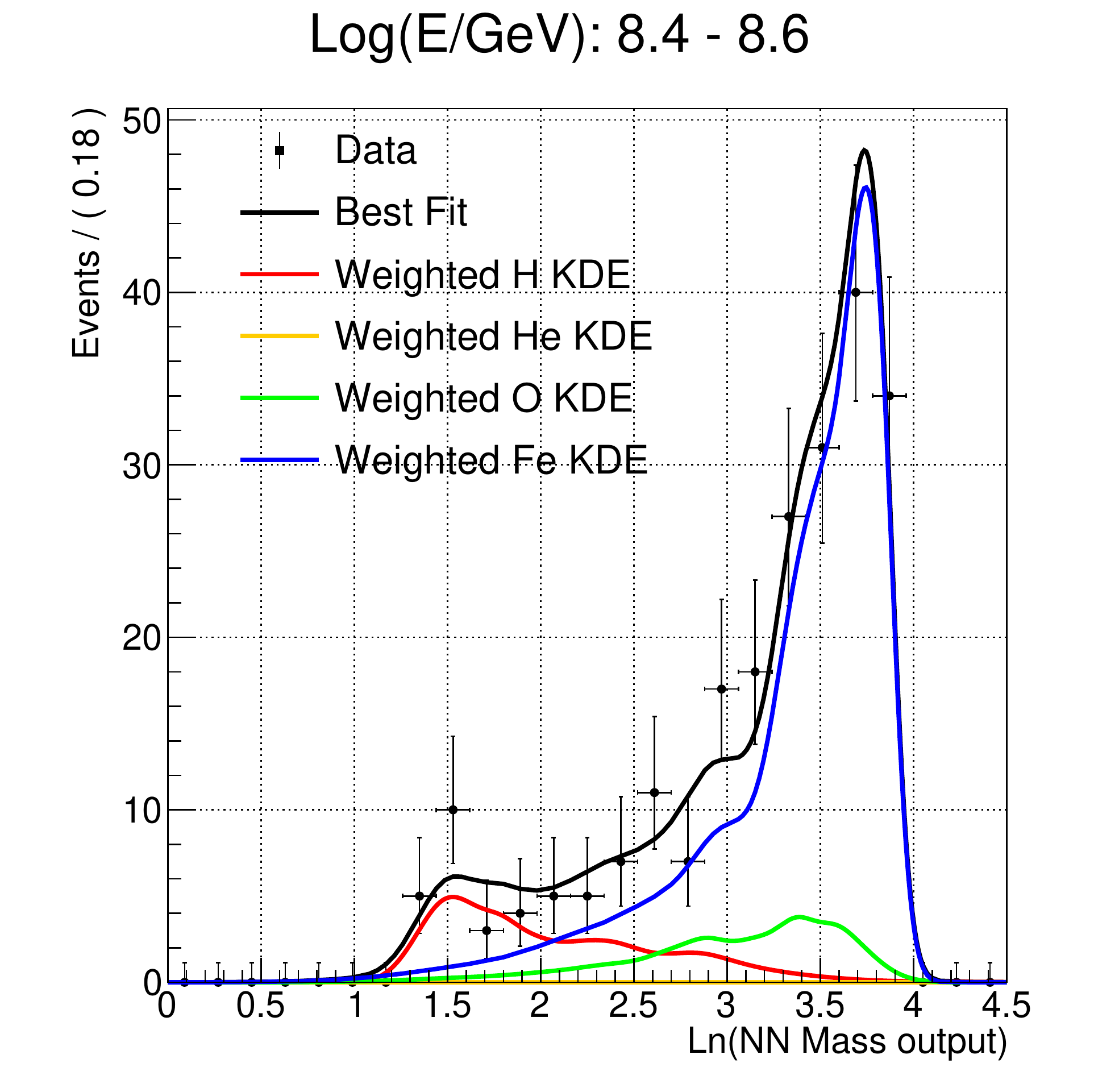}
\includegraphics[width=0.24\textwidth]{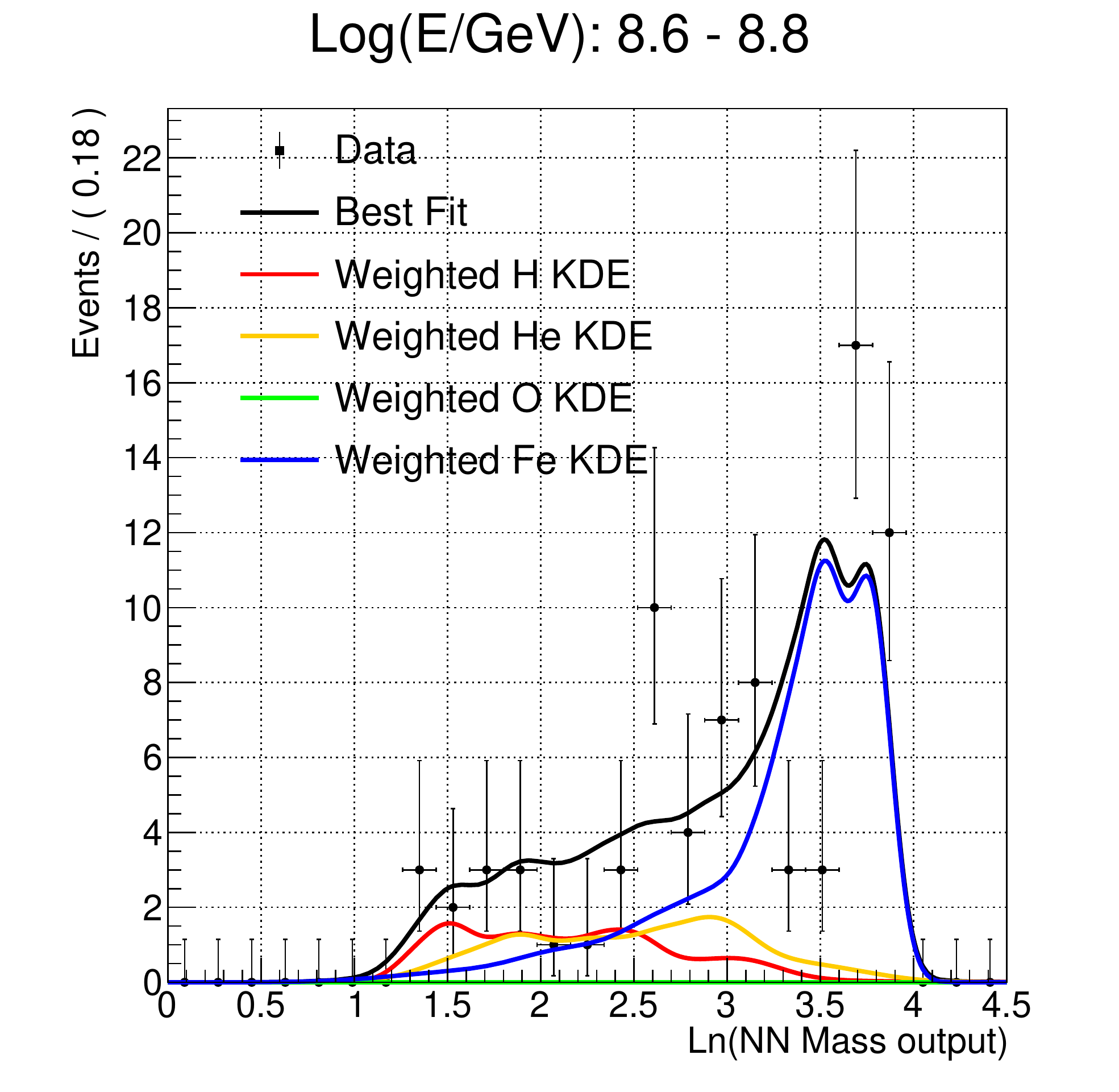}
\includegraphics[width=0.24\textwidth]{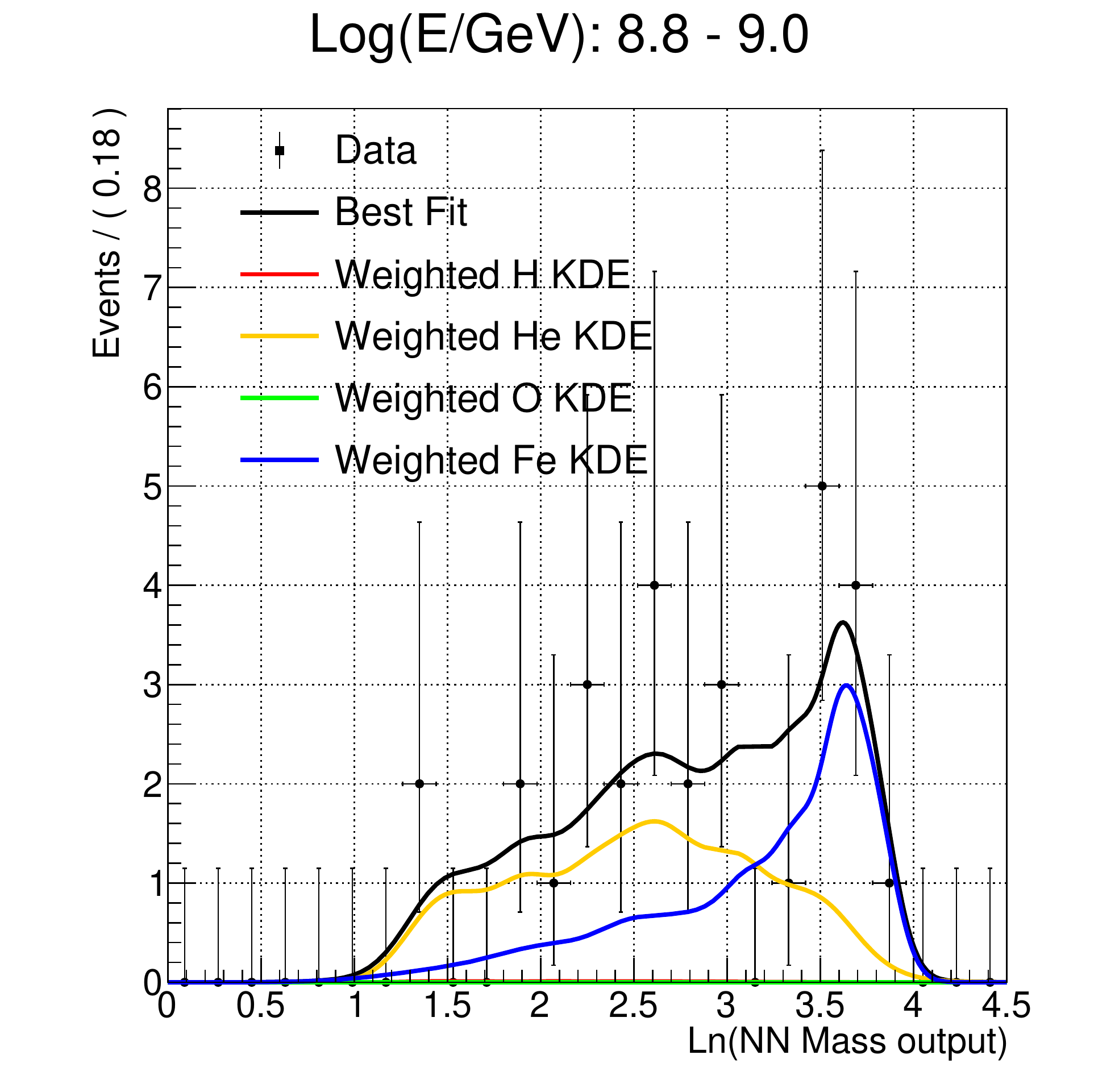}
\caption{Best fit of the experimental data histograms by the KDE templates (derived from simulation and shown independently in Figure \ref{f:nn_templatehisto}) for the coincident analysis.  Energy ranges are labeled in log$_{10}$(E$_{0,reco}$/GeV) in the titles of each figure. The y-axis represents the the number of data events and the solid lines represent the weighted KDE templates for proton (red), helium (orange), oxygen (green) and iron (blue). The solid black line represents best fit distribution. 
}
\label{f:nn_fithisto}
\end{center}
\end{figure*}

As discussed in Section \ref{kde_explan}, the individual templates for each energy bin are shown in Figure \ref{f:nn_templatehisto}, while the fit to the data for each energy bin is then shown in Figure \ref{f:nn_fithisto}.  
Correlation matrices for the fit results for all energy bins are shown in Table \ref{table:appendix_corelation_coef}.

%\clearpage
%\newpage

%\clearpage
%\newpage

%\clearpage
%\newpage

\makeatletter
\renewcommand*\env@matrix[1][r]{\hskip -\arraycolsep
  \let\@ifnextchar\new@ifnextchar
  \array{*\c@MaxMatrixCols #1}}
\makeatother

\begin{table*}[h]
\caption{Correlation coefficients from KDE fits. \label{table:appendix_corelation_coef}}

\begin{equation*}
\begin{array}{l|l}
\text{log(E/GeV)}=6.5-6.6 & \text{log(E/GeV)}=6.6-6.7\\
cor =
\begin{bmatrix}
1.000 & -0.889 & 0.660 & -0.428\\
-0.889 & 1.000 & -0.874 & 0.625\\
0.660 & -0.874 & 1.000 & -0.878\\
-0.428 & 0.625 & -0.878 & 1.000\\
\end{bmatrix} &

cor =
\begin{bmatrix}
1.000 & -0.786 & 0.544 & -0.428\\
-0.786 & 1.000 & -0.888 & 0.755\\
0.544 & -0.888 & 1.000 & -0.938\\
-0.428 & 0.755 & -0.938 & 1.000\\
\end{bmatrix}\\
\hline
\text{log(E/GeV)}=6.7-6.8 & \text{log(E/GeV)}=6.8-6.9\\
cor =
\begin{bmatrix}
1.000 & -0.866 & 0.616 & -0.401\\
-0.866 & 1.000 & -0.865 & 0.619\\
0.616 & -0.865 & 1.000 & -0.862\\
-0.401 & 0.619 & -0.862 & 1.000\\
\end{bmatrix} &

cor =
\begin{bmatrix}
1.000 & -0.877 & 0.656 & -0.455\\
-0.877 & 1.000 & -0.887 & 0.670\\
0.656 & -0.887 & 1.000 & -0.878\\
-0.455 & 0.670 & -0.878 & 1.000\\
\end{bmatrix}\\
\hline
\text{log(E/GeV)}=6.9-7.0 & \text{log(E/GeV)}=7.0-7.1\\
cor =
\begin{bmatrix}
1.000 & -0.778 & 0.466 & -0.254\\
-0.778 & 1.000 & -0.833 & 0.527\\
0.466 & -0.833 & 1.000 & -0.784\\
-0.254 & 0.527 & -0.784 & 1.000\\
\end{bmatrix} &

cor =
\begin{bmatrix}
1.000 & -0.883 & 0.669 & -0.505\\
-0.883 & 1.000 & -0.883 & 0.699\\
0.669 & -0.883 & 1.000 & -0.891\\
-0.505 & 0.699 & -0.891 & 1.000\\
\end{bmatrix}\\
\hline
\text{log(E/GeV)}=7.1-7.2 & \text{log(E/GeV)}=7.2-7.3\\
cor =
\begin{bmatrix}
1.000 & -0.829 & 0.598 & -0.378\\
-0.829 & 1.000 & -0.884 & 0.613\\
0.598 & -0.884 & 1.000 & -0.819\\
-0.378 & 0.613 & -0.819 & 1.000\\
\end{bmatrix} &

cor =
\begin{bmatrix}
1.000 & -0.773 & 0.586 & -0.427\\
-0.773 & 1.000 & -0.913 & 0.717\\
0.586 & -0.913 & 1.000 & -0.874\\
-0.427 & 0.717 & -0.874 & 1.000\\
\end{bmatrix}\\
\hline
\text{log(E/GeV)}=7.3-7.4 & \text{log(E/GeV)}=7.4-7.5\\
cor =
\begin{bmatrix}
1.000 & -0.873 & 0.653 & -0.464\\
-0.873 & 1.000 & -0.881 & 0.669\\
0.653 & -0.881 & 1.000 & -0.870\\
-0.464 & 0.669 & -0.870 & 1.000\\
\end{bmatrix} &

cor =
\begin{bmatrix}
1.000 & -0.889 & 0.660 & -0.416\\
-0.889 & 1.000 & -0.866 & 0.587\\
0.660 & -0.866 & 1.000 & -0.803\\
-0.416 & 0.587 & -0.803 & 1.000\\
\end{bmatrix}\\
\hline
\text{log(E/GeV)}=7.5-7.6 & \text{log(E/GeV)}=7.6-7.7\\
cor =
\begin{bmatrix}
1.000 & -0.885 & 0.556 & -0.283\\
-0.885 & 1.000 & -0.782 & 0.440\\
0.556 & -0.782 & 1.000 & -0.748\\
-0.283 & 0.440 & -0.748 & 1.000\\
\end{bmatrix} &

cor =
\begin{bmatrix}
1.000 & -0.688 & 0.166 & -0.014\\
-0.688 & 1.000 & -0.647 & 0.279\\
0.166 & -0.647 & 1.000 & -0.679\\
-0.014 & 0.279 & -0.679 & 1.000\\
\end{bmatrix}\\
\hline
\text{log(E/GeV)}=7.7-7.8 & \text{log(E/GeV)}=7.8-7.9\\
cor =
\begin{bmatrix}
1.000 & -0.789 & 0.438 & -0.189\\
-0.789 & 1.000 & -0.769 & 0.401\\
0.438 & -0.769 & 1.000 & -0.717\\
-0.189 & 0.401 & -0.717 & 1.000\\
\end{bmatrix} &

cor =
\begin{bmatrix}
1.000 & -0.856 & 0.596 & -0.296\\
-0.856 & 1.000 & -0.840 & 0.461\\
0.596 & -0.840 & 1.000 & -0.712\\
-0.296 & 0.461 & -0.712 & 1.000\\
\end{bmatrix}\\

\hline
\text{log(E/GeV)}=7.9-8.0 & \text{log(E/GeV)}=8.0-8.2\\
cor =
\begin{bmatrix}
1.000 & -0.895 & 0.657 & -0.386\\
-0.895 & 1.000 & -0.866 & 0.555\\
0.657 & -0.866 & 1.000 & -0.784\\
-0.386 & 0.555 & -0.784 & 1.000\\
\end{bmatrix} &

cor =
\begin{bmatrix}
1.000 & -0.716 & 0.289 & -0.035\\
-0.716 & 1.000 & -0.764 & 0.318\\
0.289 & -0.764 & 1.000 & -0.649\\
-0.035 & 0.318 & -0.649 & 1.000\\
\end{bmatrix}\\
\hline
\text{log(E/GeV)}=8.2-8.4 & \text{log(E/GeV)}=8.4-8.6\\
cor =
\begin{bmatrix}
1.000 & -0.013 & -0.419 & 0.186\\
-0.013 & 1.000 & -0.012 & 0.005\\
-0.419 & -0.012 & 1.000 & -0.727\\
0.186 & 0.005 & -0.727 & 1.000\\
\end{bmatrix} &

cor =
\begin{bmatrix}
1.000 & -0.029 & -0.406 & 0.188\\
-0.029 & 1.000 & -0.023 & 0.009\\
-0.406 & -0.023 & 1.000 & -0.790\\
0.188 & 0.009 & -0.790 & 1.000\\
\end{bmatrix}\\
\hline
\text{log(E/GeV)}=8.6-8.8 & \text{log(E/GeV)}=8.8-9.0\\
cor =
\begin{bmatrix}
1.000 & -0.835 & 0.007 & 0.254\\
-0.835 & 1.000 & -0.027 & -0.469\\
0.007 & -0.027 & 1.000 & -0.013\\
0.254 & -0.469 & -0.013 & 1.000\\
\end{bmatrix} &

cor =
\begin{bmatrix}
1.000 & -0.659 & 0.005 & 0.188\\
-0.659 & 1.000 & -0.031 & -0.570\\
0.005 & -0.031 & 1.000 & -0.024\\
0.188 & -0.570 & -0.024 & 1.000\\
\end{bmatrix}\\

\end{array}
\end{equation*}

\end{table*}

\end{document}

%% file: define.tex
\newcommand{\fix}[1]{}%{\textbf{\textcolor{red}{#1}}}
\newcommand{\fixka}[1]{}%{\textbf{\textcolor{red}{#1}}}
\newcommand{\internalcite}[1]{}%{\texttt{\textcolor{green}{(from #1)}}}
\newcommand{\fixed}[1]{}%#1} %don't show comments which have been addressed.  Uncomment to show all comments

\newcommand{\be}{\begin{equation}} 
\newcommand{\ee}{\end{equation}}
\newcommand{\bea}{\begin{eqnarray}} 
\newcommand{\eea}{\end{eqnarray}}

\newcommand{\bc}{\begin{centering}}
\newcommand{\ec}{\end{centering}}

\newcommand{\lna}{$\langle\ln A\rangle$}
\newcommand{\s }{$S_{125}$}
\newcommand{\logs }{$\log_{10}(S_{125})$}
\newcommand{\lte}{$\log_{10}(E/\mathrm{GeV})$}